\documentclass[10pt,journal]{IEEEtran}
\newif\ifonecolumn
\IEEEoverridecommandlockouts

\usepackage[T1]{fontenc}
\usepackage[utf8]{inputenc}
\usepackage{bm}
\usepackage{balance}
\usepackage{comment}
\usepackage{datetime}    

\usepackage{booktabs}  
\usepackage{tabularx}  

\usepackage{graphicx}
\usepackage{url} 

\usepackage{amsmath}  
\usepackage{amsfonts} 
\usepackage{amsthm}   
\usepackage{amsbsy}   
\usepackage{amssymb}  
\usepackage{csquotes} 

\usepackage{cite}
\usepackage{algorithm}
\usepackage{algpseudocode}
\usepackage{pgfplots}

\algnewcommand{\FForAll}[1]{\State\algorithmicforall\ #1\ \algorithmicdo}
\algnewcommand{\EEndFor}{\unskip\ \algorithmicend\ \algorithmicfor}

\usepackage[caption=false,font=footnotesize,subrefformat=parens,labelformat=parens]{subfig}

\usepackage{tikz}

\usetikzlibrary{plotmarks}
\pgfplotsset{compat=newest}
\pgfplotsset{plot coordinates/math parser=false}
\usetikzlibrary{matrix,calc,backgrounds}
\usetikzlibrary{math}

\newcommand{\sss}[1]{{\scriptstyle #1}}

\usepackage{nicefrac}   

\makeatletter
\def\squarecorner#1{
    \pgf@x=\the\wd\pgfnodeparttextbox%
    \pgfmathsetlength\pgf@xc{\pgfkeysvalueof{/pgf/inner xsep}}%
    \advance\pgf@x by 2\pgf@xc%
    \pgfmathsetlength\pgf@xb{\pgfkeysvalueof{/pgf/minimum width}}%
    \ifdim\pgf@x<\pgf@xb%
    \pgf@x=\pgf@xb%
    \fi%
    \pgf@y=\ht\pgfnodeparttextbox%
    \advance\pgf@y by\dp\pgfnodeparttextbox%
    \pgfmathsetlength\pgf@yc{\pgfkeysvalueof{/pgf/inner ysep}}%
    \advance\pgf@y by 2\pgf@yc%
    \pgfmathsetlength\pgf@yb{\pgfkeysvalueof{/pgf/minimum height}}%
    \ifdim\pgf@y<\pgf@yb%
    \pgf@y=\pgf@yb%
    \fi%
    \ifdim\pgf@x<\pgf@y%
    \pgf@x=\pgf@y%
    \else
    \pgf@y=\pgf@x%
    \fi
    \pgf@x=#1.5\pgf@x%
    \advance\pgf@x by.5\wd\pgfnodeparttextbox%
    \pgfmathsetlength\pgf@xa{\pgfkeysvalueof{/pgf/outer xsep}}%
    \advance\pgf@x by#1\pgf@xa%
    \pgf@y=#1.5\pgf@y%
    \advance\pgf@y by-.5\dp\pgfnodeparttextbox%
    \advance\pgf@y by.5\ht\pgfnodeparttextbox%
    \pgfmathsetlength\pgf@ya{\pgfkeysvalueof{/pgf/outer ysep}}%
    \advance\pgf@y by#1\pgf@ya%
}
\makeatother

\pgfdeclareshape{square}{
    \savedanchor\northeast{\squarecorner{}}
    \savedanchor\southwest{\squarecorner{-}}
    
    \inheritanchor[from=rectangle]{base east}
    \inheritanchor[from=rectangle]{base west}
    \inheritanchor[from=rectangle]{base}
    \inheritanchor[from=rectangle]{center}
    \inheritanchor[from=rectangle]{east}
    \inheritanchor[from=rectangle]{mid east}
    \inheritanchor[from=rectangle]{mid west}
    \inheritanchor[from=rectangle]{mid}
    \inheritanchor[from=rectangle]{north east}
    \inheritanchor[from=rectangle]{north west}
    \inheritanchor[from=rectangle]{north}
    \inheritanchor[from=rectangle]{south east}
    \inheritanchor[from=rectangle]{south west}
    \inheritanchor[from=rectangle]{south}
    \inheritanchor[from=rectangle]{west}
    \inheritanchorborder[from=rectangle]
    \inheritbackgroundpath[from=rectangle]
}

\newcommand{\Reals}{\mathbb{R}}
\renewcommand{\vec}[1]{\bm {#1}}
\newcommand{\Neighbours}[2][]{\mathcal{N}_{#1}\left( #2 \right)}
\newcommand{\ZZ}{\mathbb{Z}}
\newcommand{\FF}{\mathbb{F}}
\DeclareMathOperator{\supp}{supp}
\DeclareMathOperator{\IPA}{IPA}
\DeclareMathOperator{\coeff}{coef}

\newtheorem{thm}{Theorem}
\newtheorem{lemma}{Lemma}
\newtheorem{cor}{Corollary}

\newtheorem{define}{Definition}

\newtheorem{example}{Example}

\newcommand{\revone}[1]{#1}
\newcommand{\revtwo}[1]{#1}
\newcommand{\revus}[1]{#1}

\usepackage[normalem]{ulem}
\newcommand{\jjdel}[1]{}

\begin{document}

\title{Failure Analysis of the Interval-Passing Algorithm for Compressed Sensing}

\author{Yauhen Yakimenka and
  Eirik Rosnes,~\IEEEmembership{Senior~Member,~IEEE}
  \thanks{This work was partially funded by the Norwegian-Estonian Research Cooperation Programme (grant EMP133). The work of E.\ Rosnes was partially funded by the Research Council of Norway (grant 240985/F20). The calculations were carried out in part in the High Performance Computing Centre of the University of Tartu. This paper was presented in part at the 54th Annual Allerton Conference on Communication, Control, and Computing, Monticello, IL, USA, September 2016.}%
    \thanks{Y.\ Yakimenka was with Institute of Computer Science, University of Tartu, Tartu 50409, Estonia. He is now with Simula UiB, N-5020 Bergen, Norway (e-mail: yauhen@simula.no).}
  \thanks{E.\ Rosnes is with Simula UiB, N-5020 Bergen, Norway (e-mail: eirikrosnes@simula.no).} }%

%

\maketitle

\begin{abstract}

In this work, we perform a complete failure analysis of the interval-passing algorithm (IPA) for compressed sensing. The IPA is an efficient iterative algorithm for reconstructing a $k$-sparse nonnegative $n$-dimensional real signal $\vec{x}$ from a small number of linear measurements $\vec{y}$. In particular, we show that the IPA fails to recover $\vec{x}$ from $\vec{y}$ if and only if it fails to recover a corresponding binary vector of the same support, and also that only positions of nonzero values in the measurement matrix are of importance to the success of recovery. Based on this observation, we introduce \emph{termatiko sets} and show that the IPA fails to fully recover $\vec x$ if and only if the support of $\vec x$ contains a nonempty termatiko set, thus giving a complete (graph-theoretic) description of the failing sets of the IPA. Two heuristics to locate small-size termatiko sets are presented. For binary column-regular measurement matrices with no $4$-cycles, we provide a lower bound on the \emph{termatiko distance}, defined as the smallest size of a nonempty termatiko set. For measurement matrices constructed from the parity-check matrices of array low-density parity-check codes, upper bounds on the termatiko distance equal to half the best known upper bound on the minimum distance are provided for column-weight at most $7$, while for column-weight $3$, the exact termatiko distance and its corresponding multiplicity are provided. Next, we show that adding redundant rows to the measurement matrix does not create new termatiko sets, but rather potentially removes termatiko sets and thus improves performance. An algorithm is provided to efficiently search for such redundant rows.  Finally, we present numerical results for different specific measurement matrices and also for protograph-based ensembles of measurement matrices, as well as simulation results of IPA performance, showing the influence of small-size termatiko sets.
\end{abstract}


\section{Introduction}\label{sec:intro}

\IEEEPARstart{T}{he} reconstruction of a (mathematical) object from a partial set of observations in an efficient and reliable manner is of fundamental importance. Compressed sensing, motivated by the ground-breaking work of Cand\`{e}s and Tao \cite{can06,can05}, and independently by Donoho \cite{don06}, is a research area in which the object to be reconstructed is a $k$-sparse signal vector (there are at most $k$ nonzero entries in the vector) over the real numbers. The partial information provided is a linear transformation of the signal vector, the \emph{measurement vector}, and the objective is to reconstruct the object from a small number of measurements. Compressed sensing provides a mathematical framework which shows that, under some conditions, signals can be recovered from far fewer measurements than with conventional signal acquisition methods. The main idea in compressed sensing is to exploit that most interesting signals have an inherent structure or contain redundancy. The compressed sensing problem is described in more details in Section~\ref{sec:CS} below.


\revtwo{Iterative reconstruction algorithms for compressed sensing have been considered, for instance,  in \cite{zha12,RDVD12,cha10,sar06_1,pha09,Don13,don09_1}  and references therein. 
Among those, 
the interval-passing algorithm (IPA) \cite{cha10} is a low-complexity reconstruction algorithm for reconstructing nonnegative sparse signals with binary measurement matrices. The extension to nonnegative real measurement matrices was considered in \cite{RDVD12}.  In \cite{wu13}, Wu and Yang proposed to use the concept of \emph{verification}  \cite{sar06_1} to enhance reconstruction performance, and they showed 
that the enhanced algorithm performs better than the plain IPA and also better than the plain verification algorithm  for measurement matrices equal to parity-check matrices of low-density parity-check (LDPC) codes.  As a side note, there is a clear connection between the IPA and the iterative message-passing algorithm proposed for counter braids in \cite{lu08} (see also \cite{ros18}). A counter braid is a counter architecture introduced by Lu \emph{et al.} in  \cite{lu08} for per-flow measurements on high-speed links. In fact, it can easily be seen that the decoding algorithm for counter braids is a special case of the IPA (see Section~\ref{sec:ipa} below). Thus, the results derived in this work apply immediately also to iterative decoding of counter braids as described in \cite{lu08}.}

In this work, we show that the IPA fails for a nonnegative signal $\vec{x} =(x_1,\dotsc,x_n) \in \Reals_{\geq 0}^n$, $\Reals_{\geq 0}$ is the set of nonnegative real numbers, if and only if it fails for a corresponding binary vector $\vec{z}$ of the same support,  and also that only positions of nonzero values in the measurement matrix are of importance to the success of recovery. Thus, failing sets as subsets of $[n] \triangleq \{1,\dotsc,n\}$ can be defined. It has previously been shown that traditional stopping sets for belief propagation decoding of LDPC codes are failing sets of the IPA, in the sense that if the support of a signal $\vec x \in \Reals_{\geq 0}^n$ contains a nonempty stopping set, then the IPA fails to fully recover $\vec x$ \cite[Thm.~1]{RDVD12}. In this work, we extend the results in \cite{RDVD12} and define \emph{termatiko sets} (which contain stopping sets as a special case) and show that the IPA fails to fully recover a signal $\vec x \in \Reals_{\geq 0}^n$  if and only if the support of $\vec x$ contains a nonempty termatiko set, thus giving a complete (graph-theoretic) description of the failing sets of the IPA. Analogously to the stopping distance, we define the size of the smallest nonempty termatiko set as the termatiko distance. Also, two heuristics to locate small-size termatiko sets are presented.  For binary column-regular matrices with no $4$-cycles we provide a general lower bound on the termatiko distance, and for matrices equal to parity-check matrices of array LDPC codes \cite{fan00} we provide an upper bound equal to half the best known upper bound on the minimum distance for column-weight at most $7$. In the special case of column-weight $3$, the termatiko distance turns out to be exactly $3$ and a formula for the corresponding multiplicity is derived. Adding redundant rows to improve performance of iterative message-passing algorithms has been considered previously in various scenarios, and we provide an algorithm to search for redundant rows of the measurement matrix and show that this can only reduce the number of termatiko sets. Finally, we perform an extensive numerical study which includes both specific binary parity-check matrices of LDPC codes and parity-check matrices from LDPC code ensembles as measurement matrices, as well as simulation results.


The remainder of this paper is organized as follows. Notation and background, including a detailed description of the IPA, are introduced in Section~\ref{sec:notation}, while the failing sets of the IPA are analyzed in Section~\ref{sec:failing_sets}, introducing the concept of termatiko sets and showing that the IPA fails to recover a nonnegative real signal $\vec x \in \Reals^n_{\geq 0}$ if and only if the support of $\vec x$ contains a nonempty termatiko set. Two heuristics to identify small-size termatiko sets are also presented. 
In Section~\ref{sec:col-reg},  a lower bound on the termatiko distance  for column-regular measurement matrices is presented. Next, the exact termatiko distance and a formula for its multiplicity for binary measurement matrices obtained from the parity-check matrices of column-weight $3$ array LDPC codes are derived. For column-weights $4$ to $7$, upper bounds on the termatiko distance of these measurement matrices are presented by splitting minimum-weight codewords into two equal parts. Adding redundant rows to the measurement matrix in order to improve the performance of the IPA is considered in Section~\ref{sec:add-red-rows}. Numerical results for different specific measurement matrices and also for ensembles of measurement matrices, as well as simulation results of IPA performance are presented in Section~\ref{sec:numerical_results}. Conclusions are drawn in Section~\ref{sec:conclu}.

\section{Notation and Background} \label{sec:notation}
In this section, we introduce the problem formulation, revise notation from \cite{RDVD12}, and describe the IPA in detail. 

\revtwo{\subsection{Notation}}
\revtwo{We denote the set difference of two arbitrary sets $N$ and $M$ by $N \setminus M$. We also use $N_1 \setminus N_2 \setminus \dotsb \setminus N_t$ as a shorthand for $(\dotsb((N_1 \setminus N_2) \setminus N_3) \setminus \dotsb \setminus N_t)$.} \revtwo{The \emph{support} of a vector $\vec x  \in \Reals^n$,  where $\Reals$ is the field of real numbers, is the set of its nonzero coordinates, i.e., 
		\[
		\supp(\vec x) = \left\{ i \mid x_i \neq 0 \right\}.
		\]
		
	We also define the $\ell_0$- and $\ell_1$-norms as follows:
	\begin{align*}
		\lVert \vec x \rVert_0 &= \lvert \supp(\vec x) \rvert \text{ and }  
		\lVert \vec x \rVert_1 = \sum_i \lvert x_i \rvert.
	\end{align*}
	Note that the $\ell_0$-norm (as defined here) is not a norm in a strict mathematical sense.}


\subsection{Compressed Sensing} \label{sec:CS}
	Let $\vec{x} \in \Reals^n$ be an $n$-dimensional $k$-sparse signal (i.e., it has at most $k$ nonzero entries), and let $A = (a_{ji}) \in \Reals^{m \times n}$  be an  $m \times n$ real measurement matrix. We consider the recovery of $\vec{x}$ from 
	measurements $\vec{y} = A \vec{x} \in \Reals^m$, where $m < n$ and $k < n$.

	The reconstruction problem of compressed sensing is to find the sparsest $\vec{x}$ (or the one that minimizes the $\ell_0$-norm) under the constraint $\vec{y} = A \vec{x}$, which in general is \revtwo{an NP-hard problem \cite{donoho2006most,natarajan1995sparse}.} Basis pursuit is an algorithm which reconstructs $\vec{x}$ by minimizing its $\ell_1$-norm under the constraint $\vec{y} = A \vec{x}$ \cite{can05}. This is a linear program, and thus it can be solved in polynomial time. The algorithm has a remarkable performance, but its complexity is high, making it impractical for many applications that require fast reconstruction. A fast reconstruction algorithm for nonnegative real signals and measurement matrices is the IPA which is described below in Section~\ref{sec:ipa}.

\subsection{Tanner Graph Representation}\label{sec:Tanner-graph-notation}
    We associate with matrix $A$ the bipartite Tanner graph $G = (V \cup C, 
    E)$, where $V = \{ v_1, v_2, \dotsc, v_n \}$ is a set of \emph{variable 
    nodes}, $C = \{ c_1, c_2, \dotsc, c_m \}$ is a set of \emph{measurement 
    nodes}, and $E$ is a set of edges from $C$ to $V$. We will often equate $V$ with $[n]$ and $C$ with $[m]$. There is an edge in $E$ between $c \in C$ and $v \in V$ if and only if 
    $a_{cv} \neq 0$. We also denote the sets of neighbors for each node $v \in V$ and $c \in C$ as
    \begin{align*}
        &\Neighbours{v} = \{ c \in C \mid (c, v) \in E \}\revus{\,,} \\
        &\Neighbours{c} = \{ v \in V \mid (c, v) \in E \} \,,
    \end{align*}
respectively.    Furthermore, if $T \subset V$ or $T \subset C$ and $w \in V \cup C$, then define
    \begin{displaymath}
        \Neighbours{T}    = \bigcup_{t \in T} \Neighbours{t} \text{ and }
        \Neighbours[T]{w} = \Neighbours{w} \cap T \,.
    \end{displaymath}
A \emph{stopping set} \cite{di02} of the Tanner graph $G$ is defined as a subset $S$ of $V$ such that all its neighboring measurement nodes are connected at least twice to $S$. 

\subsection{Interval-Passing Algorithm} \label{sec:ipa}
The IPA is an iterative algorithm to reconstruct a nonnegative real signal $\vec x \in \Reals_{\geq 0}^n$ from a set of linear measurements $\vec y = A \vec x$, introduced by Chandar \emph{et al.} in \cite{cha10} for binary measurement matrices.  The algorithm was extended to nonnegative real  measurement matrices in  \cite{RDVD12}, and this is the case that we will consider. The IPA iteratively sends messages between variable and measurement nodes. Each message contains two real numbers, a \emph{lower bound} and an \emph{upper bound} on the value of the variable node to which it is affiliated. Let $\mu^{(\ell)}_{v \to c}$ (resp.\ $\mu^{(\ell)}_{c \to v}$) denote the lower bound of the message from variable node $v$ (resp.\ measurement node $c$) to measurement node $c$ (resp.\ variable node $v$) at iteration $\ell$.  The corresponding upper bound of the message is denoted by  $M^{(\ell)}_{v \to c}$ (resp.\ $M^{(\ell)}_{c \to v}$). It is a distinct property of the algorithm that at any iteration $\ell$, $\mu^{(\ell)}_{v \to c} \leq x_v \leq M^{(\ell)}_{v \to c}$ and $\mu^{(\ell)}_{c \to v} \leq x_v \leq M^{(\ell)}_{c \to v}$, for all $v \in V$ and $c \in \Neighbours{v}$. Also, the messages from variable to measurement nodes, $\mu^{(\ell)}_{v \to c}$ and $M^{(\ell)}_{v \to c}$, are independent of $c \in \Neighbours v$. Thus, we will often denote $\mu^{(\ell)}_{v \to c}$  by $\mu^{(\ell)}_{v \to \cdot}$ and $M^{(\ell)}_{v \to c}$ by $M^{(\ell)}_{v \to \cdot}$.

The detailed steps of the IPA are shown in Algorithm~\ref{alg:ipa} below, where we denote by IPA($\vec y$, $A$) the output of the algorithm, $\hat{\vec x}$, when provided with inputs $\vec y$ and $A$.

\revtwo{A counter braid is a  counter architecture  for per-flow measurements on high-speed links introduced in \cite{lu08}. Counter braids address the problem of cheap high-speed memory-efficient approximate counting. In particular, a single-layer counter braid can be represented by a bipartite graph with flow nodes and counter nodes. When a flow is encountered (for instance, on a high-speed link), all counter nodes connected to the flow node representing the encountered flow are incremented. The decoding operation tries to recover  the flow sizes (the values of the flow nodes or the number of observed flows of different types) from the values of the counter nodes, and this can be achieved using message passing where upper and lower bounds on the flow sizes are passed in an iterative manner on the bipartite graph representing the counter braid. See \cite[Exhibit~2]{lu08} for further details.

It can readily be seen that in the special case when setting $M^{(0)}_{v \to \cdot}=\infty$ for all $v \in V$, the IPA reduces to the iterative decoding algorithm outlined in \cite[Exhibit~2]{lu08} for counter braids with one layer. In fact, due to this initialization, only upper bounds need to be computed for odd iterations and only lower bounds for even iterations (for both \revtwo{variable/flow} nodes and measurement/counter nodes). This is the case since either the upper bound (for even iterations) or the lower bound (for odd iterations) becomes trivial. The case of multiple layers is a recursive application of the one-layer case and, therefore, reduces to that. We refer the interested reader to \cite[Sec.~4]{lu08} for further details.}

\revus{
\begin{example}\label{ex:ipa-reconstruction}
	Suppose we are given the measurement matrix
	\[
	A = 
	\begin{pmatrix}
	1 & 2 & 1 & 0 & 0 & 0 \\
	3 & 0 & 0 & 1 & 3 & 0 \\
	0 & 1 & 0 & 1 & 0 & 3 \\
	0 & 0 & 4 & 0 & 3 & 2 \\
	\end{pmatrix}
	\]
	and the signal vector $\vec x = (1, 8, 3, 0, 0, 0)^T$, where $(\cdot)^T$ denotes the transpose of its argument. The measurement vector is then $\vec y = A \vec x = (20, 3, 8, 12)^T$ and Fig.~\ref{fig:ipa-example} illustrates the iterations of the IPA.
\end{example}


\begin{figure*}
	\centering
	\begin{tikzpicture}
	\tikzmath{\W = 6.66; \H=2.75; \dy = -4; \dx= 2.75; \iter = 0; \ud = 1; \x0 = (\W+\dx)*.5; \y0 = \iter*\dy;}
	\node at (\x0-.75,\y0+\H/2) [rotate=90] {$\xleftarrow{\text{iteration \iter}}$};
	\node at (\x0+0,      \y0+\H) [circle,draw] (v1\iter\ud) {$v_1$};
	\node at (\x0+\W/5,   \y0+\H) [circle,draw] (v2\iter\ud) {$v_2$};
	\node at (\x0+2*\W/5, \y0+\H) [circle,draw] (v3\iter\ud) {$v_3$};
	\node at (\x0+3*\W/5, \y0+\H) [circle,draw] (v4\iter\ud) {$v_4$};
	\node at (\x0+4*\W/5, \y0+\H) [circle,draw] (v5\iter\ud) {$v_5$};
	\node at (\x0+5*\W/5, \y0+\H) [circle,draw] (v6\iter\ud) {$v_6$};
	
	\node at (\x0+0,      \y0) [square,draw] (c1\iter\ud) {$c_1$};
	\node at (\x0+\W/3,   \y0) [square,draw] (c2\iter\ud) {$c_2$};
	\node at (\x0+2*\W/3, \y0) [square,draw] (c3\iter\ud) {$c_3$};
	\node at (\x0+3*\W/3, \y0) [square,draw] (c4\iter\ud) {$c_4$};
	
	\draw[->] (v1\iter\ud) -- (c1\iter\ud) node [near start,below,sloped] {$\sss{[0,\bm{1}]}$};
	\draw[->] (v1\iter\ud) -- (c2\iter\ud) node [very near start,below,sloped] {$\sss{[0,\bm{1}]}$};
	
	\draw[->] (v2\iter\ud) -- (c1\iter\ud) node [very near start,below,sloped] {$\sss{[0,\bm{8}]}$};
	\draw[->] (v2\iter\ud) -- (c3\iter\ud) node [pos=0.09,below,sloped] {$\sss{[0,\bm{8}]}$};
	
	\draw[->] (v3\iter\ud) -- (c1\iter\ud) node [pos=0.07,below,sloped] {$\sss{[0,\bm{3}]}$};
	\draw[->] (v3\iter\ud) -- (c4\iter\ud) node [pos=0.05,below,sloped] {$\sss{[0,\bm{3}]}$};
	
	\draw[->] (v4\iter\ud) -- (c2\iter\ud) node [near start,below,sloped] {$\sss{[\bm{0},3]}$};
	\draw[->] (v4\iter\ud) -- (c3\iter\ud) node [pos=0.12,above,sloped] {$\sss{[\bm{0},3]}$};
	
	\draw[->] (v5\iter\ud) -- (c2\iter\ud) node [very near start,below,sloped] {$\sss{[\bm{0},1]}$};
	\draw[->] (v5\iter\ud) -- (c4\iter\ud) node [very near start,below,sloped] {$\sss{[\bm{0},1]}$};
	
	\draw[->] (v6\iter\ud) -- (c3\iter\ud) node [near start,below,sloped] {$\sss{[\bm{0},\nicefrac 83]}$};
	\draw[->] (v6\iter\ud) -- (c4\iter\ud) node [near start,above,sloped] {$\sss{[\bm{0},\nicefrac 83]}$};

	\tikzmath{\iter = 1; \ud = 0; \x0 = 0; \y0 = \iter*\dy;}
	\node at (\x0-.75,\y0+\H/2) [rotate=90] {$\xrightarrow{\text{iteration \iter}}$};
	\node at (\x0+0,      \y0+\H) [circle,draw] (v1\iter\ud) {$v_1$};
	\node at (\x0+\W/5,   \y0+\H) [circle,draw] (v2\iter\ud) {$v_2$};
	\node at (\x0+2*\W/5, \y0+\H) [circle,draw] (v3\iter\ud) {$v_3$};
	\node at (\x0+3*\W/5, \y0+\H) [circle,draw] (v4\iter\ud) {$v_4$};
	\node at (\x0+4*\W/5, \y0+\H) [circle,draw] (v5\iter\ud) {$v_5$};
	\node at (\x0+5*\W/5, \y0+\H) [circle,draw] (v6\iter\ud) {$v_6$};
	
	\node at (\x0+0,      \y0) [square,draw] (c1\iter\ud) {$c_1$};
	\node at (\x0+\W/3,   \y0) [square,draw] (c2\iter\ud) {$c_2$};
	\node at (\x0+2*\W/3, \y0) [square,draw] (c3\iter\ud) {$c_3$};
	\node at (\x0+3*\W/3, \y0) [square,draw] (c4\iter\ud) {$c_4$};
	
	\draw[->] (c1\iter\ud) -- (v1\iter\ud) node [near start,above,sloped] {$\sss{[\bm{1},20]}$};
	\draw[->] (c1\iter\ud) -- (v2\iter\ud) node [near start,above,sloped] {$\sss{[\bm{8},10]}$};
	\draw[->] (c1\iter\ud) -- (v3\iter\ud) node [pos=0.1,below,sloped] {$\sss{[\bm{3},20]}$};
	
	\draw[->] (c2\iter\ud) -- (v1\iter\ud) node [pos=0.1,below,sloped] {$\sss{[0,\bm{1}]}$};
	\draw[->] (c2\iter\ud) -- (v4\iter\ud) node [pos=0.1,above,sloped] {$\sss{[\bm{0},3]}$};
	\draw[->] (c2\iter\ud) -- (v5\iter\ud) node [pos=0.1,below,sloped] {$\sss{[\bm{0},1]}$};
	
	\draw[->] (c3\iter\ud) -- (v2\iter\ud) node [pos=0.1,below,sloped] {$\sss{[0,\bm{8}]}$};
	\draw[->] (c3\iter\ud) -- (v4\iter\ud) node [near start,below,sloped] {$\sss{[\bm{0},8]}$};
	\draw[->] (c3\iter\ud) -- (v6\iter\ud) node [pos=0.1,below,sloped] {$\sss{[\bm{0},\nicefrac 83]}$};
	
	\draw[->] (c4\iter\ud) -- (v3\iter\ud) node [pos=0.07,below,sloped] {$\sss{[\nicefrac{11}{12},\bm{3}]}$};
	\draw[->] (c4\iter\ud) -- (v5\iter\ud) node [near start,below,sloped] {$\sss{[\bm{0},4]}$};
	\draw[->] (c4\iter\ud) -- (v6\iter\ud) node [near start,below,sloped] {$\sss{[\bm{0},6]}$};

	\tikzmath{\iter = 1; \ud = 1; \x0 = \W+\dx; \y0 = \iter*\dy;}
	\node at (\x0-.75,\y0+\H/2) [rotate=90] {$\xleftarrow{\text{iteration \iter}}$};
	\node at (\x0+0,      \y0+\H) [circle,draw] (v1\iter\ud) {$v_1$};
	\node at (\x0+\W/5,   \y0+\H) [circle,draw] (v2\iter\ud) {$v_2$};
	\node at (\x0+2*\W/5, \y0+\H) [circle,draw] (v3\iter\ud) {$v_3$};
	\node at (\x0+3*\W/5, \y0+\H) [circle,draw] (v4\iter\ud) {$v_4$};
	\node at (\x0+4*\W/5, \y0+\H) [circle,draw] (v5\iter\ud) {$v_5$};
	\node at (\x0+5*\W/5, \y0+\H) [circle,draw] (v6\iter\ud) {$v_6$};
	
	\node at (\x0+0,      \y0) [square,draw] (c1\iter\ud) {$c_1$};
	\node at (\x0+\W/3,   \y0) [square,draw] (c2\iter\ud) {$c_2$};
	\node at (\x0+2*\W/3, \y0) [square,draw] (c3\iter\ud) {$c_3$};
	\node at (\x0+3*\W/3, \y0) [square,draw] (c4\iter\ud) {$c_4$};
	
	\draw[->] (v1\iter\ud) -- (c1\iter\ud) node [near start,below,sloped] {$\sss{[\bm{1},\bm{1}]}$};
	\draw[->] (v1\iter\ud) -- (c2\iter\ud) node [very near start,below,sloped] {$\sss{[\bm{1},\bm{1}]}$};;
	
	\draw[->] (v2\iter\ud) -- (c1\iter\ud) node [very near start,below,sloped] {$\sss{[\bm{8},\bm{8}]}$};
	\draw[->] (v2\iter\ud) -- (c3\iter\ud) node [pos=0.09,below,sloped] {$\sss{[\bm{8},\bm{8}]}$};
	
	\draw[->] (v3\iter\ud) -- (c1\iter\ud) node [pos=0.07,below,sloped] {$\sss{[\bm{3},\bm{3}]}$};
	\draw[->] (v3\iter\ud) -- (c4\iter\ud) node [pos=0.05,below,sloped] {$\sss{[\bm{3},\bm{3}]}$};
	
	\draw[->] (v4\iter\ud) -- (c2\iter\ud) node [near start,below,sloped] {$\sss{[\bm{0},3]}$};
	\draw[->] (v4\iter\ud) -- (c3\iter\ud) node [pos=0.12,above,sloped] {$\sss{[\bm{0},3]}$};
	
	\draw[->] (v5\iter\ud) -- (c2\iter\ud) node [very near start,below,sloped] {$\sss{[\bm{0},1]}$};
	\draw[->] (v5\iter\ud) -- (c4\iter\ud) node [very near start,below,sloped] {$\sss{[\bm{0},1]}$};
	
	\draw[->] (v6\iter\ud) -- (c3\iter\ud) node [very near start,below,sloped] {$\sss{[\bm{0},\nicefrac 83]}$};
	\draw[->] (v6\iter\ud) -- (c4\iter\ud) node [near start,above,sloped] {$\sss{[\bm{0},\nicefrac 83]}$};

	\tikzmath{\iter = 2; \ud = 0; \x0 = 0; \y0 = \iter*\dy;}
	\node at (\x0-.75,\y0+\H/2) [rotate=90] {$\xrightarrow{\text{iteration \iter}}$};
	\node at (\x0+0,      \y0+\H) [circle,draw] (v1\iter\ud) {$v_1$};
	\node at (\x0+\W/5,   \y0+\H) [circle,draw] (v2\iter\ud) {$v_2$};
	\node at (\x0+2*\W/5, \y0+\H) [circle,draw] (v3\iter\ud) {$v_3$};
	\node at (\x0+3*\W/5, \y0+\H) [circle,draw] (v4\iter\ud) {$v_4$};
	\node at (\x0+4*\W/5, \y0+\H) [circle,draw] (v5\iter\ud) {$v_5$};
	\node at (\x0+5*\W/5, \y0+\H) [circle,draw] (v6\iter\ud) {$v_6$};
	
	\node at (\x0+0,      \y0) [square,draw] (c1\iter\ud) {$c_1$};
	\node at (\x0+\W/3,   \y0) [square,draw] (c2\iter\ud) {$c_2$};
	\node at (\x0+2*\W/3, \y0) [square,draw] (c3\iter\ud) {$c_3$};
	\node at (\x0+3*\W/3, \y0) [square,draw] (c4\iter\ud) {$c_4$};
	
	\draw[->] (c1\iter\ud) -- (v1\iter\ud) node [near start,above,sloped] {$\sss{[\bm{1},\bm{1}]}$};
	\draw[->] (c1\iter\ud) -- (v2\iter\ud) node [near start,above,sloped] {$\sss{[\bm{8},\bm{8}]}$};
	\draw[->] (c1\iter\ud) -- (v3\iter\ud) node [pos=0.1,below,sloped] {$\sss{[\bm{3},\bm{3}]}$};
	
	\draw[->] (c2\iter\ud) -- (v1\iter\ud) node [pos=0.1,below,sloped] {$\sss{[0,\bm{1}]}$};
	\draw[->] (c2\iter\ud) -- (v4\iter\ud) node [pos=0.1,above,sloped] {$\sss{[\bm{0},\bm{0}]}$};
	\draw[->] (c2\iter\ud) -- (v5\iter\ud) node [pos=0.1,below,sloped] {$\sss{[\bm{0},\bm{0}]}$};
	
	\draw[->] (c3\iter\ud) -- (v2\iter\ud) node [pos=0.1,below,sloped] {$\sss{[0,\bm{8}]}$};
	\draw[->] (c3\iter\ud) -- (v4\iter\ud) node [near start,below,sloped] {$\sss{[\bm{0},\bm{0}]}$};
	\draw[->] (c3\iter\ud) -- (v6\iter\ud) node [pos=0.1,below,sloped] {$\sss{[\bm{0},\bm{0}]}$};
	
	\draw[->] (c4\iter\ud) -- (v3\iter\ud) node [pos=0.07,below,sloped] {$\sss{[\nicefrac{11}{12},\bm{3}]}$};
	\draw[->] (c4\iter\ud) -- (v5\iter\ud) node [near start,below,sloped] {$\sss{[\bm{0},\bm{0}]}$};
	\draw[->] (c4\iter\ud) -- (v6\iter\ud) node [near start,below,sloped] {$\sss{[\bm{0},\bm{0}]}$};

	\tikzmath{\iter = 2; \ud = 1; \x0 = \W + \dx; \y0 = \iter*\dy;}
	\node at (\x0-.75,\y0+\H/2) [rotate=90] {$\xleftarrow{\text{iteration \iter}}$};
	\node at (\x0+0,      \y0+\H) [circle,draw] (v1\iter\ud) {$v_1$};
	\node at (\x0+\W/5,   \y0+\H) [circle,draw] (v2\iter\ud) {$v_2$};
	\node at (\x0+2*\W/5, \y0+\H) [circle,draw] (v3\iter\ud) {$v_3$};
	\node at (\x0+3*\W/5, \y0+\H) [circle,draw] (v4\iter\ud) {$v_4$};
	\node at (\x0+4*\W/5, \y0+\H) [circle,draw] (v5\iter\ud) {$v_5$};
	\node at (\x0+5*\W/5, \y0+\H) [circle,draw] (v6\iter\ud) {$v_6$};
	
	\node at (\x0+0,      \y0) [square,draw] (c1\iter\ud) {$c_1$};
	\node at (\x0+\W/3,   \y0) [square,draw] (c2\iter\ud) {$c_2$};
	\node at (\x0+2*\W/3, \y0) [square,draw] (c3\iter\ud) {$c_3$};
	\node at (\x0+3*\W/3, \y0) [square,draw] (c4\iter\ud) {$c_4$};
	
	\draw[->] (v1\iter\ud) -- (c1\iter\ud) node [near start,below,sloped] {$\sss{[\bm{1},\bm{1}]}$};
	\draw[->] (v1\iter\ud) -- (c2\iter\ud) node [very near start,below,sloped] {$\sss{[\bm{1},\bm{1}]}$};;
	
	\draw[->] (v2\iter\ud) -- (c1\iter\ud) node [very near start,below,sloped] {$\sss{[\bm{8},\bm{8}]}$};
	\draw[->] (v2\iter\ud) -- (c3\iter\ud) node [pos=0.09,below,sloped] {$\sss{[\bm{8},\bm{8}]}$};
	
	\draw[->] (v3\iter\ud) -- (c1\iter\ud) node [pos=0.07,below,sloped] {$\sss{[\bm{3},\bm{3}]}$};
	\draw[->] (v3\iter\ud) -- (c4\iter\ud) node [pos=0.05,below,sloped] {$\sss{[\bm{3},\bm{3}]}$};
	
	\draw[->] (v4\iter\ud) -- (c2\iter\ud) node [near start,below,sloped] {$\sss{[\bm{0},\bm{0}]}$};
	\draw[->] (v4\iter\ud) -- (c3\iter\ud) node [pos=0.12,above,sloped] {$\sss{[\bm{0},\bm{0}]}$};
	
	\draw[->] (v5\iter\ud) -- (c2\iter\ud) node [very near start,below,sloped] {$\sss{[\bm{0},\bm{0}]}$};
	\draw[->] (v5\iter\ud) -- (c4\iter\ud) node [very near start,below,sloped] {$\sss{[\bm{0},\bm{0}]}$};
	
	\draw[->] (v6\iter\ud) -- (c3\iter\ud) node [very near start,below,sloped] {$\sss{[\bm{0},\bm{0}]}$};
	\draw[->] (v6\iter\ud) -- (c4\iter\ud) node [near start,above,sloped] {$\sss{[\bm{0},\bm{0}]}$};
	\end{tikzpicture}
	\caption[IPA reconstruction example]{\revus{IPA reconstruction example. The original signal vector is $\vec x = (1,8,3,0,0,0)^T$ and the measurement vector is $\vec y = (20,3,8,12)^T$. Numbers in bold correspond to exact bounds. The last iteration is omitted because the signal has already been reconstructed.}}
	\label{fig:ipa-example}
\end{figure*}
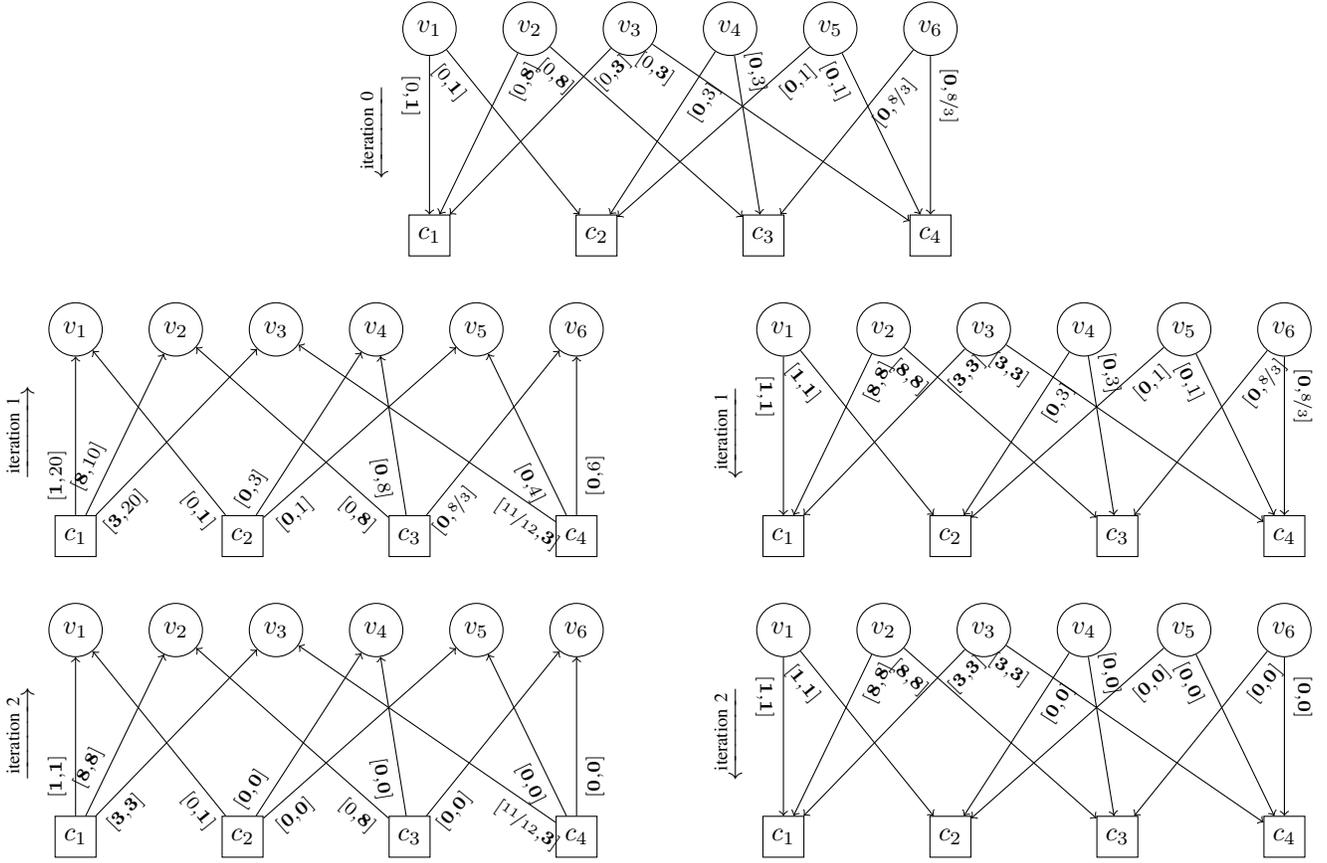
}

\begin{algorithm}
	\caption{Interval-Passing Algorithm (cf.\ \cite[Alg.~1]{RDVD12})}
	\label{alg:ipa}
	
	\begin{algorithmic}[1]
		\Function {IPA}{$\vec{y}$, $A$}
		\Statex \textit{Initialization}
		\ForAll {$v \in V$} 
		\State $\begin{aligned}
		\mu^{(0)}_{v \to \cdot} \leftarrow 0 \text{ and }
		M^{(0)}_{v \to \cdot} \leftarrow \min_{c \in \Neighbours{v}} \left( y_{c} / a_{cv} \right) 
		\end{aligned}$ \label{alg:ipa:muM0}
		\EndFor
		
		\Statex \textit{Iterations}
		\State $\ell \leftarrow 0$
		\Repeat
		\State $\ell \leftarrow \ell+1$
		
		\ForAll {$c \in C$, $v \in \Neighbours{c}$}
		\State $\begin{aligned}
		\mu^{(\ell)}_{c \to v} \leftarrow \frac{1}{a_{cv}}\left(y_c - \sum_{v' \in 
			\Neighbours{c}, v' \neq v} a_{cv'} M^{(\ell-1)}_{v' \to 
			\cdot}\right)\end{aligned}$ \label{alg:ipa:mucv}
		\If {$\mu^{(\ell)}_{c \to v} < 0$}
		\State $\mu^{(\ell)}_{c \to v} \leftarrow 0$
		\EndIf
		\State $\begin{aligned}
		M^{(\ell)}_{c \to v} \leftarrow \frac{1}{a_{cv}}\left( y_c - \sum_{v' \in \Neighbours{c}, v' \neq v} a_{cv'} \mu^{(\ell-1)}_{v' \to \cdot} \right)
		\end{aligned}$ \label{alg:ipa:Mcv}
		\EndFor
		
		\ForAll {$v \in V$}
		\State $\begin{aligned}
		\mu^{(\ell)}_{v \to \cdot} \leftarrow \max_{c \in \Neighbours{v}} 
		\mu^{(\ell)}_{c \to v}  
		\end{aligned}$ \label{alg:ipa:muvc}
		\State $\begin{aligned}
		M^{(\ell)}_{v \to \cdot} \leftarrow \min_{c \in \Neighbours{v}} 
		M^{(\ell)}_{c \to v} 
		\end{aligned}$ \label{alg:ipa:Mvc}
		\EndFor

		\Until{$\mu^{(\ell)}_{v \to \cdot} = \mu^{(\ell-1)}_{v \to \cdot}$ and $M^{(\ell)}_{v \to \cdot} = M^{(\ell-1)}_{v \to \cdot}$, \: $\forall v \in V$}
		
		\Statex \textit{Result}
		\FForAll {$v \in V$}
		$\begin{aligned}
		\hat{x}_v \leftarrow \mu^{(\ell)}_{v \to \cdot}
		\end{aligned}$ 
		\EEndFor \label{alg:hatxv}
		\State \textbf{return} $\hat{\vec x}$
		
		\EndFunction
	\end{algorithmic}
\end{algorithm}

\section{Failing Sets of the Interval-Passing Algorithm} \label{sec:failing_sets}
In this section, we present several results related to the failure of the IPA. In particular, in Section~\ref{sec:thm:supp-converge}, we show that the IPA fails to recover $\vec{x}$ from $\vec{y}$ if and only if it fails to recover a corresponding binary vector of the same support, and also that only positions of nonzero values in the matrix $A$ are of importance for success of recovery (see Theorem~\ref{thm:0-1-x} below). Based on Theorem~\ref{thm:0-1-x}, we introduce the concept of termatiko sets in Section~\ref{sec:termatiko_sets} and give a complete (graph-theoretic) description of the failing sets of the IPA in Section~\ref{sec:criterion}. In Section~\ref{sec:counter-example}, a counter-example  to \cite[Thm.~2]{RDVD12} is provided, while two heuristic approaches to locate small-size termatiko sets from a list of stopping sets is outlined in Section~\ref{sec:alg_small_size}.





\subsection{Signal Support Recovery} \label{sec:thm:supp-converge}
    Consider the two related problems IPA($\vec y$, $A$) and IPA($\vec s$, $B$), \revtwo{where $\vec x \in \Reals_{\geq 0}^n$, $A = (a_{ji}) \in \Reals_{\geq 0}^{m \times n}$, $\vec y = A \vec x$, $B = (b_{ji}) \in \{0,1\}^{m \times n}$,} $\vec s = B \vec z$, and $\vec z \in \{0,1\}^n$ has support $\supp(\vec z) = \supp(\vec x)$, i.e., $\vec x$ and $\vec z$ have the same support. The binary matrix $B$ contains ones exactly in the positions where $A$ has nonzero values. We will show below (see Theorem~\ref{thm:0-1-x}) that these two problems behave identically, namely, they 
    recover exactly the same positions of $\vec x$ and $\vec z$. However, note that this is true if the identical algorithm (Algorithm~\ref{alg:ipa}) is applied to both problems, i.e., the binary nature of $\vec z$ is not exploited.

    \begin{thm} \label{thm:0-1-x}
        Let $A = (a_{ji}) \in \Reals_{\geq 0}^{m \times n}$, $\vec x \in \Reals_{\geq 0}^n$, $B = (b_{ji}) \in \{0,1\}^{m \times n}$, and $\vec z \in \{0,1\}^n$, where $\supp(\vec z) = \supp(\vec x)$ and
        \[
       b_{ji} = \begin{cases}
       0\,, &\text{if $a_{ji} = 0$} \,,\\
       1\,, &\text{otherwise} \,.
       \end{cases}
       \]
       Further, denote $\vec y = A \vec x$, $\vec s = B \vec z$, $\hat{\vec x} = 
        \mathrm{IPA}(\vec y, A)$, and $\hat{\vec z} = \mathrm{IPA}(\vec s, B)$.  
        Then, for all $v \in V$, 
        \[
            \hat{x}_v = x_v \quad \text{if and only if} \quad \hat{z}_v = z_v 
            \,.
        \]
    \end{thm}

    \begin{IEEEproof}
        Define subsets of $V$ in which either the lower or the upper bound of a variable-to-measurement message, at a given iteration $\ell$, is equal to $x_v$ or $z_v$ as follows: 
        \begin{align*}
            \gamma_x^{(\ell)} &= \left\{ v \in V \mid \mu^{(\ell)}_{v \to \cdot} = x_v 
            \right\},\\
            \Gamma_x^{(\ell)} &= \left\{ v \in V \mid M^{(\ell)}_{v \to \cdot} = x_v  
            \right\},\\
            \gamma_z^{(\ell)} &= \left\{ v \in V \mid \lambda^{(\ell)}_{v \to \cdot} = 
            z_v \right\},\\
            \Gamma_z^{(\ell)} &= \left\{ v \in V \mid \Lambda^{(\ell)}_{v \to \cdot} = 
            z_v \right\},
        \end{align*}
        where $\lambda^{(\ell)}_{v \to \cdot}$ and $\Lambda^{(\ell)}_{v \to \cdot}$ denote, respectively, the lower and the upper bound of the variable-to-measurement message from variable node $v$ to any measurement node  $c \in \Neighbours{v}$ at iteration $\ell$ for IPA($\vec s$, $B$) (analogously to $\mu^{(\ell)}_{v \to \cdot}$ and $M^{(\ell)}_{v \to \cdot}$ for IPA($\vec y$, $A$)).
        
        To prove the \revtwo{theorem}, it is enough to show that at each iteration $\ell$, $\gamma_x^{(\ell)} = \gamma_z^{(\ell)}$ and $\Gamma_x^{(\ell)} = \Gamma_z^{(\ell)}$. We demonstrate this by induction on $\ell$. 
        
        {\flushleft\emph{Base Case}.}
        \ifonecolumn
        \begin{align*}
        \gamma_x^{(0)} &= \{ v \in V \mid x_v = 0 \} = \{ v \in V \mid z_v 
        = 0 \} = \gamma_z^{(0)}\,,  \\
        \Gamma_x^{(0)} &= \{ v \in V \mid \exists c \in \Neighbours{v} 
        \text{ s.t. } y_c = a_{cv} x_v \} = \{ v \in V \mid \exists c \in \Neighbours{v} \text{ 
        	s.t. } s_c = z_v \} = \Gamma_z^{(0)} \,.
        \end{align*}
        \else
        \begin{align*}
            \gamma_x^{(0)} &= \{ v \in V \mid x_v = 0 \} = \{ v \in V \mid z_v 
            = 0 \} = \gamma_z^{(0)}\,,  \\
            \Gamma_x^{(0)} &= \{ v \in V \mid \exists c \in \Neighbours{v} 
            \text{ s.t. } y_c = a_{cv} x_v \} \\
            &= \{ v \in V \mid \exists c \in \Neighbours{v} \text{ 
            	s.t. } s_c = z_v \} = \Gamma_z^{(0)} \,.
        \end{align*}
        \fi
        
        {\flushleft \emph{Inductive Step}.}
        
        Consider iteration $\ell \geq 1$. First note that all $v \in V$ with $x_v = 0$ (and hence $z_v = 0$) belong to both $\gamma_x^{(\ell)}$ and $\gamma_z^{(\ell)}$.
        
        If $x_v > 0$ (and hence $z_v = 1$) then from Line~\ref{alg:ipa:muvc} of Algorithm~\ref{alg:ipa} and the definition of $\gamma_x^{(\ell)}$, we have $v \in \gamma_x^{(\ell)}$ if and only if there exists $c \in \Neighbours{v}$ such that $\mu_{c \to v}^{(\ell)} = x_v$. More precisely:
        \ifonecolumn
        \[
        a_{cv} x_v = y_c - \sum_{\substack{v' \in \Neighbours[]{c} \\ v' \neq v}} a_{cv'} M_{v' \to \cdot}^{(\ell-1)}
        =a_{cv} x_v + \sum_{\substack{v' \in \Neighbours[]{c} \\ v' \neq v}} a_{cv'} \left( x_{v'} - M_{v' \to \cdot}^{(\ell-1)} \right) \leq a_{cv} x_v \,.
        \]
        \else
        \begin{align*}
        a_{cv} x_v &= y_c - \sum_{\substack{v' \in \Neighbours[]{c} \\ v' \neq v}} a_{cv'} M_{v' \to \cdot}^{(\ell-1)} \\
        &=a_{cv} x_v + \sum_{\substack{v' \in \Neighbours[]{c} \\ v' \neq v}} a_{cv'} \left( x_{v'} - M_{v' \to \cdot}^{(\ell-1)} \right) \leq a_{cv} x_v \,.
        \end{align*}
        \fi
        Equality holds if and only if $M_{v' \to \cdot}^{(\ell-1)} = x_{v'}$ for all $v' \in \Neighbours[]{c} \setminus \{v\}$ or, in our notation, $\Neighbours[]{c} \setminus \{v\} \subset \Gamma_x^{(\ell-1)}$. However, by inductive assumption $\Gamma_z^{(\ell-1)} = \Gamma_x^{(\ell-1)}$ and hence $\Lambda_{v' \to \cdot}^{(\ell-1)} = z_{v'}$ for all $v' \in 
            \Neighbours{c} \setminus \{ v \}$. This is equivalent to
            $\lambda_{c \to v}^{(\ell)} = z_v$ and thus $v \in \gamma_z^{(\ell)}$.
        
        Hence, for all $v \in V$, $v$ either belongs to both $\gamma_x^{(\ell)}$ and 
        $\gamma_z^{(\ell)}$, or to none of them.
        
        Analogously, we can show that $\Gamma_x^{(\ell)} = \Gamma_z^{(\ell)}$. Details are omitted for brevity.
    \end{IEEEproof}
    
    Theorem~\ref{thm:0-1-x} gives a powerful tool for analysis of IPA performance. Instead of considering $A \in \Reals_{\geq 0}^{m \times n}$ and $\vec x \in \Reals^n_{\geq 0}$ we need only to work with binary $A$ and $\vec x$ (although all operations are still performed over $\Reals$). Thus, in the rest of the paper, we assume that $A$ is binary.
    

\subsection{Termatiko Sets} \label{sec:termatiko_sets}
We define termatiko sets through failures of the IPA.

\begin{define}\label{def:termatiko}
    We call $T \subset V$ a \emph{termatiko set} if and only if $\IPA(A \vec{x}_T, A) = \vec 0$, where $\vec x_T$ is a binary vector with support $\supp(\vec x_T) = T$.
\end{define}
From Theorem~\ref{thm:0-1-x}, it follows that the IPA completely fails to recover $\vec x \in \Reals^n_{\geq 0}$ if and only if $\supp(\vec x) = T$, where $T$ is a nonempty termatiko set.

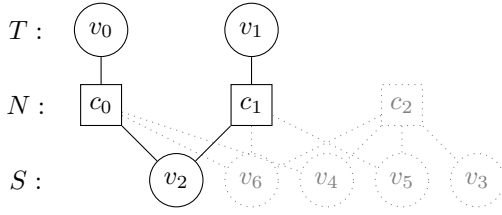
\begin{figure}
    \centering
    \begin{tikzpicture}
    \node at (1.0,+1) [circle,draw] (v0) {$v_{0}$};
    \node at (3.0,+1) [circle,draw] (v1) {$v_{1}$};
    \node at (2.0,-1) [circle,draw] (v2) {$v_{2}$};
    \node at (6.0,-1) [circle,draw,dotted,gray] (v3) {$v_{3}$};
    \node at (4.0,-1) [circle,draw,dotted,gray] (v4) {$v_{4}$};
    \node at (5.0,-1) [circle,draw,dotted,gray] (v5) {$v_{5}$};
    \node at (3.0,-1) [circle,draw,dotted,gray] (v6) {$v_{6}$};
    
    \node at (1.0,0) [square,draw] (c0) {$c_{0}$};
    \node at (3.0,0) [square,draw] (c1) {$c_{1}$};
    \node at (5.0,0) [square,draw,dotted,gray] (c2) {$c_{2}$};
    
    \node at (0,+1) {$T:$};
    \node at (0,0) {$N:$};
    \node at (0,-1) {$S:$};
    
    \draw (c0) -- (v0);
    \draw (c0) -- (v2);
    \draw[dotted,gray] (c0) -- (v4);
    \draw[dotted,gray] (c0) -- (v6);
    
    \draw (c1) -- (v1);
    \draw (c1) -- (v2);
    \draw[dotted,gray] (c1) -- (v5);
    \draw[dotted,gray] (c1) -- (v6);
    
    \draw[dotted,gray] (c2) -- (v3);
    \draw[dotted,gray] (c2) -- (v4);
    \draw[dotted,gray] (c2) -- (v5);
    \draw[dotted,gray] (c2) -- (v6);
    \end{tikzpicture}
    \caption{Example of a termatiko set $T$ with all measurement nodes in $N$ connected to both $T$ and $S$ (cf.\ Theorem~\ref{thm:termatikos}). The rest of the Tanner graph is drawn dotted.}
    \label{fig:term-class1-ex}
\end{figure}

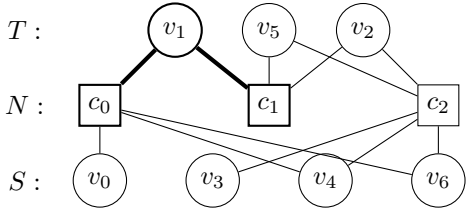
\begin{figure}
    \centering
    \begin{tikzpicture}
    \node at (1.0,-1) [circle,draw] (v0) {$v_{0}$};
    \node at (3.25,+1) [circle,draw] (v5) {$v_{5}$};
    \node at (2.0,+1) [circle,draw,thick] (v1) {$v_{1}$};
    \node at (2.5,-1) [circle,draw] (v3) {$v_{3}$};
    \node at (4.0,-1) [circle,draw] (v4) {$v_{4}$};
    \node at (4.5,+1) [circle,draw] (v2) {$v_{2}$};
    \node at (5.5,-1) [circle,draw] (v6) {$v_{6}$};
    
    \node at (1.0,0) [square,draw,thick] (c0) {$c_{0}$};
    \node at (3.25,0) [square,draw,thick] (c1) {$c_{1}$};
    \node at (5.5,0) [square,draw] (c2) {$c_{2}$};
    
    \draw (c0) -- (v0);
    \draw[ultra thick] (c0) -- (v1);
    \draw (c0) -- (v4);
    \draw (c0) -- (v6);
    
    \node at (0,+1) {$T:$};
    \node at (0,0) {$N:$};
    \node at (0,-1) {$S:$};
    
    \draw (c1) -- (v5);
    \draw[ultra thick] (c1) -- (v1);
    \draw (c1) -- (v2);
    
    \draw (c2) -- (v2);
    \draw (c2) -- (v3);
    \draw (c2) -- (v4);
    \draw (c2) -- (v5);
    \draw (c2) -- (v6);
    \end{tikzpicture}
    \caption{Example of a termatiko set $T$ with a measurement node $c_1$ connected to $T$ only (cf.\ Theorem~\ref{thm:termatikos}). Highlighted is the connection to a measurement node $c_0$, which is connected to $T$ only once.}
    \label{fig:term-class2-ex}
\end{figure}

\begin{thm} \label{thm:termatikos}
    Let $T$ be a subset of the set of variable nodes $V$. We denote by $N = \Neighbours T$ the set of measurement nodes connected to $T$ and also denote by $S$ the other variable nodes connected only to $N$ as follows:
    \begin{equation} \notag 
    S = \left\{ v \in V \setminus T \,:\, \Neighbours[N] v = \Neighbours v \right\} \,.
    \end{equation}
    Then, $T$ is a termatiko set if and only if for each $c \in N$ one of the following two conditions holds (cf.\ Figs.~\ref{fig:term-class1-ex} and \ref{fig:term-class2-ex}):
    \begin{itemize}
        \item $c$ is connected to $S$ (this implies $S \neq \varnothing$);
        \item $c$ is not connected to $S$ and
        \begin{gather*}
            \Big|\left\{
                v \in \Neighbours[T]{c} \,:\, \forall c' \in \Neighbours v ,\, |\Neighbours[T]{c'}| \geq 2
            \right\}\Big| \geq 2 \,.
        \end{gather*}
    \end{itemize}
\end{thm}

\begin{IEEEproof}
    Consider the problem $\IPA(A \vec{x}_T, A)$, where $\vec x_T$ is a binary vector with support $\supp(\vec x_T) = T$ and $T$ satisfies the conditions of the theorem.
    
    We first note that measurement nodes in $C \setminus N$ have value zero and hence all variable nodes connected to them (i.e., $v \in V \setminus (T \cup S)$) are recovered with zeros at the initialization step of Algorithm~\ref{alg:ipa}. As a consequence, they can be safely pruned and w.l.o.g.\ we can assume that $C = N$ and $V = T \cup S$.
    
We show by induction that for all $v \in T \cup S$ at each iteration $\ell \geq 0$ it holds that $\mu_{v \to \cdot}^{(\ell)} = 0$ and $M_{v \to \cdot}^{(\ell)} \geq 1$. Moreover, each measurement node $c \in N$ that is not connected to $S$ has at least two different neighbors $v_1, v_2 \in T$ with $M_{v_1 \to \cdot}^{(\ell)} \geq 2$ and $M_{v_2 \to \cdot}^{(\ell)} \geq 2$.

    We will use the fact that
    \[
    x_v = \begin{cases}
    1\,, & \text{if } v \in T \,, \\
    0\,, & \text{if } v \in S \,.
    \end{cases}
    \]
    Also we note that $y_c = | \Neighbours[T]{c} |$ for all $c \in N$.
    
    {\flushleft \emph{Base Case.}}
    
    For $\ell = 0$ we immediately obtain from Algorithm~\ref{alg:ipa} that $\mu_{v \to \cdot}^{(0)} = 0$ and, as each $c \in N$ has at least one nonzero neighbor, $M_{v \to \cdot}^{(0)} \geq 1$. In addition, consider $c \in N$ that is not connected to $S$. It has at least two different neighbors $v_1, v_2 \in T$, each connected only to measurement nodes with not less than two neighbors in $T$. Therefore, \revtwo{$M_{v_1 \to \cdot}^{(0)} \geq 2$ and $M_{v_2 \to \cdot}^{(0)} \geq 2$}. 
    
    {\flushleft \emph{Inductive Step.}}
    
    Consider $\ell \geq 1$. For all $c \in N$ and all $v \in \Neighbours c$,
    \[
        M_{c \to v}^{(\ell)} = y_c - \sum_{v' \in \Neighbours{c} \,, v' \neq v} \mu_{v' \to \cdot}^{(\ell-1)} = y_c \,.
    \]
    Hence, upper bounds are exactly the same as for $l = 0$ and the same inequalities hold for them.
    
    In order to find lower bounds, we consider two cases for $c \in N$. If $c$ is connected to $S$, then
    	\ifonecolumn
    	\[
    	y_c - \sum_{\substack{v' \in \Neighbours{c} \\ v' \neq v}} M_{v' \to \cdot}^{(\ell-1)} 
    	\leq \left( |\Neighbours{c}| - 1 \right) - \sum_{\substack{v' \in \Neighbours{c} \\ v' \neq v}} 1 = 0
    	\]
    	\else
        \begin{multline*}
            y_c - \sum_{\substack{v' \in \Neighbours{c} \\ v' \neq v}} M_{v' \to \cdot}^{(\ell-1)} 
            \leq \left( |\Neighbours{c}| - 1 \right) - \sum_{\substack{v' \in \Neighbours{c} \\ v' \neq v}} 1 = 0
        \end{multline*}
        \fi
    and therefore $\mu_{c \to v}^{(\ell)} = 0$. If $c$ is connected to $T$ only, then
    	\ifonecolumn
        \[
            y_c - \sum_{\substack{v' \in \Neighbours{c} \\ v' \neq v}} M_{v' \to \cdot}^{(\ell-1)} 
            \leq |\Neighbours[T]{c}| - \Bigg( 1 + \sum_{\substack{v' \in \Neighbours[T]{c} \\ v' \neq v}} 1 \Bigg) = 0
        \]
        \else
        \begin{multline*}
        y_c - \sum_{\substack{v' \in \Neighbours{c} \\ v' \neq v}} M_{v' \to \cdot}^{(\ell-1)} 
        \leq |\Neighbours[T]{c}| - \Bigg( 1 + \sum_{\substack{v' \in \Neighbours[T]{c} \\ v' \neq v}} 1 \Bigg) = 0
        \end{multline*}
        \fi
    and again $\mu_{c \to v}^{(\ell)} = 0$. Here, the extra $1$ inside the parenthesis indicates the fact that for at least one $v'$ we have $M_{v' \to \cdot}^{(\ell-1)} \geq 2$. Thus, at each iteration of the IPA for each $v \in V$ the lower bound is equal to zero, and the algorithm will return $\hat{\vec x} = \vec 0$.
     
    We have demonstrated that if $T$ satisfies the conditions of the theorem, it is a termatiko set. What remains to be proven is that if $T$ does not satisfy the conditions of the theorem, the IPA can recover at least some of the nonzero values. 
    
    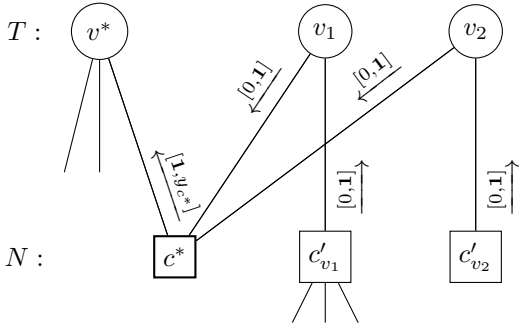
\begin{figure}
        \centering
        \begin{tikzpicture}
        \node at (-1, 3) {$T:$};
        \node at (-1, 0) {$N:$};
        
        \node at (0, 3) [circle,draw] (vstar) {$v^*$};
        \node at (3, 3) [circle,draw] (v1) {$v_1$};
        \node at (5, 3) [circle,draw] (v2) {$v_2$};
        
        \node at (1, 0) [square,draw,thick] (cstar) {$c^*$};
        \node at (3, 0) [square,draw] (c1) {$c'_{v_1}$};
        \node at (5, 0) [square,draw] (c2) {$c'_{v_2}$};
        
        \draw (cstar) -- (vstar);
        \draw (cstar) -- (vstar) node [near start,above,sloped] (mcsvs) {$\xleftarrow{[\bm 1,y_{c^*}]}$};
        \draw (v1) -- (cstar);
        \draw (v1) -- (cstar) node [near start,above,sloped] (mv1cs) {$\xleftarrow{[0,\bm 1]}$};
        \draw (v2) -- (cstar);
        \draw (v2) -- (cstar) node [near start,above,sloped] (mv2cs) {$\xleftarrow{[0,\bm 1]}$};
        
        \draw (c1) -- (v1);
        \draw (c1) -- (v1) node [near start,below,sloped] (mc1v1) {$\xrightarrow{[0,\bm 1]}$};

        \draw (c2) -- (v2);
        \draw (c2) -- (v2) node [near start,below,sloped] (mc2v2) {$\xrightarrow{[0,\bm 1]}$};
        
        \node at (-.5, 1) (f1) {};
        \node at (0, 1)   (f2) {};
        \draw (vstar) -- (f1);
        \draw (vstar) -- (f2);
        
        \node at (2.5, -1) (f3) {};
        \node at (3, -1) (f4) {};
        \node at (3.5, -1) (f5) {};
        \draw (c1) -- (f3);
        \draw (c1) -- (f4);
        \draw (c1) -- (f5);
        \end{tikzpicture}
        \caption{Exact bounds propagation in a nontermatiko set. Here $[\mu,M]$ denotes sending a lower bound of $\mu$ and an upper bound of $M$ in the direction given by the corresponding arrow. Numbers in bold are exact bounds.}
        \label{fig:non-termatiko}
    \end{figure}
    
    Assume that there exists $c^* \in N$ connected to $T$ only (i.e., $\Neighbours[T]{c^*} = \Neighbours[]{c^*}$) and such that 
    \[
        \Big|\left\{
            v \in \Neighbours[T]{c^*} \,:\, \forall c' \in \Neighbours v ,\, |\Neighbours[T]{c'}| \geq 2
        \right\}\Big| \leq 1 \,.
    \]
    If this set has one element, denote it by $v^*$. If it is empty, let $v^*$ be any element of $\Neighbours[T]{c^*}$.
    
    A special case when $|\Neighbours[T]{c^*}| = 1$ is trivial. Otherwise, for any $v \in \Neighbours[T]{c^*} \setminus \{ v^* \}$, there exists $c'_v \in \Neighbours{v}$ such that $|\Neighbours[T]{c'_v}| \leq 1$, which in truth means that $\Neighbours[T]{c'_v} = \{ v \}$.
    
    Hence, at the initialization step of the IPA, for all $v \in \Neighbours[T]{c^*} \setminus \{ v^* \}$ we will have $\mu_{v \to \cdot}^{(0)} = 0$ and $M_{v \to \cdot}^{(0)} = 1$. Therefore, at iteration $\ell = 1$:
    \[
    \mu^{(1)}_{c^* \to v^*} \leftarrow y_{c^*} - \sum_{\substack{v \in \Neighbours[T]{c^*} \\ v \neq v^*}} M^{(0)}_{v \to \cdot} = y_{c^*} - \sum_{\substack{v \in \Neighbours[T]{c^*} \\ v \neq v^*}} 1 = 1 \,.
    \]
    Thus, the IPA will output 1 for position $v^* \in T$, which means that $T$ is not a termatiko set. See Fig.~\ref{fig:non-termatiko} for illustration.
\end{IEEEproof}

Theorem~\ref{thm:termatikos} gives a precise graph-theoretic description of termatiko sets. In fact, it defines two important subclasses of termatiko sets; stopping sets and sets with all $c \in N$ connected to both $T$ and $S$. Also, it is worth noting that $T \cup S$ is a stopping set. Thus, a termatiko set is always a subset of some stopping set. We define the size of the smallest nonempty termatiko set as the \emph{termatiko distance}.

\subsection{General Failing Sets} \label{sec:criterion}

In Section~\ref{sec:termatiko_sets}, we defined termatiko sets as supports of binary vectors that avert the IPA from recovering any of the ones. However, the algorithm can recover only some of the positions of ones. 

Before proceeding further, we prove the following lemma.

\begin{lemma}\label{lem:supp-subset}
    Let $\vec x, \vec x' \in \{ 0,1 \}^n$ such that $\supp(\vec x) \subset \supp(\vec x')$ and denote $D = \supp(\vec x') \setminus \supp(\vec x)$. Let $\mu^{(\ell)}$ and $M^{(\ell)}$ be respectively lower and upper bounds at the $\ell$-th step of Algorithm~\ref{alg:ipa} on input $(A \vec x, A)$. Also, let $\lambda^{(\ell)}$ and $\Lambda^{(\ell)}$ be respectively lower and upper bounds at the $\ell$-th step of Algorithm~\ref{alg:ipa} on input $(A \vec x', A)$. Then, the following holds:
    \[
    \begin{array}{llllr}
    \lambda_{v \to \cdot}^{(\ell)} 
    &\leq \mu_{v \to \cdot}^{(\ell)} 
    &\leq M_{v \to \cdot}^{(\ell)} 
    &\leq \Lambda_{v \to \cdot}^{(\ell)} \,, 
    & \forall v \notin D \,, \\
    \lambda_{v \to \cdot}^{(\ell)} 
    &\leq \mu_{v \to \cdot}^{(\ell)} + 1 
    &\leq M_{v \to \cdot}^{(\ell)} + 1 
    &\leq \Lambda_{v \to \cdot}^{(\ell)} \,, 
    & \forall v \in D \,.
    \end{array}
    \]
\end{lemma}
\begin{IEEEproof}
    Denote $\vec y = A \vec x$ and $\vec y' = A \vec x'$. Obviously, for any $c \in C$, $y'_c = y_c + |\Neighbours{c} \cap D| \geq y_c$. In particular, for any $c \in \Neighbours[]{D}$, $y'_c \geq y_c + 1$, and for all $c \notin \Neighbours[]{D}$, $y'_c = y_c$.
    
    We prove the lemma by induction. 
    
    {\flushleft \emph{Base Case.}}
    
    Obviously, $\lambda_{v \to \cdot}^{(0)} = \mu_{v \to \cdot}^{(0)} = 0$ for all $v \in V$. Next, if $v \in D$, then $c \in \Neighbours[]{v}$ implies $c \in \Neighbours{D}$ and hence $\Lambda_{v \to \cdot}^{(0)} \geq \min_{c \in \Neighbours{v}} (y_c + 1) = M_{v \to \cdot}^{(0)} + 1$. Analogously, if $v \notin D$, then $\Lambda_{v \to \cdot}^{(0)} \geq M^{(0)}_{v \to \cdot}$.
    
    {\flushleft \emph{Inductive Step.}}
    
    Consider step $\ell \geq 1$. From Line~\ref{alg:ipa:mucv} of Algorithm~\ref{alg:ipa} we have:
    \begin{align*}
    \lambda_{c \to v}^{(\ell)} &= y'_c - \sum_{\substack{v' \in \Neighbours[]{c} \\ v' \neq v}} \Lambda_{v' \to \cdot}^{(\ell - 1)}\\
    &= y_c + |\Neighbours[]{c} \cap D|\\[5pt]
    &\qquad - \sum_{\substack{v' \in \Neighbours[]{c} \cap D \\ v' \neq v}} \Lambda_{v' \to \cdot}^{(\ell - 1)} - \sum_{\substack{v' \in \Neighbours[]{c} \setminus D \\ v' \neq v}} \Lambda_{v' \to \cdot}^{(\ell - 1)} \\
    &\leq y_c + |\Neighbours[]{c} \cap D| - \sum_{\substack{v' \in \Neighbours[]{c} \cap D \\ v' \neq v}} \left( M_{v' \to \cdot}^{(\ell - 1)} + 1 \right)\\
    &\qquad - \sum_{\substack{v' \in \Neighbours[]{c} \setminus D \\ v' \neq v}} M_{v' \to \cdot}^{(\ell - 1)} 
    = \begin{cases}
    \mu_{c \to v}^{(\ell)} \,, & v \notin D \,, \\
    \mu_{c \to v}^{(\ell)} + 1 \,, & v \in D \,.
    \end{cases}
    \end{align*}
    One can show in a similar manner that $\Lambda_{c \to v}^{(\ell)} \geq M_{c \to v}^{(\ell)} + 1$ for $v \in D$ and $\Lambda_{c \to v}^{(\ell)} \geq M_{c \to v}^{(\ell)}$ for $v \notin D$.
    
    Finally, from Lines~\ref{alg:ipa:muvc} and \ref{alg:ipa:Mvc} of Algorithm~\ref{alg:ipa} we obtain
    \begin{align*}
    \lambda_{v \to \cdot}^{(\ell)} &= \max_{c' \in \Neighbours[]{v}} \lambda_{c' \to v}^{(\ell)} \leq \max_{c' \in \Neighbours[]{v}} \mu_{c' \to v}^{(\ell)} = \mu_{v \to \cdot}^{(\ell)} \,, \text{ for } v \notin D \,, \\
    \lambda_{v \to \cdot}^{(\ell)} &= \max_{c' \in \Neighbours[]{v}} \lambda_{c' \to v}^{(\ell)} \leq \mu_{v \to \cdot}^{(\ell)} + 1 \,, \text{ for } v \in D \,, \\
    \Lambda_{v \to \cdot}^{(\ell)} &= \min_{c' \in \Neighbours[]{v}} \Lambda_{c' \to v}^{(\ell)} \geq M_{v \to \cdot}^{(\ell)} \,, \text{ for } v \notin D \,, \\
    \Lambda_{v \to \cdot}^{(\ell)} &= \min_{c' \in \Neighbours[]{v}} \Lambda_{c' \to v}^{(\ell)} \geq M_{v \to \cdot}^{(\ell)} + 1\,, \text{ for } v \in D \,. \qedhere
    \end{align*}

    This concludes the proof.
\end{IEEEproof}

The next theorem presents a connection between (partial) failures of the IPA and termatiko sets. In particular, it shows that the IPA fails on any signal in $\Reals^n_{\geq 0}$ if and only if its support contains a nonempty termatiko set.

\begin{thm}
    The IPA fails to recover a nonnegative real signal $\vec x \in \Reals^n_{\geq 0}$ if and only if the support of $\vec x$ contains a nonempty termatiko set.
\end{thm}

\begin{IEEEproof}
    Assume that $\vec x' \in \{0,1\}^n$ is a binary signal and $T$ is a nonempty termatiko set such that $T \subset \supp(\vec x')$. We also consider a binary $\vec x \in \{0,1\}^n$ with $\supp(\vec x) = T$.
    
    Since $T$ is a termatiko set, on each step of $\IPA(A \vec x, A)$ lower bounds on variable nodes in $T$ will be zeros. Further application of Lemma~\ref{lem:supp-subset} to $\vec x$ and $\vec x'$ shows that lower bounds on variable nodes in $T$ will be zeros also on each step of $\IPA(A \vec x', A)$ and, therefore, these positions will be incorrectly recovered as zeros.
\end{IEEEproof}

\subsection{Counter-Example to \cite[Thm.~2]{RDVD12}} \label{sec:counter-example}


\revtwo{In \cite[Thm.~2]{RDVD12}, a condition for full recovery of $\vec x$ is stated. For convenience, the result is stated below in Theorem~\ref{thm:4} using our notation.}

\revtwo{
	\begin{thm}[\hspace{-.01pt}{{\cite[Thm.~2]{RDVD12}}}] \label{thm:4}
		Let $A \in \{0,1\}^{m \times n}$ be a binary measurement matrix and $V_S = 	\{ v_1, v_2, \dotsc, v_k \}$ be a subset of variable nodes forming a minimal stopping set. Let $\vec x \in \Reals_{\geq 0}^n$ be a signal with at most $k-2$ nonzero values, i.e., $\lVert \vec x \rVert_0 \le k-2$, such that $\supp(\vec x) \subset V_S$. Then, the IPA can recover $\vec x$ if there exists at least one zero measurement node among the neighbors of $V_S$.
	\end{thm}
}

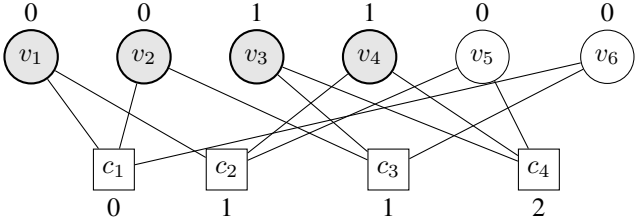
\begin{figure}
	\centering
	\begin{tikzpicture}
	\node at (0.0,+1.5) [circle,draw,thick,fill=black!10] (v1) {$v_{1}$};
	\node at (0.0,2.1) {0};
	\node at (1.5,+1.5) [circle,draw,thick,fill=black!10] (v2) {$v_{2}$};
	\node at (1.5,2.1) {0};
	\node at (3.0,+1.5) [circle,draw,thick,fill=black!10] (v3) {$v_{3}$};
	\node at (3.0,2.1) {1};
	\node at (4.5,+1.5) [circle,draw,thick,fill=black!10] (v4) {$v_{4}$};
	\node at (4.5,2.1) {1};
	\node at (6.0,+1.5) [circle,draw] (v5) {$v_{5}$};
	\node at (6.0,2.1) {0};
	\node at (7.66,+1.5) [circle,draw] (v6) {$v_{6}$};
	\node at (7.66,2.1) {0};
	
	\node at (1.1,0) [square,draw] (c1) {$c_{1}$};
	\node at (1.1,-.5) {0};
	\node at (2.6,0) [square,draw] (c2) {$c_{2}$};
	\node at (2.6,-.5) {1};
	\node at (4.75,0) [square,draw] (c3) {$c_{3}$};
	\node at (4.75,-.5) {1};
	\node at (6.75,0) [square,draw] (c4) {$c_{4}$};
	\node at (6.75,-.5) {2};
	
	\draw (c1) -- (v1);
	\draw (c1) -- (v2);
	\draw (c1) -- (v6);
	
	\draw (c2) -- (v1);
	\draw (c2) -- (v4);
	\draw (c2) -- (v5);
	
	\draw (c3) -- (v2);
	\draw (c3) -- (v3);
	\draw (c3) -- (v6);
	
	\draw (c4) -- (v3);
	\draw (c4) -- (v4);
	\draw (c4) -- (v5);
	
	\end{tikzpicture}
	\caption{Counter-example to \cite[Thm.~2]{RDVD12}. The set of variable nodes is $V = \{v_1,\dotsc,v_6\}$ (circles) and the set of measurement nodes is $C = \{c_1,\dotsc,c_4\}$ (squares).  The integer attached to a node is its corresponding value ($x_{v_i}$ for variable node $v_i$ and $y_{c_i}$ for measurement node $c_i$). $V_S = \{ v_1,v_2,v_3,v_4\} \subset V$  (shaded in gray) is a minimal stopping set and $c_1$ is a zero-valued ($y_{c_1}=0$) measurement node connected to $V_S$.  Note that $v_5$ is not in $V_S$, but exactly because of it, the IPA cannot correctly recover $v_4$.}
	\label{fig:counter-th2} 
\end{figure}

 However, in Fig.~\ref{fig:counter-th2}, we provide a counter-example to this theorem. Note that the Tanner graph of Fig.~\ref{fig:counter-th2} is $(2,3)$-regular (only regular Tanner graphs with variable node degree at least two were considered in \cite{RDVD12}) and satisfies the conditions of \cite[Thm.~2]{RDVD12}. In particular, there are at most $|V_S|-2=2$ nonzero-valued variable nodes which are both in $V_S$ ($V_S$ is a minimal stopping set contained in $V$); and there is at least one zero-valued measurement node among the neighbors of $V_S$. However, it can be readily seen that the IPA will output $\hat{\vec x} = (0,0,1,0,0,0)$, i.e., it recovers only one nonzero variable node ($v_4$ and $v_5$ are both connected to $c_2$ and $c_4$ and thus indistinguishable; hence, the IPA will definitely fail).
We believe that the main problematic issue in the proof given in \cite{RDVD12} is that variable nodes 
outside of the minimal stopping set $V_S$ are not considered. Despite the fact that such
nodes will be recovered as zeros in the end (because of the specific implementation of the IPA, see Algorithm~\ref{alg:hatxv}), during iterations they still can ``disturb'' the values inside of the stopping set.

Finally, we remark that since the statement of \cite[Thm.~2]{RDVD12} is used in the  proof of \cite[Thm.~3]{RDVD12}, the latter should be further verified. \revtwo{The correctness of \cite[Thm.~3]{RDVD12} is left as an open problem.}

\subsection{Heuristics to Find Small-Size Termatiko Sets} \label{sec:alg_small_size}

As shown above, 
stopping sets may contain termatiko sets as proper subsets (and every stopping set is a termatiko set by itself). Thus, one way to locate termatiko sets is to first enumerate all stopping sets of size at most $\tau$ (for a given binary measurement matrix and threshold $\tau$) and \revone{then check which of their subsets are termatiko sets}. 
%
%
For a given binary measurement matrix $A$, small-size stopping sets can be identified using the algorithm from \cite{ros12,ros09}. \revone{In the rest of this paper, we refer to this method as Heuristic 1. Note that termatiko sets of size smaller than $\tau$ can be missed by Heuristic 1. This is the case when they are proper subsets of stopping sets of size larger than $\tau$ (and such stopping sets are not considered by the heuristic). To find the \emph{exact} number of termatiko sets of a given size one would have to run exhaustively through all coordinate subsets of that particular size. } 

Next, we present another heuristic approach that targets the subclass of termatiko sets mentioned in Section~\ref{sec:termatiko_sets}, namely, the case when all $c \in N$ are connected to both $T$ and $S$.  
This symmetry leads to the observation that both $T$ and $S$ are termatiko sets. Therefore, we can try to split a stopping set into two disjoint termatiko sets, $T$ and $S$. We call stopping sets that allow such a split \emph{splittable}.

Consider a stopping set $D \subset V$. Our goal is to split the variable nodes from $D$ into two disjoint sets $T$ and $S$ such that $D = T \cup S$ and each $c \in N = \Neighbours[]{D}$ is connected to both $T$ and $S$. 
The heuristic greedy algorithm outlined in Algorithm~\ref{alg:SPLIT} tries to find such a split by painting (green or red) the variable nodes in $D$. The whole algorithm is based on the following idea.\footnote{In some sense, this algorithm is similar to iterative decoding of LDPC codes over the binary erasure channel (BEC), where the algorithm looks for check nodes that have all-but-one neighboring variable node known, thus making the recovery of such a variable node trivial.} If there is a $c \in N$ such that all its neighbors in $D$ except exactly one have already been painted to the same color, then the remaining node should be painted the color opposite to other neighbors of $c$.  In the algorithm, the color of variable node $v \in D$ is denoted by $col_v$.  It starts with a random node, paints it green (Line~\ref{alg:split:init-green}), and puts it into a working set $Q$ of ``freshly-painted'' nodes. Further, at each iteration, it takes a random variable node $v$ from $Q$ and constructs the set of variable nodes $Opp$. A node $u \in D$ is included in $Opp$ if it is not colored and also connected via some $c$ to $v$ and all the neighbors of $c$ in $D$ except $u$ have the same color (Line~\ref{alg:split:Opp}). By our heuristic assumption, we paint all the variable nodes in $Opp$ the color opposite to the color of $v$ (Line~\ref{alg:split:paint}). Further, all the elements of $Opp$ are added to $Q$ for further processing (Line~\ref{alg:split:add-to-Q}). If at some point $Q$ becomes empty but not all the variable nodes from $D$ have been painted yet, the algorithm has nothing better to do than just randomly guess a color of some variable node that has not been painted yet (Line~\ref{alg:split:RG-start} to Line~\ref{alg:split:RG-end}). Algorithm~\ref{alg:SPLIT} finishes when $Q$ becomes empty and all the variable nodes from $D$ have been painted. After that, in Line~\ref{alg:split:exists_start} to Line~\ref{alg:split:exists_end}, the algorithm verifies the obtained solution for correctness to the stated goal, i.e., whether each $c \in N$ is connected both to $T$ and $S$. In turn, from this it follows that both $T$ and $S$ are termatiko sets. If so, it returns the pair $(T,S)$, otherwise it returns $\textsf{FAIL}$.

We remark that by changing the  randomized steps of Algorithm~\ref{alg:SPLIT} into a branching step, one can get an exhaustive search algorithm that outputs all the splits $(T,S)$ with the stated property (each $c \in N$ is connected to both $T$ and $S$).

\begin{algorithm}
	\caption{Splitting a Stopping Set $D \subset V$}
	\label{alg:SPLIT}
	
	\begin{algorithmic}[1]
		\Function {\textsf{SPLIT}}{$D \subset V$}
		
		\Statex \textit{Initialization}
		
		\State $N \leftarrow \Neighbours[]{D}$
		\FForAll {$v \in D$} $col_v \leftarrow \mathsf{?}$ \EEndFor
		\State $v \xleftarrow{\mathsf{rnd}} D$
		\State $col_v \leftarrow \mathsf{GREEN}$	\label{alg:split:init-green}
		\State $Q \leftarrow \{ v \}$

		\Statex \textit{Iterations}
		
		\While {$Q \neq \varnothing$}
		\State $v \xleftarrow{\mathsf{rnd}} Q$
		\State $Q \leftarrow Q \setminus \{ v \}$
		\If {$col_v = \mathsf{GREEN}$} 
		\State $OppCol \leftarrow \mathsf{RED}$
		\Else {} 
		\State $OppCol \leftarrow \mathsf{GREEN}$
		\EndIf
		\State $Opp \leftarrow \{ u \in D \,:\, col_u = \mathsf{?} \text{ and } \exists c \in \Neighbours[]{u} \cap \Neighbours[]{v} \text{ s.t. } \forall v' \in \Neighbours[D]{c} \setminus \{ u \}, col_{v'} = col_v \}$ \label{alg:split:Opp}
		\FForAll {$u \in Opp$} $col_u \leftarrow OppCol$ \EEndFor \label{alg:split:paint}
		\State $Q \leftarrow Q \cup Opp$ \label{alg:split:add-to-Q}
		
		\If {$Q = \varnothing$ and $\{ u \in D \,:\, col_u = \mathsf{?} \} \neq \varnothing$}
		\State $v \xleftarrow{\mathsf{rnd}} \{ u \in D \,:\, col_u = \mathsf{?} \}$ \label{alg:split:RG-start}
		\State $OppCol \xleftarrow{\mathsf{rnd}} \{\mathsf{GREEN}, \mathsf{RED}\}$
		\State $col_v \leftarrow OppCol$
		\State $Q \leftarrow \{ v \}$ \label{alg:split:RG-end}
		\EndIf
		\EndWhile
		
		\Statex \textit{Check if the result is correct}
		
		\If {$\exists c \in N$ s.t. $|\{ col_v \,:\, v \in \Neighbours[D]{c} \}| = 1$} \label{alg:split:exists_start}
		\State \textbf{return} \textsf{FAIL}
		\EndIf \label{alg:split:exists_end}
		
		\Statex \textit{Result}
		
		\State $T \leftarrow \text{ variable nodes painted } \mathsf{GREEN}$
		\State $S \leftarrow  \text{ variable nodes painted } \mathsf{RED}$
		\State \textbf{return} $(T,S)$
		
		\EndFunction
	\end{algorithmic}
\end{algorithm}

\section{Column-Regular Measurement Matrices} \label{sec:col-reg}
In this section, we present some results for column-regular measurement matrices, i.e., those having the same amount of nonzero entires in each column. The first result is a lower bound on the termatiko distance $h_{\rm min}$.

\begin{thm} \label{thm:col-a-reg}
	The termatiko distance of a column $a$-regular measurement matrix with no cycles of length $4$ is at least $a$.
\end{thm}
\begin{IEEEproof}
	Assume to the contrary that we have a termatiko set $T = \{v_1, v_2, \dotsc, v_t\}$ of size $t \leq a-1$. Define $N$ and $S$ as in Theorem~\ref{thm:termatikos}.
	
	First assume that $S \neq \varnothing$. Take any $u \in S$. Also, split $N$ into $t$ nonintersecting subsets $N_1, \dotsc,N_t$ such that $N = N_1 \cup N_2 \cup \dotsb \cup N_t$, where
	\[
	\begin{array}{rl}
	N_1 =& \Neighbours[]{v_1} \,, \\
	N_2 =& \Neighbours[]{v_2} \setminus N_1 \,, \\
	N_3 =& \Neighbours[]{v_3} \setminus N_2 \,, \\
	\dots \\
	N_t =& \Neighbours[]{v_t} \setminus N_{t-1} \,.
	\end{array}
	\]
	
	As the measurement matrix has no cycles of length $4$, no variable nodes can share more than one measurement node. In particular, $u$ cannot share more than one measurement node with any of $v_1, v_2, \dotsc, v_t$. Therefore, $u$ is connected not more than once to each of the sets $N_1, N_2, \dotsc, N_t$, and thus $|\Neighbours[]{u}| \leq t \leq a-1$, which contradicts the fact that the degree of each variable node is $a$, and it follows that $S = \varnothing$.
	
	Since $S = \varnothing$, each measurement node in $N$ should be connected to $T$ at least twice. Furthermore, since the degree of each variable node is $a$, we have $|N| \leq a t / 2$. On the other hand, by definition, we have $|\Neighbours[]{v_j}| = a$  and $\Neighbours[]{v_j}$ \revtwo{does not share} more than one element with each of $\Neighbours[]{v_{j-1}}, \Neighbours[]{v_{j-2}}, \dotsc, \Neighbours[]{v_{1}}$. Therefore,
	\[
	|N_j| = \Big| \Neighbours[]{v_j} \setminus \Neighbours[]{v_{j-1}} \setminus \Neighbours[]{v_{j-2}} \setminus \dotsb \setminus \Neighbours[]{v_{1}} \Big| \geq a-j+1 \,,
	\]
	and we get
	\[
	|N| = \left| \bigcup_{j=1}^t N_j \right| \geq at - \frac{t(t-1)}{2} \,.
	\]
	It follows that
	\[
	at - \frac{t(t-1)}{2} \leq |N| \leq \frac{a t}{2} \,,
	\]
	from which we get that $t \geq a+1$. This is a contradiction since we have assumed that $t \leq a-1$.
\end{IEEEproof}

As each stopping set is a termatiko set and each codeword support is a stopping set, we have that $h_{\rm min} \leq s_{\rm min} \leq d_{\rm min}$ (for any measurement matrix). Hence, the following result can be seen a corollary of Theorem~\ref{thm:col-a-reg}.

\begin{cor}
	For a column $a$-regular parity-check matrix, $d_{\rm min} \geq s_{\rm min} \geq a$.
\end{cor}

\subsection{Measurement Matrices From Array Low-Density Parity-Check Codes} \label{sec:array_based}
\revtwo{Next, we consider a particular case of column $a$-regular measurement matrices, namely, the parity-check matrices of array LDPC codes \cite{fan00}.} For a prime $q > 2$ and an integer $a < q$ the array LDPC code $\mathcal C(q,a)$ has length $q^2$ and can be defined by the parity-check matrix
\[
H(q,a) = 
\begin{pmatrix}
I      & I       & I          & \cdots & I \\
I      & P       & P^2        & \cdots & P^{q-1} \\
I      & P^2     & P^4        & \cdots & P^{2(q-1)} \\
\vdots & \vdots  & \vdots     & \ddots & \vdots \\
I      & P^{a-1} & P^{2(a-1)} & \cdots & P^{(a-1)(q-1)}
\end{pmatrix}\,,
\]
where $I$ is the $q \times q$ identity matrix and $P$ is a $q \times q$ permutation matrix defined by
\[
P =
\begin{pmatrix}
0      & 0      & \cdots & 0      & 1 \\
1      & 0      & \cdots & 0      & 0 \\
\vdots & \vdots & \ddots & \vdots & \vdots \\
0      & 0      & \cdots & 1      & 0 \\
\end{pmatrix} \,.
\]
It is easy to see that $\mathcal C(q,a)$ is an $(a,q)$-regular code of dimension $q^2-qa+a-1$, and its minimum distance will be denoted by $d(q,a)$. 

In \cite{yan03}, a new representation of $H(q,a)$ was introduced. In particular, since each column of the parity-check matrix $H(q,a)$ has $a$ blocks and each block is a permutation of $(1,0,0,\dotsc,0)^T$, we can represent each column as a length-$a$ column vector of elements from $\FF_q$, the field of integers modulo $q$. More precisely, $i \in \FF_q$ is bijectively mapped to a vector
\[
\left( \overbrace{0,\dotsc,0}^{i}, 1, \overbrace{0,\dotsc,0}^{q-i-1} \right)^T  \,,
\]
and any column in $H(q,a)$ is of the form
\begin{equation} \label{eq:arrcode-cw-form}
(i, i+j, i+2j, \dotsc, i+(a-1)j)^T \quad (\operatorname{mod} \,q)
\end{equation}
for some $i, j \in \FF_q$. Note that in \eqref{eq:arrcode-cw-form} the field elements $i$ and $j$ are considered as integers and the operations (addition and multiplication) are standard integer operations, while $(\operatorname{mod} \,q)$ denotes integer reduction modulo $q$.  In the following, with some abuse of notation, a field element from $ \FF_q$ and its integer representation are used interchangeably. Furthermore,  addition, subtraction, and multiplication might be either standard integer addition, integer subtraction, and integer multiplication, or denote field operations. However,  this will become clear from the context. Also, note that since there are $q^2$ distinct columns in $H(q,a)$, any pair $(i,j) \in \FF_q^2$ specifies a valid column. Therefore, the columns of $H(q,a)$ (or variable nodes $V$) can be identified with pairs $(i, j) \in \FF_q^2$.

Further, as rows of the matrix can be split into $a$ blocks of $q$ rows each, it is convenient to identify rows of $H(q, a)$ (or measurement nodes $C$) with pairs in $\ZZ_a \times \FF_q$, so that the $j$-th row ($1 \leq j \leq aq$) is identified (or indexed) by\footnote{$\ZZ_a$ denotes the ring of integers modulo $a$, and we use angular brackets for measurement nodes to clearly differentiate between $C$ and $V$.} 
\[
\left\langle \left\lfloor (j-1)/q \right\rfloor, (j-1) \; (\operatorname{mod} q) \right\rangle \,.
\]
In other words, row $1$ is indexed by $\langle 0,0 \rangle$, row $2$ by $\langle 0,1 \rangle$, up to row $q$ which is indexed by $\langle 0,q-1 \rangle$, row $q+1$ by $\langle 1,0 \rangle$, and so on. With this notation, variable node $(i,j) \in V = \FF_q^2$ is connected to \revtwo{each measurement node in the set} $\{\langle 0,i \rangle, \langle 1,i+j \rangle, \langle 2,i+2j\rangle, \dotsc, \langle a-1,i+(a-1)j \rangle\} = \{\langle s,i+sj \rangle \,|\, s \in \ZZ_a \}$. 

For $s \in \ZZ_a$, we call the $q$ consecutive rows (or, equivalently, measurement nodes) indexed by $\{\langle s,0 \rangle, \langle s,1 \rangle, \dotsc, \langle s,q-1 \rangle\}$ the \emph{$s$-th strip}. We will extensively use the fact that every variable node has exactly one neighboring measurement node in each of the strips.

Define the permutations $\varphi : \FF_q^2 \mapsto \FF_q^2$ and $\psi : \ZZ_a \times \FF_q \mapsto \ZZ_a \times \FF_q$, with parameters $\alpha \in \FF_q \setminus \{0\}$, $\beta_1,\beta_2 \in \FF_q$, by\footnote{$\varphi(i,j)$ and $\psi(s, t)$ are shorthand notations for $\varphi((i,j))$ and $\psi(\langle  s,t \rangle)$, respectively.}
\begin{align*}
\varphi(i,j) &= (\alpha i + \beta_1, \alpha j + \beta_2) \,, \\
\psi(s,t) &=\left\langle s,\alpha t + (\beta_1 + s \beta_2) \right\rangle \,.
\end{align*}
It is well-known (cf.\ \cite[Lem.~2]{yan03}) that $\mathcal C(q,a)$ is invariant under the doubly transitive group of ``affine'' permutations  defined above. In other words, such a pair of transformations is an automorphism on the Tanner graph of an array LDPC code, i.e., $\langle s,t \rangle \in \Neighbours{(i,j)}$ if and only if $\psi(s,t) \in \Neighbours{\varphi(i,j)}$ for all choices of $\alpha,\beta_1,\beta_2$. In particular, $T = \{v_1, v_2, \dotsc, v_{|T|}\}$ is a termatiko set if and only if $\{\varphi(v_1), \varphi(v_2), \dotsc, \varphi(v_{|T|})\}$ is a termatiko set. The number of choices for $\alpha, \beta_1, \beta_2$ is $q^2(q-1)$ and this is the number of different automorphisms of this particular type, one of them being the identity (when $\alpha=1$, $\beta_1=\beta_2=0$). Furthermore, it is also well-known that there are no cycles of length $4$ in the Tanner graph corresponding to the parity-check matrix of an array LDPC code \cite{fan00}. 

In the following, the \textit{support matrix} of a subset of variable nodes, $U \subset V$, will be the submatrix of $H(q,a)$ consisting of the columns indexed by $U$. Furthermore, the \textit{support matrix of a codeword} is the support matrix of the support of the codeword. We will mostly write the support matrix in a compact form using the representation in \eqref{eq:arrcode-cw-form}, i.e., as an $a \times |U|$ matrix over $\FF_q$. For example, the support matrix of some subset $\{ (i_1, j_1), (i_2, j_2), (i_3, j_3) \} \subset V$ of three variable nodes  is written as\footnote{Recall that we equate $V$ with $\FF_q^2$.} 
\[
\begin{bmatrix}
i_1            & i_2            & i_3 \\
i_1 + j_1      & i_2 + j_2      & i_3 + j_3 \\
i_1 + 2 j_1    & i_2 + 2 j_2    & i_3 + 2 j_3 \\
\cdots         & \cdots         & \cdots \\
i_1 + (a-1)j_1 & i_2 + (a-1)j_2 & i_3 + (a-1)j_3 \\
\end{bmatrix} \,.
\]

\subsection{Termatiko Distance Multiplicity of $H(q,3)$}

Consider the array LDPC code $\mathcal C(q,3)$. It is $(3, q)$-reqular and each column of its parity-check matrix $H(q,3)$ can be represented by the vector $(i, i+j, i+2j)^T \in \FF_q^3$, from which it follows that if $v \in V$ is connected to $c_1 = \langle 0,s_1 \rangle$, $c_2 = \langle 1,s_2 \rangle$, and $c_3 = \langle 2,s_3 \rangle$, then $2 s_2 = s_1 + s_3$ (i.e., $s_1, s_2, s_3$ form an arithmetic progression).

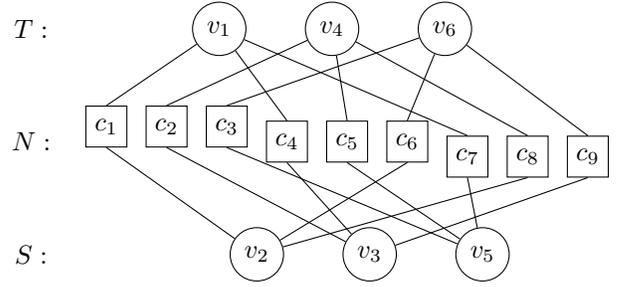
\begin{figure}
	\centering
	\begin{tikzpicture}
	\node at (2.5,+1.5) [circle,draw] (v1) {$v_1$};
	\node at (4.0,+1.5) [circle,draw] (v4) {$v_4$};
	\node at (5.5,+1.5) [circle,draw] (v6) {$v_6$};
	
	\node at (3.0,-1.5) [circle,draw] (v2) {$v_2$};
	\node at (4.5,-1.5) [circle,draw] (v3) {$v_3$};
	\node at (6.0,-1.5) [circle,draw] (v5) {$v_5$};
	
	\node at (1.0,0.2) [square,draw] (c1) {$c_1$};
	\node at (1.8,0.2) [square,draw] (c2) {$c_2$};
	\node at (2.6,0.2) [square,draw] (c3) {$c_3$};
	\node at (3.4,0.0) [square,draw] (c4) {$c_4$};
	\node at (4.2,0.0) [square,draw] (c5) {$c_5$};
	\node at (5.0,0.0) [square,draw] (c6) {$c_6$};
	\node at (5.8,-0.2) [square,draw] (c7) {$c_7$};5.8
	\node at (6.6,-0.2) [square,draw] (c8) {$c_8$};6.6
	\node at (7.4,-0.2) [square,draw] (c9) {$c_9$};7.4
	
	\draw (v1) -- (c1.north);
	\draw (v1) -- (c4.north);
	\draw (v1) -- (c7.north);
	\draw (v2) -- (c1.south);
	\draw (v2) -- (c6.south);
	\draw (v2) -- (c8.south);
	\draw (v3) -- (c2.south);
	\draw (v3) -- (c4.south);
	\draw (v3) -- (c9.south);
	\draw (v4) -- (c2.north);
	\draw (v4) -- (c5.north);
	\draw (v4) -- (c8.north);
	\draw (v5) -- (c3.south);
	\draw (v5) -- (c5.south);
	\draw (v5) -- (c7.south);
	\draw (v6) -- (c3.north);
	\draw (v6) -- (c6.north);
	\draw (v6) -- (c9.north);
	
	\node at (0,+1.5) {$T:$};
	\node at (0,0) {$N:$};
	\node at (0,-1.5) {$S:$};
	\end{tikzpicture}
	\caption{Termatiko set of size $3$ in $H(q,3)$. Measurement nodes $c_1, c_2, \dotsc, c_9$ are grouped according to being in the first, second, and third strip in $H(q,3)$.}
	\label{fig:arr-T3}
\end{figure}

\begin{thm} \label{thm:Hq3-T3}
	There are $q^2 (q-1)(q-2)/3$ termatiko sets of minimum size $3$ in $H(q,3)$ for any $q \geq 5$ and their support matrices have (up to automorphisms) one of the forms
	\[
		\begin{bmatrix}
		0 & 2    & -2-2j \\
		0 & 2+j  & 1     \\
		0 & 2+2j & 4+2j  \\
		\end{bmatrix}
		\text{ or }
		\begin{bmatrix}
		0 & 2    & 4+2j \\
		0 & 2+j  & 1+j  \\
		0 & 2+2j & -2   \\
		\end{bmatrix}\,,
	\]
	for any $j \in \FF_q \setminus \{ q-1, q-2 \}$.
\end{thm}

\begin{IEEEproof}
See the appendix.
\end{IEEEproof}

We remark that this formula is similar to the formula for the number of weight-$6$ codewords in $\mathcal{C}(q,3)$ provided in \cite[Thm.~2]{liu10}. In fact, \revtwo{we observe that} the number of termatiko sets of size $3$ is twice the number of codewords of weight $6$. Fig.~\ref{fig:arr-T3} provides an illustration of a termatiko set of size $3$ in $H(q,3)$.

\subsection{Upper Bound on the Termatiko Distance of $H(q,a)$}

For $H(q, a)$\revus{,} it follows from Theorem~\ref{thm:col-a-reg} that the termatiko distance $h_{\rm min} \geq a$, and from Theorem~\ref{thm:Hq3-T3} it follows that  the bound is indeed tight for $a=3$. In this subsection, we derive upper bounds on the termatiko distance when $4 \leq a\leq 7$. 
The approach is inspired by the following observation. 


It was shown in \cite{mit02} that $d(q,3)=6$, and in \cite{yan03} the authors derived the explicit support matrix 
\begin{equation*} \label{eq:cw6}
	\begin{bmatrix}
	\bm{0} & 0     & 2i-2j  & \bm {2i-2j}  & -2i & \bm {-2i}    \\
	\bm{0} & -2i+j & 0      & \bm {-i}     & -i  & \bm {-2i+j}  \\
	\bm{0} & -4i+2j & -2i+2j & \bm {-4i+2j} & 0   & \bm{-2i+2j}
	\end{bmatrix}
\end{equation*}
(up to equivalence under the aforementioned automorphisms) for codewords of weight $6$, 
where $i \in \FF_q \setminus \{0\}$ and $j \in \FF_q$ with $j \neq i, 2i$. 
It is worth noting that the columns $1$, $4$, and $6$ (marked in bold) of the support matrix above form a termatiko set. The same is true for the columns $2$, $3$, and $5$. Hence, the support of each minimum-weight codeword in $H(q,3)$ can be split into two size-$3$ termatiko sets.

Deriving upper bounds on the minimum distance of array LDPC codes has attracted some attention, and tight bounds have been derived for $4 \leq a\leq 7$ in \cite{sug08,ros14}. In these works, explicit support matrices of codewords have been tabulated.  A further exploration of these support matrices shows that a  half-and-half split into two termatiko sets is possible; the connected measurement nodes are connected to both termatiko sets. We can now successfully apply Algorithm~\ref{alg:SPLIT} to some known cases.

\newlength{\btwnmtrxwidth}
\setlength{\btwnmtrxwidth}{6pt}
\newlength{\submtrxwidth}
\setlength{\submtrxwidth}{0.47\textwidth}

\begin{figure*}
	\centering
	\begin{minipage}[b]{\submtrxwidth}
		\[
		\left[
		\arraycolsep=3pt\def\arraystretch{0.75}\footnotesize
		\begin{array}{ccc|ccc}
		0 & 2 i-2 j & -2 i & 0 & 2 i-2 j & -2 i \\
		0 & -i & -2i+j  & -2 i +j& 0 & -i \\
		0 & -4 i +2j& -2 i +2j& -4i +2j & -2i +2j & 0 \\
		\end{array}
		\right]
		\]
		\caption{Codeword support matrix of a weight-$6$ codeword of $H(q,3)$. The vertical line illustrates how to split the codeword support into two distinct  termatiko sets each of half the size.}
		\label{fig:gen_form}
	\end{minipage}
	\hspace{\columnsep}
	\begin{minipage}[b]{\submtrxwidth}
		\[
		\left[
		\arraycolsep=3pt\def\arraystretch{0.75}\footnotesize
		\begin{array}{ccccc|ccccc}
		0 & -6 & -24 & -12 & -30 & 0 & -12 & -24 & -6 & -30 \\
		0 & 3 & -13 & -4 & -12 & 3 & 0 & -12 & -4 & -13 \\
		0 & 12 & -2 & 4 & 6 & 6 & 12 & 0 & -2 & 4 \\
		0 & 21 & 9 & 12 & 24 & 9 & 24 & 12 & 0 & 21 \\
		\end{array}
		\right]
		\]
		\caption{Codeword support matrix of a weight-$10$ codeword of $H(q,4)$ for $q \geq 11$. The vertical line illustrates how to split the codeword support into two distinct  termatiko sets each of half the size.}
		\label{fig:split-Hq4}
	\end{minipage}

	~\\[20pt]

	\begin{minipage}[b]{\submtrxwidth}
		\[
		\left[
		\arraycolsep=3pt\def\arraystretch{0.75}\footnotesize
		\begin{array}{cccc|cccc}
		0 & 3 k+3 z & 2 k+4 z & 2 z & 0 & 3 k+3 z & 2 k+4 z & 2 z \\
		0 & 3 z & k+4 z & k+2 z & k+4 z & 0 & k+2 z & 3 z  \\
		0 & 2 k+3 z & 4 z & 2 k+2 z & 2 k+3 z & 2 k+2 z & 0 & 4 z  \\
		0 & 4 k+3 z & 4 k+4 z & 3 k+2 z & 3 k+2 z & 4 k+4 z & 4 k+3 z & 0 \\
		\end{array}
		\right]
		\]
		\caption{Codeword support matrix of a weight-$8$ codeword of $H(5,4)$ for $z \in \FF_5 \setminus \{0\}$ and $k \in \{0, 2z\}$. The vertical line illustrates how to split the codeword support into two distinct  termatiko sets each of half the size.}
		\label{fig:split-Hq5a4}
	\end{minipage}
	\hspace{\columnsep}
	\begin{minipage}[b]{\submtrxwidth}
		\[
		\left[
		\arraycolsep=3pt\def\arraystretch{0.75}\footnotesize
		\begin{array}{cccc|cccc}
		0 & 2 k+5 z & 2 k+z & 4 z & 0 & 2 k+5 z & 2 k+z & 4 z \\
		0 & k+2 z & 5 z & k+4 z & k+2 z & 0 & k+4 z & 5 z \\
		0 & 6 z & 5 k+2 z & 2 k+4 z & 2 k+4 z & 5 k+2 z & 0 & 6 z \\
		0 & 6 k+3 z & 3 k+6 z & 3 k+4 z & 3 k+6 z & 3 k+4 z & 6 k+3 z & 0 \\
		\end{array}
		\right]
		\]
		\caption{Codeword support matrix of a weight-$8$ codeword of $H(7,4)$ for $z \in \FF_7 \setminus \{0\}$ and $k \in \{0, 2z, 4z, 6z\}$. The vertical line illustrates how to split the codeword support into two distinct  termatiko sets each of half the size.}
		\label{fig:split-Hq7a4}
	\end{minipage}

	~\\[20pt]

	\begin{minipage}[b]{\submtrxwidth}
		\[
		\left[
		\arraycolsep=3pt\def\arraystretch{0.75}\footnotesize
		\begin{array}{cccccc|cccccc}
		0 & -4 & -18 & -22 & -6 & -16 & 0 & -6 & -22 & -18 & -4 & -16 \\
		0 & 1 & -8 & -12 & -3 & -11 & 1 & 0 & -11 & -12 & -3 & -8 \\
		0 & 6 & 2 & -2 & 0 & -6 & 2 & 6 & 0 & -6 & -2 & 0 \\
		0 & 11 & 12 & 8 & 3 & -1 & 3 & 12 & 11 & 0 & -1 & 8 \\
		0 & 16 & 22 & 18 & 6 & 4 & 4 & 18 & 22 & 6 & 0 & 16 \\
		\end{array}
		\right]
		\]
		\caption{Codeword support matrix of a weight-$12$ codeword of $H(q,5)$. The vertical line illustrates how to split the codeword support into two distinct  termatiko sets each of half the size.}
		\label{fig:split-Hq5}
	\end{minipage}
	\hspace{\columnsep}
	\begin{minipage}[b]{\submtrxwidth}
		\[
		\left[
		\arraycolsep=3pt\def\arraystretch{0.75}\footnotesize
		\begin{array}{ccccc|ccccc}
		0 & 5 & 4 & 7 & 6 & 7 & 4 & 0 & 5 & 6 \\
		1 & 0 & 10 & 8 & 3 & 1 & 3 & 10 & 8 & 0 \\
		2 & 6 & 5 & 9 & 0 & 6 & 2 & 9 & 0 & 5 \\
		3 & 1 & 0 & 10 & 8 & 0 & 1 & 8 & 3 & 10 \\
		4 & 7 & 6 & 0 & 5 & 5 & 0 & 7 & 6 & 4 \\
		\end{array}
		\right]
		\]
		\caption{Codeword support matrix of a weight-$10$ codeword of $H(11,5)$. The vertical line illustrates how to split the codeword support into two distinct  termatiko sets each of half the size.}
		\label{fig:split-Hq11a5}
	\end{minipage}

	~\\[20pt]
	\begin{minipage}[b]{\textwidth}
	\[
	\left[
	\arraycolsep=3pt\def\arraystretch{0.75}\footnotesize
	\begin{array}{cccccccccc|cccccccccc}
	0 & -22 & -2 & -20 & 10 & -8 & 12 & -10 & -32 & 22 & -10 & -2 & 10 & -32 & 22 & -20 & 0 & -8 & -22 & 12 \\
	0 & -16 & 8 & -8 & 9 & -7 & 17 & 1 & -15 & 16 & -8 & 0 & 16 & -16 & 17 & -15 & 9 & 1 & -7 & 8 \\
	0 & -10 & 18 & 4 & 8 & -6 & 22 & 12 & 2 & 10 & -6 & 2 & 22 & 0 & 12 & -10 & 18 & 10 & 8 & 4 \\
	0 & -4 & 28 & 16 & 7 & -5 & 27 & 23 & 19 & 4 & -4 & 4 & 28 & 16 & 7 & -5 & 27 & 19 & 23 & 0 \\
	0 & 2 & 38 & 28 & 6 & -4 & 32 & 34 & 36 & -2 & -2 & 6 & 34 & 32 & 2 & 0 & 36 & 28 & 38 & -4 \\
	0 & 8 & 48 & 40 & 5 & -3 & 37 & 45 & 53 & -8 & 0 & 8 & 40 & 48 & -3 & 5 & 45 & 37 & 53 & -8 \\
	\end{array}
	\right]
	\]
	\caption{Codeword support matrix of a weight-$20$ codeword of $H(q,6)$. The vertical line illustrates how to split the codeword support into two distinct  termatiko sets each of half the size. }
	\label{fig:split-Hq6}
	\end{minipage}

	~\\[20pt]

		\begin{minipage}[b]{\submtrxwidth}
		\[
		\left[
		\arraycolsep=3pt\def\arraystretch{0.75}\footnotesize
		\begin{array}{cccccc|cccccc}
		0 & 3 & 6 & 2 & 5 & 4 & 2 & 6 & 5 & 4 & 0 & 3 \\
		0 & 6 & 5 & 4 & 3 & 1 & 3 & 0 & 6 & 5 & 1 & 4 \\
		0 & 2 & 4 & 6 & 1 & 5 & 4 & 1 & 0 & 6 & 2 & 5 \\
		0 & 5 & 3 & 1 & 6 & 2 & 5 & 2 & 1 & 0 & 3 & 6 \\
		0 & 1 & 2 & 3 & 4 & 6 & 6 & 3 & 2 & 1 & 4 & 0 \\
		0 & 4 & 1 & 5 & 2 & 3 & 0 & 4 & 3 & 2 & 5 & 1 \\
		\end{array}
		\right]
		\]
		\caption{Codeword support matrix of a weight-$12$ codeword of $H(7,6)$. The vertical line illustrates how to split the codeword support into two distinct  termatiko sets each of half the size.}
		\label{fig:split-Hq7a6}
	\end{minipage}
	\hspace{\columnsep}
	\begin{minipage}[b]{\submtrxwidth}
		\[
		\left[
		\arraycolsep=3pt\def\arraystretch{0.75}\footnotesize
		\begin{array}{cccccccc|cccccccc}
		0 & 10 & 1 & 5 & 7 & 6 & 6 & 0 & 6 & 10 & 5 & 1 & 0 & 7 & 0 & 6 \\
		0 & 4 & 7 & 10 & 2 & 6 & 9 & 8 & 7 & 0 & 8 & 9 & 10 & 6 & 2 & 4 \\
		0 & 9 & 2 & 4 & 8 & 6 & 1 & 5 & 8 & 1 & 0 & 6 & 9 & 5 & 4 & 2 \\
		0 & 3 & 8 & 9 & 3 & 6 & 4 & 2 & 9 & 2 & 3 & 3 & 8 & 4 & 6 & 0 \\
		0 & 8 & 3 & 3 & 9 & 6 & 7 & 10 & 10 & 3 & 6 & 0 & 7 & 3 & 8 & 9 \\
		0 & 2 & 9 & 8 & 4 & 6 & 10 & 7 & 0 & 4 & 9 & 8 & 6 & 2 & 10 & 7 \\
		\end{array}
		\right]
		\]
		\caption{Codeword support matrix of a weight-$16$ codeword of $H(11,6)$. The vertical line illustrates how to split the codeword support into two distinct  termatiko sets each of half the size.}
		\label{fig:split-Hq11a6}
	\end{minipage}

	~\\[20pt]
	
	\begin{minipage}[b]{\textwidth}
		\[
		\left[
		\arraycolsep=3pt\def\arraystretch{0.75}\footnotesize
		\begin{array}{cccccccccccc|cccccccccccc}
		0 & -18 & -14 & -20 & -8 & -4 & 8 & 2 & 6 & -12 & 10 & -22 & 6 & 0 & -4 & -22 & 8 & -20 & 10 & -12 & -8 & -18 & 2 & -14 \\
		0 & -14 & -10 & -12 & -7 & 1 & 6 & 4 & 8 & -6 & 9 & -15 & 6 & 4 & 0 & -14 & 9 & -15 & 8 & -10 & -6 & -12 & 1 & -7 \\
		0 & -10 & -6 & -4 & -6 & 6 & 4 & 6 & 10 & 0 & 8 & -8 & 6 & 8 & 4 & -6 & 10 & -10 & 6 & -8 & -4 & -6 & 0 & 0 \\
		0 & -6 & -2 & 4 & -5 & 11 & 2 & 8 & 12 & 6 & 7 & -1 & 6 & 12 & 8 & 2 & 11 & -5 & 4 & -6 & -2 & 0 & -1 & 7 \\
		0 & -2 & 2 & 12 & -4 & 16 & 0 & 10 & 14 & 12 & 6 & 6 & 6 & 16 & 12 & 10 & 12 & 0 & 2 & -4 & 0 & 6 & -2 & 14 \\
		0 & 2 & 6 & 20 & -3 & 21 & -2 & 12 & 16 & 18 & 5 & 13 & 6 & 20 & 16 & 18 & 13 & 5 & 0 & -2 & 2 & 12 & -3 & 21 \\
		0 & 6 & 10 & 28 & -2 & 26 & -4 & 14 & 18 & 24 & 4 & 20 & 6 & 24 & 20 & 26 & 14 & 10 & -2 & 0 & 4 & 18 & -4 & 28 \\
		\end{array}
		\right]
		\]
		\caption{Codeword support matrix of a weight-$24$ codeword of $H(q,7)$. The vertical line illustrates how to split the codeword support into two distinct  termatiko sets each of half the size. }
		\label{fig:split-Hqa7}
	\end{minipage}
\end{figure*}

\subsubsection{$H(q,3)$}
Applying Algorithm~\ref{alg:SPLIT} to the aforementioned support matrix we obtain the (correct) split, as depicted in Fig.~\ref{fig:gen_form}. Note that the columns there are reordered so that both the first three and the last three form termatiko sets.

If we set $i = -1$, then we get the first general form from Theorem~\ref{thm:Hq3-T3} (with columns reordered) in the left part of Fig.~\ref{fig:gen_form}. Moreover,  if we set $i=-1$, but also apply an automorphism with $\alpha=1$, $\beta_1=0$, $\beta_2=-2-j$, and finally substitute $j \mapsto -3-j$, then we get the second general form from Theorem~\ref{thm:Hq3-T3} (with columns reordered) in the right part of Fig.~\ref{fig:gen_form}. 

\subsubsection{$H(q,4)$}
In \cite[Fig.~3]{sug08}, the authors presented the support matrix of a weight-$10$ codeword for $H(q,4)$ for $q > 7$. Since $\alpha = 12$ is co-prime with any prime $q > 4$, each matrix entry  in the matrix from  \cite{sug08} can be multiplied by $\alpha = 12$, which is equivalent to applying a doubly transitive automorphism. The resulting matrix becomes 
%
\[
\left[
\arraycolsep=3pt\def\arraystretch{0.75}\footnotesize
\begin{array}{cccccccccc}
0 & 0 & -12 & -24 & -6 & -6 & -24 & -12 & -30 & -30 \\
0 & 3 & 0 & -12 & -4 & 3 & -13 & -4 & -13 & -12 \\
0 & 6 & 12 & 0 & -2 & 12 & -2 & 4 & 4 & 6 \\
0 & 9 & 24 & 12 & 0 & 21 & 9 & 12 & 21 & 24 \\
\end{array}
\right]\,.
\]
Applying Algorithm~\ref{alg:SPLIT} gives the split indicated in Fig.~\ref{fig:split-Hq4} where the columns have been re-ordered. For $q=11$, we exhaustively checked all the $4$-subsets of $\FF_q^2$ and did not find any termatiko sets among them, therefore $h(11,4)=5$. For the special cases $H(5,4)$ and $H(7,4)$, weight-$8$ codeword support matrices were presented in \cite[Thms.~7 and 8]{yan03}. These can be split, and the results of the splits are shown in Figs.~\ref{fig:split-Hq5a4} and \ref{fig:split-Hq7a4}.

\subsubsection{$H(q,5)$}
In \cite[Fig.~4]{sug08}, an explicit support matrix of weight-$12$ codewords from $H(q,5)$ is presented for $q \neq 11$.\footnote{It seems the authors did not verify that the columns of the support matrix are different. However, for $q=11$, two columns are identical. Therefore, we treat $H(11,5)$ as a special case.} Multiplying each entry of the matrix by $\alpha = 6$, which is co-prime with $q > 5$, and applying Algorithm~\ref{alg:SPLIT}  to the resulting matrix results in a half-and-half split (see Fig.~\ref{fig:split-Hq5}). For $q=7$, we verified exhaustively that the bound is tight, i.e., $h(7,5)=6$. Furthermore, for $q=11$, there exists a weight-$10$ codeword and the result of its split is shown in Fig.~\ref{fig:split-Hq11a5}.

\subsubsection{$H(q,6)$}
In \cite[Eq.~(13)]{ros14}, the authors presented a support matrix of codewords of weight $20$ for $H(q,6)$. We multiply its entries by $\alpha = 2$ and apply Algorithm~\ref{alg:SPLIT} to the resulting matrix. The algorithm succeeds to create a half-and-half split and the result is presented in Fig.~\ref{fig:split-Hq6}. The authors proved in \cite{ros14} that there are no repetitive columns in the matrix for $q > 11$. For the special cases $H(7,6)$ and $H(11,6)$, they provided particular support matrices which we also are able to split half-and-half with our algorithm (see Figs.~\ref{fig:split-Hq7a6} and \ref{fig:split-Hq11a6}). \revtwo{For $H(11,6)$, we also ran a brute-force exhaustive search confirming that there are no termatiko sets of size less than $8$ and larger than or equal to the lower bound of $6$ from Theorem~\ref{thm:col-a-reg}.}

\subsubsection{$H(q,7)$}
Again, in \cite[Eq.~(17)]{ros14}, the authors presented a support matrix for codewords of weight $24$ for $H(q,7)$. We multiply its entries by $\alpha = 4$ and successfully split it using Algorithm~\ref{alg:SPLIT} (see Fig.~\ref{fig:split-Hqa7}). \revone{For $H(11,7)$, the stopping distance is $15$ (see \cite{ros14}). We applied  Heuristic 1 to all stopping sets of size $15$ in $H(11,7)$, but did not find any termatiko sets of size smaller than $12$.  
Hence, Heuristic 1 was not able to tighten the upper bound of $12$ computed from Algorithm~\ref{alg:SPLIT}. Moreover,  we also ran a brute-force exhaustive search confirming that there are no termatiko sets of size less than $9$ and larger than or equal to the lower bound of $7$ from Theorem~\ref{thm:col-a-reg}.}

\subsubsection{$H(q,a>7)$}

From the previous subsections it appears (\revone{at least for small $a$ and $q$}) that the termatiko distance is half the minimum  distance for array LDPC codes, \revone{since the upper bound obtained by splitting half-and-half a minimum-weight codeword matches either the lower bound from Theorem~\ref{thm:col-a-reg} or a lower bound obtained by brute-force exhaustive search.} However, proving this in general might be difficult. \revone{Moreover, not all stopping sets can be split half-and-half.} For instance, for $q=7$ and $a=4$ we have found a \revus{stopping set (which is also a minimal codeword)} of weight $20$ that cannot be split into two termatiko sets\revus{,} each of size $10$. \revus{We verified this by exhaustive search.} The support matrix of \revus{this} codeword is
\[
	\left[
	\arraycolsep=3pt\def\arraystretch{0.75}\footnotesize
	\begin{array}{cccccccccccccccccccc}
	2 & 3 & 4 & 1 & 2 & 3 & 5 & 6 & 0 & 1 & 2 & 5 & 6 & 5 & 4 & 5 & 5 & 0 & 2  &5 \\
	2 & 3 & 4 & 2 & 3 & 4 & 6 & 0 & 2 & 3 & 4 & 0 & 1 & 1 & 1 & 2 & 3 & 6 & 1  &4 \\
	2 & 3 & 4 & 3 & 4 & 5 & 0 & 1 & 4 & 5 & 6 & 2 & 3 & 4 & 5 & 6 & 1 & 5 & 0  &3 \\
	2 & 3 & 4 & 4 & 5 & 6 & 1 & 2 & 6 & 0 & 1 & 4 & 5 & 0 & 2 & 3 & 6 & 4 & 6  &2 \\
	\end{array}
	\right]\,.
	\]

We gather the results for the termatiko distances of array LDPC codes in Table~\ref{tab:arrcode-termatiko-distances}. We additionally put results for measurement matrices $H(5,5)$ and $H(7,7)$, although usually $a < q$ is required for array LDPC codes.\footnote{Having $a=q$ gives array LDPC codes of strictly positive rate since $H(q,a)$ has redundant rows.} The exact termatiko distances for these two cases where obtained by splitting small-size stopping sets using Algorithm~\ref{alg:SPLIT}. This procedure gave termatiko sets of size $5$ and $7$, respectively, and from Theorem~\ref{thm:col-a-reg} it follows that these values give the exact termatiko distance in these two cases. Alternatively, for $a=5$, one can remove the $5$-th and the last column from the matrix in Fig.~\ref{fig:split-Hq5} (they are identical for $q=5$) and get a valid codeword support matrix of a weight-$10$ codeword that also is splittable in two termatiko sets of size $5$.


\begin{table}[t]
	\caption{Termatiko distances of array LDPC code matrices $H(q,a)$.}
	\label{tab:arrcode-termatiko-distances}
	\centering
	\vskip -2.0ex
	\begin{tabular}{@{}cccccc@{}}
		\toprule
		~           & $a=3$ & $a=4$  & $a=5$  & $a=6$ & $a=7$ \\
		\midrule
		$q=5$       & 3     & 4      & 5      & --    & -- \\
		$q=7$       & 3     & 4      & 6      & 6     & 7 \\
		$q=11$      & 3     & 5      & 5      & \revone{8}  & \revone{9}..12 \\
		$q \geq 13$ & 3     & 4 or 5 & 5 or 6 & 6..10 & 7..12 \\
		\bottomrule
	\end{tabular}
\end{table}

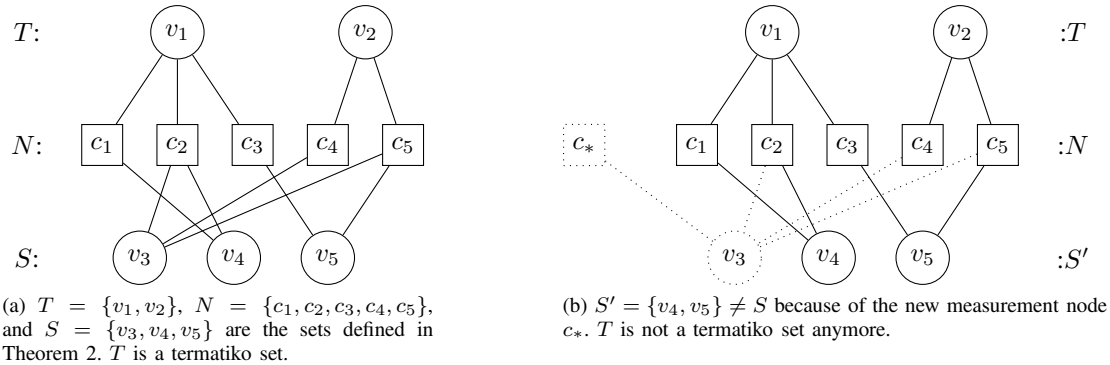
\begin{figure*}
	\centering
	
	\null\hfill
	\subfloat[$T = \{v_1, v_2\}$, $N = \{c_1, c_2, c_3, c_4, c_5\}$, and $S = \{v_3, v_4, v_5\}$ are the sets defined in Theorem~\ref{thm:termatikos}. $T$ is a termatiko set.]{%
		\begin{tikzpicture}
		\node at (1,+1.5) [circle,draw] (v1) {$v_{1}$};
		\node at (3.5,+1.5) [circle,draw] (v2) {$v_{2}$};
		\node at (0.5,-1.5) [circle,draw] (v3) {$v_{3}$};
		\node at (1.75,-1.5) [circle,draw] (v4) {$v_{4}$};
		\node at (3.0,-1.5) [circle,draw] (v5) {$v_{5}$};
		
		\node at (0.0,0) [square,draw] (c1) {$c_{1}$};
		\node at (1.0,0) [square,draw] (c2) {$c_{2}$};
		\node at (2.0,0) [square,draw] (c3) {$c_{3}$};
		\node at (3.0,0) [square,draw] (c4) {$c_{4}$};
		\node at (4.0,0) [square,draw] (c5) {$c_{5}$};
		
		\node at (-1,+1.5) {$T$:};
		\node at (-1,0) {$N$:};
		\node at (-1,-1.5) {$S$:};
		
		\draw (c1) -- (v1);
		\draw (c1) -- (v4);
		
		\draw (c2) -- (v1);
		\draw (c2) -- (v3);
		\draw (c2) -- (v4);
		
		\draw (c3) -- (v1);
		\draw (c3) -- (v5);
		
		\draw (c4) -- (v2);
		\draw (c4) -- (v3);
		
		\draw (c5) -- (v2);
		\draw (c5) -- (v3);
		\draw (c5) -- (v5);
		
		
		\end{tikzpicture}
		\label{fig:add1red-a-upd}
	}\hfill
	\subfloat[$S' = \{v_4, v_5\} \neq S$ because of the new measurement node $c_*$. $T$ is not a termatiko set anymore.]{%
		\begin{tikzpicture}
		\node at (1,+1.5) [circle,draw] (v1) {$v_{1}$};
		\node at (3.5,+1.5) [circle,draw] (v2) {$v_{2}$};
		\node at (0.5,-1.5) [circle,draw,dotted] (v3) {$v_{3}$};
		\node at (1.75,-1.5) [circle,draw] (v4) {$v_{4}$};
		\node at (3,-1.5) [circle,draw] (v5) {$v_{5}$};

		\node at (-1.5, 0) [square,draw,dotted] (c2m1) {$c_*$};
		\node at (0.0,0) [square,draw] (c1) {$c_{1}$};
		\node at (1.0,0) [square,draw] (c2) {$c_{2}$};
		\node at (2.0,0) [square,draw] (c3) {$c_{3}$};
		\node at (3.0,0) [square,draw] (c4) {$c_{4}$};
		\node at (4.0,0) [square,draw] (c5) {$c_{5}$};
		
		\node at (5,+1.5) {:$T$};
		\node at (5,0) {:$N$};
		\node at (5,-1.5) {:$S'$};
		
		\draw[dotted] (c2m1) -- (v3);
		
		\draw (c1) -- (v1);
		\draw (c1) -- (v4);
		
		\draw (c2) -- (v1);
		\draw[dotted] (c2) -- (v3);
		\draw (c2) -- (v4);
		
		\draw (c3) -- (v1);
		\draw (c3) -- (v5);
		
		\draw (c4) -- (v2);
		\draw[dotted] (c4) -- (v3);
		
		\draw (c5) -- (v2);
		\draw[dotted] (c5) -- (v3);
		\draw (c5) -- (v5);
		
		
		\end{tikzpicture}
		\label{fig:add1red-b-upd}
	}\hfill
	\null\hfill

	\caption{Adding a redundant measurement $c_*$ corresponding to the difference of rows $c_2$ and $c_1$ for the example matrix of Example~\ref{ex:1}.}
\end{figure*}

\section{Adding Redundant Rows}\label{sec:add-red-rows}
It is well-known that for iterative (peeling) decoding  over the BEC one can add redundant rows to the parity-check matrix in order to decrease the number of stopping sets \cite{sch06}. This is also the case for relaxed linear programming decoding of binary linear codes on any symmetric channel \cite{fel05}. In this section,  we aim to improve the recovery performance of the IPA by adding redundant rows to the measurement matrix, inspired by the success on the BEC. 
However, there is one fundamental difference in the sense that the real linear combinations that are added to the measurement matrix should contain nonnegative entries only. Furthermore, we would like to stress that redundant rows that we add to the measurement matrix are not used to provide new measurements, but rather used in the recovery process, which means that also measurements need to be linearly combined at the receiver. Thus, this procedure does not decrease the compression rate of the scheme, but rather potentially improve the recovery performance.

The following lemma shows that adding redundant rows to the measurement matrix does not harm the IPA reconstruction performance, namely, that it does not create new termatiko sets.

\begin{lemma} \label{lem:no_new_termatiko_sets}
	Adding redundant measurements does not create new termatiko sets.
\end{lemma}
\begin{IEEEproof}
Let the original measurement matrix be denoted by $A$. Its extended version with nonnegative redundant rows is denoted by $A'$. The matrix  $A'$ is constructed such that the first rows of $A'$ are exactly the rows of $A$ and the remaining rows are real-valued linear combinations of the rows of $A$ with nonnegative entries.\footnote{Nonnegativity of matrix entries is important for the correctness of the $\IPA$.} Denote also by $V'$, $C'$, $E'$, $\mathcal N'$, and $\mathcal N'_T$ the entities corresponding to $A'$, similarly to Section~\ref{sec:Tanner-graph-notation}. Consider some signal vector $\vec x$ and two problems, $\IPA(\vec y, A)$ and $\IPA(\vec y', A')$, where $\vec y = A \vec x$ and $\vec y' = A' \vec x$.
	
	The set of variable nodes is the same, i.e., $V = V'$, but the set of measurement nodes is now a superset of the original set, i.e., $C \subset C'$. The same is true for the set of edges, $E \subset E'$. Also, it holds for all $v \in V$ that $\mathcal N'(v) = \mathcal N'_C(v) \cup \mathcal N'_{C' \setminus C}(v) = \Neighbours{v} \cup \mathcal N'_{C' \setminus C}(v)$ and $\mathcal N'(c) = \Neighbours{c}$ for all $c \in C$. This in turn means that $y_c = y_c'$ for $c \in C$.
	
	Let $\mu'$ and $M'$ (with corresponding indices) be bounds in the iterations of $\IPA(\vec y',A')$. Then to prove the statement of the lemma, it is enough to show that for all iterations $\ell \geq 0$, $\mu'^{(\ell)}_{v \to \cdot} \geq \mu^{(\ell)}_{v \to \cdot}$ and $M'^{(\ell)}_{v \to \cdot} \leq M^{(\ell)}_{v \to \cdot}$. In other words, we show that the intervals $[\mu',M']$ are at least as tight as $[\mu,M]$. We show this by induction on $\ell$ (the number of iterations).
	
	{\flushleft \emph{Base Case.}}
	\begin{align*}
	\mu'^{(0)}_{v \to \cdot} &= 0 = \mu_{v \to \cdot}^{(0)} \\
	M'^{(0)}_{v \to \cdot} &= \min_{c \in \mathcal N'(v)} (y'_c / a'_{cv}) 
	\leq \min_{c \in \Neighbours{v}} (y'_c / a'_{cv}) \\
	&= \min_{c \in \Neighbours{v}} (y_c / a_{cv}) = M^{(0)}_{v \to \cdot} \,.
	\end{align*}
	
	{\flushleft \emph{Inductive Step.}}
	
	Consider iteration $\ell \geq 1$.  At each step $\ell$ of the $\IPA$ and for all $c \in C$ and $v \in \mathcal N'(c) = \Neighbours[]{c}$, we have
	\ifonecolumn
	\begin{align*}
		\mu'^{(\ell)}_{c \to v} &= \frac{1}{a'_{cv}}\left(y'_c - \sum_{v' \in 
			\mathcal N'(c), v' \neq v} a'_{cv'} M'^{(\ell-1)}_{v' \to 
			\cdot}\right) 
		= \frac{1}{a_{cv}}\left(y_c - \sum_{v' \in 
			\Neighbours[]{c}, v' \neq v} a_{cv'} M'^{(\ell-1)}_{v' \to 
			\cdot}\right) \\
		&\geq \frac{1}{a_{cv}}\left(y_c - \sum_{v' \in 
			\Neighbours[]{c}, v' \neq v} a_{cv'} M^{(\ell-1)}_{v' \to 
			\cdot}\right) = \mu^{(\ell)}_{c \to v} \,.
	\end{align*}
	\else
	\begin{align*}
		\mu'^{(\ell)}_{c \to v} &= \frac{1}{a'_{cv}}\left(y'_c - \sum_{v' \in 
			\mathcal N'(c), v' \neq v} a'_{cv'} M'^{(\ell-1)}_{v' \to 
			\cdot}\right) \\
		&= \frac{1}{a_{cv}}\left(y_c - \sum_{v' \in 
			\Neighbours[]{c}, v' \neq v} a_{cv'} M'^{(\ell-1)}_{v' \to 
			\cdot}\right) \\
		&\geq \frac{1}{a_{cv}}\left(y_c - \sum_{v' \in 
			\Neighbours[]{c}, v' \neq v} a_{cv'} M^{(\ell-1)}_{v' \to 
			\cdot}\right) = \mu^{(\ell)}_{c \to v} \,.
	\end{align*}
	\fi
	In the same manner, we get that for all $c \in C$, $M'^{(\ell)}_{c \to v} \leq M^{(\ell)}_{c \to v}$. We further apply these inequalities to Lines~\ref{alg:ipa:muvc} and \ref{alg:ipa:Mvc} of Algorithm~\ref{alg:ipa} and, recalling properties of the operators $\min(\cdot)$ and $\max(\cdot)$, we obtain the desired result.
\end{IEEEproof}

From Lemma~\ref{lem:no_new_termatiko_sets} it follows that adding redundant rows to the measurement matrix cannot harm the IPA. The following example shows that adding such rows can indeed improve the performance of the IPA by removing termatiko sets.

\begin{example} \label{ex:1}
	Consider the binary measurement matrix
	\[
	A =
	\begin{pmatrix}
		\mathbf{1} & \mathbf{0} & 0 & 1 & 0 \\
		\mathbf{1} & \mathbf{0} & 1 & 1 & 0 \\
		\mathbf{1} & \mathbf{0} & 0 & 0 & 1 \\
		\mathbf{0} & \mathbf{1} & 1 & 0 & 0 \\
		\mathbf{0} & \mathbf{1} & 1 & 0 & 1 \\
	\end{pmatrix} \,.
	\]
	The corresponding Tanner graph is shown in Fig.~\subref*{fig:add1red-a-upd}. Note that the set $\{v_1, v_2\}$ is a termatiko set in this matrix (the corresponding columns of $A$ are marked in bold). However, if we add a redundant row $c_*$ equal to the difference of rows $c_2$ and $c_1$, $\{ v_1, v_2 \}$ is not a termatiko set for the extended matrix\footnote{Recall that operations are performed over $\mathbb R$.}
	\[
	A' =
	\begin{pmatrix}
		\mathbf{1} & \mathbf{0} & 0 & 1 & 0 \\
		\mathbf{1} & \mathbf{0} & 1 & 1 & 0 \\
		\mathbf{1} & \mathbf{0} & 0 & 0 & 1 \\
		\mathbf{0} & \mathbf{1} & 1 & 0 & 0 \\
		\mathbf{0} & \mathbf{1} & 1 & 0 & 1 \\
		\mathbf{0} & \mathbf{0} & 1 & 0 & 0
	\end{pmatrix} \,,
	\]
	since $c_4$ violates conditions in Theorem~\ref{thm:termatikos} with the updated matrix as explained below.
	\begin{itemize}
		\item $c_4$ is not connected to $S'$, and
		\item $\Neighbours[T]{c_4} = \{v_2\}$, $\Neighbours[]{v_2} = \{c_4, c_5\}$, and each of $c_4, c_5$ is connected to $T$ only once -- therefore 
		\[
				\Big| \{ v \in \Neighbours[T]{c_4} \,:\, \forall c' \in \Neighbours[]{v}, |\Neighbours[T]{c'}| \geq 2 \} \Big| = 0\,.
		\]
	\end{itemize}
	Fig.~\subref*{fig:add1red-b-upd} illustrates the differences.
\end{example}

\begin{table*}[!t]
     \scriptsize \centering \caption{Estimated termatiko set size spectra (initial part) of measurement matrices from Section~\ref{sec:numerical_results}, where $\hat{h}_{\rm min}$ denotes the estimated termatiko distance. $\mathfrak T_1$ corresponds to termatiko sets with all measurement nodes in $N$ connected to both $T$ and $S$, and $\mathfrak T_2$ corresponds to all the remaining termatiko sets. Also shown are the exact stopping distances and stopping set size spectra (initial part). Entries in bold are exact values. For  $A^{(1)}$, \revone{Heuristic 1} gives a multiplicity of $5875518$ for size $5$, while the exact number is $6318378$ (an underestimation of about $7.5\%$).} 
    \label{table_of_codes}
    \vskip -2.0ex 
    \begin{tabular}{@{}ccccc@{}}
        \toprule
        Measurement matrix & $\hat{h}_{\min}$ & Initial estimated termatiko set size spectrum & $s_{\min}$ & Initial stopping set size spectrum \\
        \midrule
        $A^{(1)}$  & $\bm 3$ & $\mathfrak T_1$: $(\bm{3630},\bm{93775},\bm{6318378},48548225,71709440,$ & $\bm 6$ & $(\bm{1815},\bm{605},\bm{45375},\bm{131890},\bm{3550382},\bm{28471905})$\\
        & & $36514170,7969060,856801,41745)$ \\
        & & $\mathfrak T_2$: $(0,0,0,410190,18610405,71153445,86844725,$\\
        & & $58849681,28430160)$ \\
        \addlinespace
        $A^{(2)}$ & $9$ & $\mathfrak T_1$: $(465, 3906, 12555, 8835, 0, 0, \dotsc)$  & $\bm{18}$ & $(\bm{465},\bm{2015},\bm{9548},\bm{23715},\bm{106175})$ \\
        & & $\mathfrak T_2$: $(0,0,0,1860, 5115, 10695, 2325, 5580, 2325, 6045$ \\
        & & $10850, 22103, 39990, 106175)$ \\
        \addlinespace
        $A^{(3)}$  & $8$ & $\mathfrak T_1$: $(228, 0, 0, \dotsc)$ & $\bm 9$ & $(\bm{76},\bm 0,\bm 0,\bm 0,\bm{76},\bm{76},\bm{304},\bm{1520})$\\
        & & $\mathfrak T_2$: $(0, 76, 0, 76, 684, 532, 152, 532, 1520)$ \\
        \addlinespace
        $A^{(4)}$  & $8$ & $\mathfrak T_1$: $(184, 598, 1242, 391, 0, 0)$ & $\bm{15}$ & $(\bm{46},\bm{161},\bm{391},\bm{897},\bm{2093},\bm{5796})$\\
        & & $\mathfrak T_2$: $(0, 0, 0, 69, 23, 0, 23, 46, 161, 391, 1012, 2300, 5796)$ \\
        \addlinespace
        $A^{(5)}$  & $7$ & $\mathfrak T_1$: $(106, 0, 0, 53, 901, 3233, 954, 53, 0, 0, \dotsc)$ & $\bm{14}$ & $(\bm{53},\bm 0,\bm 0,\bm 0,\bm 0,\bm{53},\bm{106},\bm{583},\bm{1484},\bm{3922},\bm{9964})$\\
        & & $\mathfrak T_2$: $(0, 0, 0, 0, 0, 0, 106, 265, 106, 636, 689, 477, $ \\
        & & $583, 371, 1325, 2915, 5830, 9964)$ \\
        \bottomrule
    \end{tabular}
    \vskip -3.5ex 
\end{table*}

Now, the question is which redundant rows to add in order to remove the largest number of harmful small-size termatiko sets. We propose the following heuristic approach. First, fix some list of small-size termatiko sets for the original measurement matrix $A$ and generate a pool of redundant rows which (hopefully) help to remove at least one termatiko set from the list as follows.

Consider a termatiko set $T$ from the list and its corresponding set $S$. A redundant row $\vec r = (r_1, r_2, \dotsc, r_n)$ for the measurement matrix $A$ can be defined uniquely by the coefficients $\alpha_1, \alpha_2, \dotsc, \alpha_m \in \mathbb R$ by the linear combinations $r_v = \sum_{c \in C} a_{cv} \alpha_c$. 
However, since in real calculations floating-point numbers are effectively rational numbers, by multiplying all $\alpha$'s by some common multiplier of their denominators, we can make them all integer, and they still give a redundant row $\vec r$ with the same support. Therefore, w.o.l.g.,  we assume that the $\alpha$'s are integers.  If original matrix $A$ has integer entries, the resulting extended matrix has integer entries as well, which allows for a faster IPA in applications where the signal $\bm x$ is over the integers.

There are two types of redundant rows that will be collected in the pool. The first type ``breaks'' the termatiko set $T$ for sure. It has one nonzero entry in the positions in $T$ and zeroes in entries indexed by $S$. The other entries of $\vec r$ can be chosen arbitrarily. More precisely, for a fixed $v_0 \in T$ we solve the (integer) linear programming problem
\begin{align*}
	\text{minimize }&\sum_{v \in V \setminus \{ T \cup S \}} r_v  =   \sum_{v \in V \setminus \{ T \cup S \}}  \sum_{c \in C} a_{cv} \alpha_c   \\ 
	\text{s.t.}\quad &r_v \geq 0 \,, \quad v \notin T \cup S \,,\\
                     &r_v= 0 \,, \quad v \in T \cup S \setminus \{v_0\} \,,\\
                     &r_{v_0} \geq 1 \,,
\end{align*}
 where $\alpha_1, \alpha_2, \dotsc, \alpha_m$ are integer variables. 
Minimization here is not essential and is used just to get smaller coefficients in a redundant row. In fact, for any feasible solution, the corresponding redundant row eliminates the termatiko set $T$. A redundant row can potentially be obtained for each $v_0 \in T$. As a final remark, relaxing the $\alpha$'s to be real numbers turns the program into a standard linear program that can be solved using the simplex method. However, as noted above, having integers (of moderate size) in the measurement matrix has some potential benefits. Thus, when the size of the program is not too large and can be solved using a standard solver in a reasonable time (which is the case in our examples), we keep the integer constraint on the $\alpha$'s.

Redundant rows of the second type do not necessarily ``break'' $T$ always, but they have good chances for doing exactly that. The basic idea is to try to make variable nodes in $S$ not satisfy Theorem~\ref{thm:termatikos}, hence not being included in $S$ for the extended matrix and, hopefully, this eliminates $T$ as a termatiko set for the extended matrix. Note that having several nonzero entries in positions in $S$ is better, since all of them will disappear from $S$ (and we do not add new ones to $S$). This will increase the probability of removing $T$. The corresponding (integer) linear program is
\begin{align*}
&r_v \geq 0 \,, \quad v \notin T \,,\\
&r_v \leq 1000 \,, \quad v \notin T \,,\\
&r_v = 0 \,, \quad v \in T \,,\\
&\sum_{v \in S} r_v \geq 10 |S| \,,
\end{align*}
where the constants $10$ and $1000$ are chosen rather arbitrarily; $10$ is used in order to make nonzero entries in $\vec r_S$ more likely, and the upper bounds of $1000$ make sure the entries in $\vec r$ are of limited size. Note that no objective function is specified, since any feasible solution will do. For each termatiko set $T$, this approach produces at most one redundant row.

Finally, after constructing the pool of redundant rows as described above, we start adjoining them to the matrix $A$ one by one in a greedy manner as follows. Let the list of termatiko sets be denoted by $\mathsf{LIST}$ and the pool of redundant rows by $\mathsf{POOL}$. For each row $\vec r \in \mathsf{POOL}$, we calculate the score
\[
\operatorname{score}(\vec r) = \sum_{T \in \mathsf{RMV}(\mathsf{LIST},\vec r)} |T| \,,
\]
where $\mathsf{RMV}(\mathsf{LIST},\vec r)$ is the subset of $\mathsf{LIST}$ consisting of the termatiko sets that are not termatiko sets after adjoining row $\vec r$ to the current measurement matrix. The row $\vec r^*$ with the maximum score is adjoined to the measurement matrix, the termatiko sets in $\mathsf{RMV}(\mathsf{LIST}, \vec r)$ are removed from $\mathsf{LIST}$, and the scores are re-calculated for the updated $\mathsf{LIST}$ and measurement matrix. The procedure is continued until $\mathsf{LIST}$ is empty or all scores are zero (which means that no more termatiko sets can be removed).

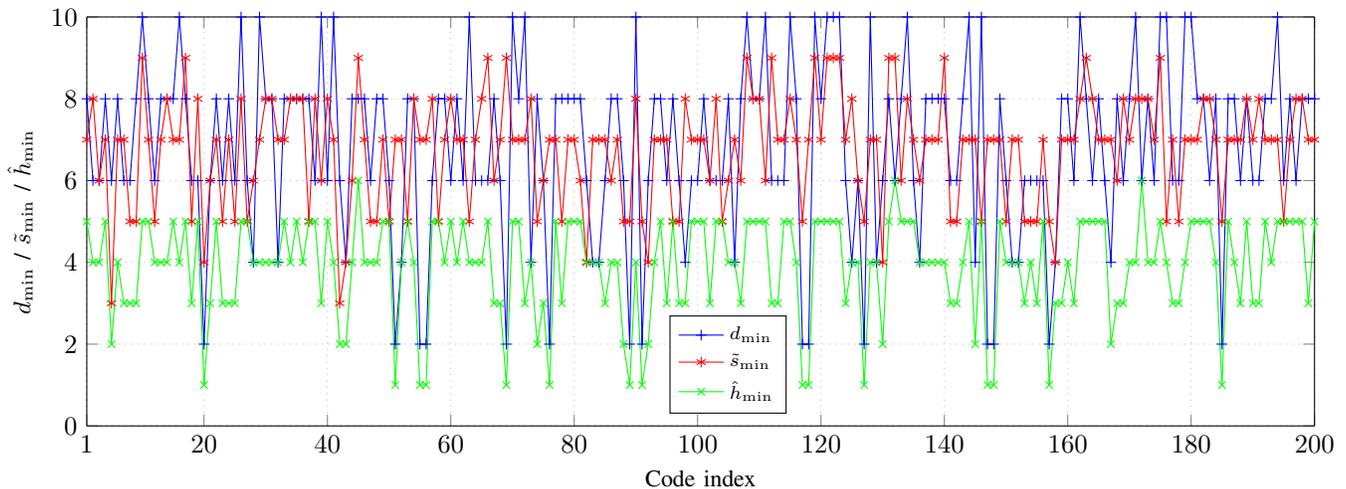
\begin{figure*}[tbp]
\centering
\begin{tikzpicture}[font=\small]

\begin{axis}[%
width=0.9\textwidth,
height=0.3\textwidth,
at={(0.808889in,0.513333in)},
scale only axis,
separate axis lines,
every outer x axis line/.append style={black},
every x tick label/.append style={font=\color{black}},
xmin=1,
xmax=200,
xtick={1,20,40,60,80,100,120,140,160,180,200},
xlabel={Code index},
xmajorgrids,
grid style={gray,opacity=0.5,dotted},
every outer y axis line/.append style={black},
every y tick label/.append style={font=\color{black}},
ymin=0,
ymax=10,
ylabel={$d_{\rm min}$ / $\tilde{s}_{\rm min}$ / $\hat{h}_{\rm min}$},
ymajorgrids,
legend style={at={(0.57,0.03)},anchor=south east,legend cell align=left,align=left,draw=black,font=\scriptsize}
]
\addplot [color=blue,solid,mark=+,mark options={solid}]
table[row sep=crcr]{%
	1	8\\
	2	6\\
	3	6\\
	4	8\\
	5	6\\
	6	8\\
	7	6\\
	8	6\\
	9	8\\
	10	10\\
	11	8\\
	12	6\\
	13	8\\
	14	8\\
	15	8\\
	16	10\\
	17	8\\
	18	6\\
	19	6\\
	20	2\\
	21	6\\
	22	8\\
	23	6\\
	24	8\\
	25	6\\
	26	10\\
	27	6\\
	28	4\\
	29	10\\
	30	8\\
	31	8\\
	32	4\\
	33	8\\
	34	8\\
	35	8\\
	36	8\\
	37	8\\
	38	6\\
	39	10\\
	40	6\\
	41	10\\
	42	6\\
	43	4\\
	44	8\\
	45	8\\
	46	8\\
	47	6\\
	48	8\\
	49	8\\
	50	6\\
	51	2\\
	52	4\\
	53	8\\
	54	8\\
	55	2\\
	56	2\\
	57	6\\
	58	8\\
	59	8\\
	60	6\\
	61	8\\
	62	6\\
	63	10\\
	64	6\\
	65	6\\
	66	6\\
	67	8\\
	68	6\\
	69	2\\
	70	10\\
	71	8\\
	72	10\\
	73	4\\
	74	8\\
	75	6\\
	76	2\\
	77	8\\
	78	8\\
	79	8\\
	80	8\\
	81	8\\
	82	6\\
	83	4\\
	84	4\\
	85	6\\
	86	8\\
	87	8\\
	88	6\\
	89	2\\
	90	10\\
	91	2\\
	92	6\\
	93	8\\
	94	8\\
	95	6\\
	96	8\\
	97	6\\
	98	4\\
	99	6\\
	100	6\\
	101	8\\
	102	6\\
	103	6\\
	104	6\\
	105	8\\
	106	4\\
	107	8\\
	108	10\\
	109	8\\
	110	8\\
	111	10\\
	112	6\\
	113	6\\
	114	6\\
	115	10\\
	116	8\\
	117	2\\
	118	2\\
	119	10\\
	120	8\\
	121	10\\
	122	10\\
	123	10\\
	124	6\\
	125	4\\
	126	6\\
	127	2\\
	128	10\\
	129	4\\
	130	6\\
	131	8\\
	132	6\\
	133	8\\
	134	10\\
	135	6\\
	136	4\\
	137	8\\
	138	8\\
	139	8\\
	140	8\\
	141	6\\
	142	6\\
	143	8\\
	144	10\\
	145	4\\
	146	10\\
	147	2\\
	148	2\\
	149	8\\
	150	6\\
	151	4\\
	152	4\\
	153	6\\
	154	6\\
	155	6\\
	156	6\\
	157	2\\
	158	4\\
	159	8\\
	160	8\\
	161	6\\
	162	10\\
	163	8\\
	164	6\\
	165	8\\
	166	6\\
	167	4\\
	168	8\\
	169	6\\
	170	8\\
	171	10\\
	172	6\\
	173	8\\
	174	6\\
	175	10\\
	176	10\\
	177	6\\
	178	6\\
	179	10\\
	180	10\\
	181	8\\
	182	8\\
	183	6\\
	184	8\\
	185	2\\
	186	8\\
	187	8\\
	188	6\\
	189	8\\
	190	6\\
	191	6\\
	192	8\\
	193	8\\
	194	10\\
	195	6\\
	196	8\\
	197	6\\
	198	8\\
	199	8\\
	200	8\\
};
\addlegendentry{$d_{\rm min}$};

\addplot [color=red,solid,mark=asterisk,mark options={solid}]
table[row sep=crcr]{%
	1	7\\
	2	8\\
	3	6\\
	4	7\\
	5	3\\
	6	7\\
	7	7\\
	8	5\\
	9	5\\
	10	9\\
	11	7\\
	12	5\\
	13	7\\
	14	8\\
	15	7\\
	16	7\\
	17	9\\
	18	5\\
	19	8\\
	20	4\\
	21	6\\
	22	7\\
	23	5\\
	24	7\\
	25	5\\
	26	8\\
	27	5\\
	28	6\\
	29	7\\
	30	8\\
	31	8\\
	32	7\\
	33	7\\
	34	8\\
	35	8\\
	36	8\\
	37	5\\
	38	8\\
	39	6\\
	40	8\\
	41	7\\
	42	3\\
	43	4\\
	44	6\\
	45	9\\
	46	7\\
	47	5\\
	48	5\\
	49	7\\
	50	5\\
	51	7\\
	52	7\\
	53	5\\
	54	8\\
	55	7\\
	56	7\\
	57	8\\
	58	5\\
	59	7\\
	60	8\\
	61	7\\
	62	7\\
	63	5\\
	64	7\\
	65	8\\
	66	9\\
	67	6\\
	68	7\\
	69	9\\
	70	7\\
	71	7\\
	72	7\\
	73	8\\
	74	5\\
	75	6\\
	76	7\\
	77	7\\
	78	5\\
	79	7\\
	80	7\\
	81	6\\
	82	4\\
	83	7\\
	84	7\\
	85	7\\
	86	6\\
	87	7\\
	88	5\\
	89	5\\
	90	8\\
	91	5\\
	92	4\\
	93	7\\
	94	7\\
	95	7\\
	96	5\\
	97	5\\
	98	8\\
	99	7\\
	100	7\\
	101	7\\
	102	6\\
	103	8\\
	104	5\\
	105	6\\
	106	7\\
	107	6\\
	108	9\\
	109	8\\
	110	8\\
	111	6\\
	112	9\\
	113	7\\
	114	7\\
	115	8\\
	116	7\\
	117	5\\
	118	7\\
	119	9\\
	120	7\\
	121	9\\
	122	9\\
	123	9\\
	124	7\\
	125	8\\
	126	6\\
	127	5\\
	128	7\\
	129	7\\
	130	4\\
	131	9\\
	132	9\\
	133	6\\
	134	8\\
	135	7\\
	136	6\\
	137	7\\
	138	7\\
	139	7\\
	140	9\\
	141	5\\
	142	5\\
	143	7\\
	144	7\\
	145	7\\
	146	5\\
	147	7\\
	148	7\\
	149	7\\
	150	5\\
	151	7\\
	152	7\\
	153	5\\
	154	5\\
	155	5\\
	156	7\\
	157	5\\
	158	4\\
	159	7\\
	160	7\\
	161	7\\
	162	8\\
	163	9\\
	164	8\\
	165	7\\
	166	7\\
	167	7\\
	168	6\\
	169	8\\
	170	7\\
	171	8\\
	172	8\\
	173	8\\
	174	7\\
	175	9\\
	176	5\\
	177	7\\
	178	5\\
	179	7\\
	180	7\\
	181	7\\
	182	8\\
	183	8\\
	184	7\\
	185	5\\
	186	7\\
	187	7\\
	188	7\\
	189	8\\
	190	7\\
	191	8\\
	192	7\\
	193	7\\
	194	7\\
	195	5\\
	196	7\\
	197	8\\
	198	8\\
	199	7\\
	200	7\\
};
\addlegendentry{$\tilde{s}_{\rm min}$};

\addplot [color=green,solid,mark=x,mark options={solid}]
table[row sep=crcr]{%
	1	5\\
	2	4\\
	3	4\\
	4	5\\
	5	2\\
	6	4\\
	7	3\\
	8	3\\
	9	3\\
	10	5\\
	11	5\\
	12	4\\
	13	4\\
	14	4\\
	15	5\\
	16	4\\
	17	5\\
	18	3\\
	19	5\\
	20	1\\
	21	3\\
	22	5\\
	23	3\\
	24	3\\
	25	3\\
	26	5\\
	27	5\\
	28	4\\
	29	4\\
	30	4\\
	31	4\\
	32	4\\
	33	5\\
	34	4\\
	35	5\\
	36	4\\
	37	5\\
	38	5\\
	39	3\\
	40	5\\
	41	4\\
	42	2\\
	43	2\\
	44	4\\
	45	6\\
	46	4\\
	47	4\\
	48	4\\
	49	5\\
	50	5\\
	51	1\\
	52	4\\
	53	5\\
	54	4\\
	55	1\\
	56	1\\
	57	5\\
	58	5\\
	59	4\\
	60	5\\
	61	4\\
	62	5\\
	63	4\\
	64	4\\
	65	4\\
	66	5\\
	67	3\\
	68	3\\
	69	1\\
	70	5\\
	71	5\\
	72	3\\
	73	4\\
	74	2\\
	75	3\\
	76	1\\
	77	5\\
	78	3\\
	79	5\\
	80	5\\
	81	5\\
	82	4\\
	83	4\\
	84	4\\
	85	3\\
	86	4\\
	87	4\\
	88	2\\
	89	1\\
	90	4\\
	91	1\\
	92	2\\
	93	4\\
	94	5\\
	95	3\\
	96	5\\
	97	3\\
	98	3\\
	99	5\\
	100	5\\
	101	5\\
	102	3\\
	103	5\\
	104	5\\
	105	4\\
	106	4\\
	107	3\\
	108	5\\
	109	5\\
	110	5\\
	111	5\\
	112	3\\
	113	3\\
	114	5\\
	115	5\\
	116	4\\
	117	1\\
	118	1\\
	119	5\\
	120	5\\
	121	5\\
	122	5\\
	123	5\\
	124	3\\
	125	4\\
	126	4\\
	127	1\\
	128	4\\
	129	4\\
	130	2\\
	131	5\\
	132	6\\
	133	5\\
	134	5\\
	135	5\\
	136	4\\
	137	4\\
	138	4\\
	139	4\\
	140	4\\
	141	3\\
	142	3\\
	143	4\\
	144	5\\
	145	2\\
	146	5\\
	147	1\\
	148	1\\
	149	5\\
	150	4\\
	151	4\\
	152	4\\
	153	3\\
	154	4\\
	155	3\\
	156	5\\
	157	1\\
	158	3\\
	159	3\\
	160	4\\
	161	3\\
	162	5\\
	163	5\\
	164	5\\
	165	5\\
	166	5\\
	167	2\\
	168	3\\
	169	3\\
	170	4\\
	171	4\\
	172	6\\
	173	4\\
	174	4\\
	175	5\\
	176	4\\
	177	3\\
	178	3\\
	179	4\\
	180	5\\
	181	5\\
	182	5\\
	183	5\\
	184	4\\
	185	1\\
	186	5\\
	187	4\\
	188	3\\
	189	5\\
	190	3\\
	191	3\\
	192	5\\
	193	4\\
	194	5\\
	195	5\\
	196	5\\
	197	5\\
	198	5\\
	199	3\\
	200	5\\
};
\addlegendentry{$\hat{h}_{\rm min}$};

\end{axis}
\end{tikzpicture}%
\caption{Minimum distance $d_{\rm min}$, minimum size of a noncodeword stopping set $\tilde{s}_{\rm min}$, and estimated termatiko distance $\hat{h}_{\rm min}$ versus code index for randomly generated binary measurement matrices from a protograph-based $(3,6)$-regular LDPC code ensemble.}
\label{fig:pg36_ensemble}
\end{figure*}

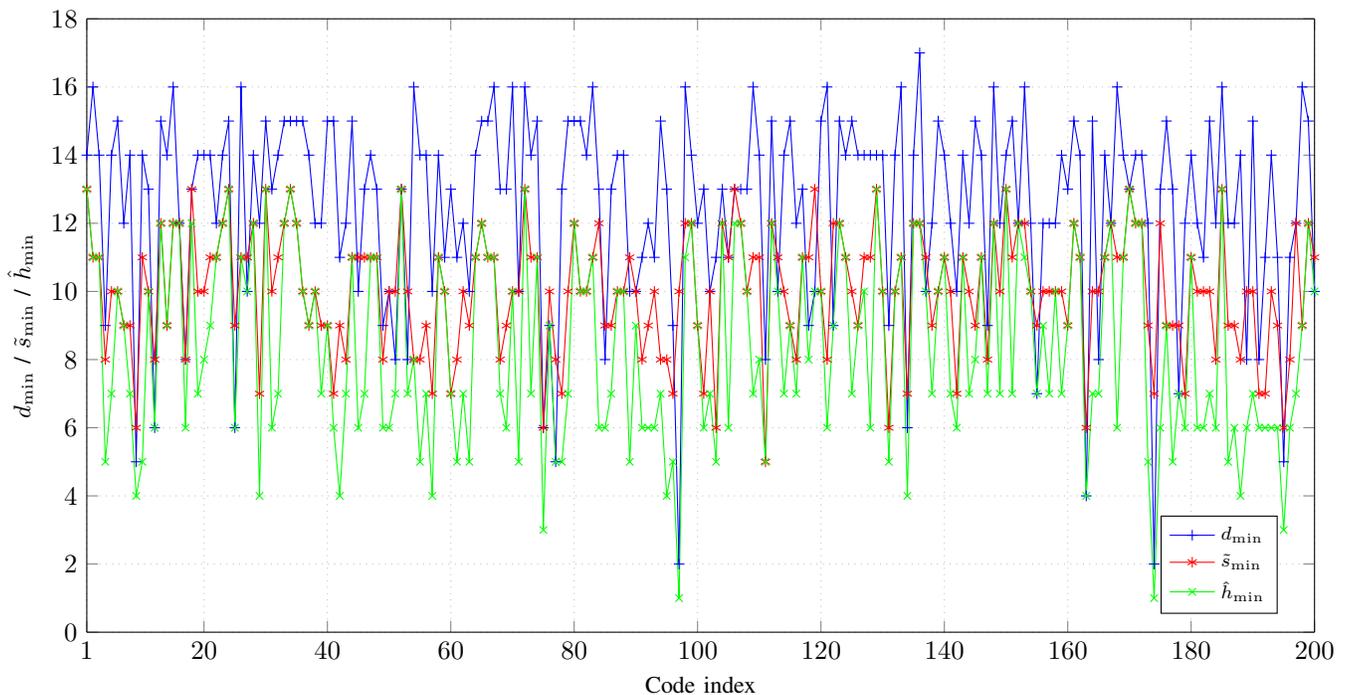
\begin{figure*}[tbp]
\centering
\begin{tikzpicture}[font=\small]

\begin{axis}[%
width=0.9\textwidth,
height=0.45\textwidth,
at={(0.808889in,0.513333in)},
scale only axis,
separate axis lines,
grid style={gray,opacity=0.5,dotted},
every outer x axis line/.append style={black},
every x tick label/.append style={font=\color{black}},
xmin=1,
xmax=200,
xlabel={Code index},
xmajorgrids,
xtick={1,20,40,60,80,100,120,140,160,180,200},
every outer y axis line/.append style={black},
every y tick label/.append style={font=\color{black}},
ymin=0,
ymax=18,
ylabel={$d_{\rm min}$ / $\tilde{s}_{\rm min}$ / $\hat{h}_{\rm min}$},
ymajorgrids,
ytick={0,2,4,6,8,10,12,14,16,18},
legend style={at={(0.97,0.03)},anchor=south east,legend cell align=left,align=left,draw=black,font=\scriptsize}
]
\addplot [color=blue,solid,mark=+,mark options={solid},unbounded coords=jump]
table[row sep=crcr]{%
	1	14\\
	2	16\\
	3	14\\
	4	9\\
	5	14\\
	6	15\\
	7	12\\
	8	14\\
	9	5\\
	10	14\\
	11	13\\
	12	6\\
	13	15\\
	14	14\\
	15	16\\
	16	12\\
	17	8\\
	18	13\\
	19	14\\
	20	14\\
	21	14\\
	22	12\\
	23	14\\
	24	15\\
	25	6\\
	26	16\\
	27	10\\
	28	14\\
	29	12\\
	30	15\\
	31	13\\
	32	14\\
	33	15\\
	34	15\\
	35	15\\
	36	15\\
	37	14\\
	38	12\\
	39	12\\
	40	15\\
	41	15\\
	42	11\\
	43	12\\
	44	15\\
	45	10\\
	46	13\\
	47	14\\
	48	13\\
	49	9\\
	50	10\\
	51	8\\
	52	13\\
	53	8\\
	54	16\\
	55	14\\
	56	14\\
	57	10\\
	58	14\\
	59	11\\
	60	13\\
	61	11\\
	62	12\\
	63	10\\
	64	14\\
	65	15\\
	66	15\\
	67	16\\
	68	13\\
	69	13\\
	70	16\\
	71	10\\
	72	16\\
	73	14\\
	74	15\\
	75	6\\
	76	9\\
	77	5\\
	78	13\\
	79	15\\
	80	15\\
	81	15\\
	82	14\\
	83	16\\
	84	13\\
	85	8\\
	86	13\\
	87	14\\
	88	14\\
	89	10\\
	90	10\\
	91	11\\
	92	12\\
	93	11\\
	94	15\\
	95	13\\
	96	9\\
	97	2\\
	98	16\\
	99	14\\
	100	12\\
	101	13\\
	102	10\\
	103	11\\
	104	13\\
	105	11\\
	106	13\\
	107	13\\
	108	13\\
	109	16\\
	110	14\\
	111	8\\
	112	15\\
	113	10\\
	114	14\\
	115	15\\
	116	12\\
	117	13\\
	118	9\\
	119	10\\
	120	15\\
	121	16\\
	122	9\\
	123	15\\
	124	14\\
	125	15\\
	126	14\\
	127	14\\
	128	14\\
	129	14\\
	130	14\\
	131	9\\
	132	14\\
	133	16\\
	134	6\\
	135	14\\
	136	17\\
	137	10\\
	138	12\\
	139	15\\
	140	14\\
	141	12\\
	142	10\\
	143	14\\
	144	12\\
	145	15\\
	146	14\\
	147	9\\
	148	16\\
	149	12\\
	150	14\\
	151	15\\
	152	12\\
	153	16\\
	154	12\\
	155	7\\
	156	12\\
	157	12\\
	158	12\\
	159	14\\
	160	13\\
	161	15\\
	162	14\\
	163	4\\
	164	15\\
	165	8\\
	166	14\\
	167	12\\
	168	16\\
	169	14\\
	170	13\\
	171	14\\
	172	14\\
	173	12\\
	174	2\\
	175	13\\
	176	15\\
	177	13\\
	178	7\\
	179	12\\
	180	14\\
	181	12\\
	182	11\\
	183	15\\
	184	12\\
	185	16\\
	186	12\\
	187	12\\
	188	14\\
	189	8\\
	190	15\\
	191	8\\
	192	11\\
	193	14\\
	194	11\\
	195	5\\
	196	11\\
	197	12\\
	198	16\\
	199	15\\
	200	10\\
};
\addlegendentry{$d_{\rm min}$};

\addplot [color=red,solid,mark=asterisk,mark options={solid}]
table[row sep=crcr]{%
	1	13\\
	2	11\\
	3	11\\
	4	8\\
	5	10\\
	6	10\\
	7	9\\
	8	9\\
	9	6\\
	10	11\\
	11	10\\
	12	8\\
	13	12\\
	14	9\\
	15	12\\
	16	12\\
	17	8\\
	18	13\\
	19	10\\
	20	10\\
	21	11\\
	22	11\\
	23	12\\
	24	13\\
	25	9\\
	26	11\\
	27	11\\
	28	12\\
	29	7\\
	30	13\\
	31	10\\
	32	11\\
	33	12\\
	34	13\\
	35	12\\
	36	10\\
	37	9\\
	38	10\\
	39	9\\
	40	9\\
	41	7\\
	42	9\\
	43	8\\
	44	11\\
	45	11\\
	46	11\\
	47	11\\
	48	11\\
	49	8\\
	50	10\\
	51	10\\
	52	13\\
	53	10\\
	54	8\\
	55	8\\
	56	9\\
	57	7\\
	58	11\\
	59	10\\
	60	7\\
	61	8\\
	62	10\\
	63	9\\
	64	11\\
	65	12\\
	66	11\\
	67	11\\
	68	8\\
	69	9\\
	70	10\\
	71	10\\
	72	13\\
	73	11\\
	74	11\\
	75	6\\
	76	10\\
	77	8\\
	78	7\\
	79	10\\
	80	12\\
	81	10\\
	82	10\\
	83	11\\
	84	12\\
	85	9\\
	86	9\\
	87	10\\
	88	10\\
	89	11\\
	90	10\\
	91	8\\
	92	9\\
	93	10\\
	94	8\\
	95	8\\
	96	7\\
	97	10\\
	98	12\\
	99	12\\
	100	9\\
	101	7\\
	102	10\\
	103	6\\
	104	12\\
	105	11\\
	106	13\\
	107	12\\
	108	10\\
	109	11\\
	110	11\\
	111	5\\
	112	12\\
	113	11\\
	114	10\\
	115	9\\
	116	8\\
	117	11\\
	118	11\\
	119	13\\
	120	10\\
	121	8\\
	122	12\\
	123	12\\
	124	11\\
	125	10\\
	126	9\\
	127	11\\
	128	11\\
	129	13\\
	130	10\\
	131	6\\
	132	10\\
	133	11\\
	134	7\\
	135	12\\
	136	12\\
	137	11\\
	138	9\\
	139	10\\
	140	11\\
	141	10\\
	142	7\\
	143	11\\
	144	10\\
	145	9\\
	146	11\\
	147	8\\
	148	12\\
	149	10\\
	150	13\\
	151	11\\
	152	12\\
	153	12\\
	154	10\\
	155	9\\
	156	10\\
	157	10\\
	158	10\\
	159	10\\
	160	9\\
	161	12\\
	162	11\\
	163	6\\
	164	10\\
	165	10\\
	166	11\\
	167	12\\
	168	11\\
	169	11\\
	170	13\\
	171	12\\
	172	12\\
	173	9\\
	174	7\\
	175	12\\
	176	9\\
	177	9\\
	178	9\\
	179	7\\
	180	11\\
	181	10\\
	182	10\\
	183	10\\
	184	8\\
	185	13\\
	186	9\\
	187	9\\
	188	8\\
	189	10\\
	190	10\\
	191	7\\
	192	7\\
	193	10\\
	194	9\\
	195	6\\
	196	8\\
	197	12\\
	198	9\\
	199	12\\
	200	11\\
};
\addlegendentry{$\tilde{s}_{\rm min}$};

\addplot [color=green,solid,mark=x,mark options={solid}]
table[row sep=crcr]{%
	1	13\\
	2	11\\
	3	11\\
	4	5\\
	5	7\\
	6	10\\
	7	9\\
	8	7\\
	9	4\\
	10	5\\
	11	10\\
	12	6\\
	13	12\\
	14	9\\
	15	12\\
	16	12\\
	17	6\\
	18	12\\
	19	7\\
	20	8\\
	21	9\\
	22	11\\
	23	12\\
	24	13\\
	25	6\\
	26	11\\
	27	10\\
	28	12\\
	29	4\\
	30	13\\
	31	6\\
	32	7\\
	33	12\\
	34	13\\
	35	12\\
	36	10\\
	37	9\\
	38	10\\
	39	7\\
	40	9\\
	41	6\\
	42	4\\
	43	7\\
	44	11\\
	45	6\\
	46	7\\
	47	11\\
	48	11\\
	49	6\\
	50	6\\
	51	7\\
	52	13\\
	53	7\\
	54	8\\
	55	5\\
	56	7\\
	57	4\\
	58	11\\
	59	10\\
	60	7\\
	61	5\\
	62	7\\
	63	5\\
	64	11\\
	65	12\\
	66	11\\
	67	11\\
	68	7\\
	69	6\\
	70	10\\
	71	5\\
	72	13\\
	73	7\\
	74	11\\
	75	3\\
	76	9\\
	77	5\\
	78	5\\
	79	7\\
	80	12\\
	81	10\\
	82	10\\
	83	11\\
	84	6\\
	85	6\\
	86	7\\
	87	10\\
	88	10\\
	89	5\\
	90	9\\
	91	6\\
	92	6\\
	93	6\\
	94	7\\
	95	4\\
	96	5\\
	97	1\\
	98	11\\
	99	12\\
	100	9\\
	101	6\\
	102	7\\
	103	5\\
	104	12\\
	105	6\\
	106	12\\
	107	12\\
	108	10\\
	109	7\\
	110	8\\
	111	5\\
	112	12\\
	113	10\\
	114	7\\
	115	9\\
	116	7\\
	117	11\\
	118	8\\
	119	10\\
	120	10\\
	121	6\\
	122	9\\
	123	12\\
	124	11\\
	125	7\\
	126	9\\
	127	10\\
	128	6\\
	129	13\\
	130	10\\
	131	5\\
	132	10\\
	133	11\\
	134	4\\
	135	12\\
	136	12\\
	137	10\\
	138	7\\
	139	10\\
	140	11\\
	141	7\\
	142	6\\
	143	11\\
	144	7\\
	145	8\\
	146	11\\
	147	7\\
	148	12\\
	149	7\\
	150	13\\
	151	7\\
	152	12\\
	153	11\\
	154	10\\
	155	7\\
	156	9\\
	157	7\\
	158	10\\
	159	7\\
	160	9\\
	161	12\\
	162	11\\
	163	4\\
	164	7\\
	165	7\\
	166	11\\
	167	12\\
	168	6\\
	169	11\\
	170	13\\
	171	12\\
	172	12\\
	173	5\\
	174	1\\
	175	6\\
	176	9\\
	177	5\\
	178	7\\
	179	6\\
	180	11\\
	181	6\\
	182	6\\
	183	7\\
	184	6\\
	185	13\\
	186	5\\
	187	6\\
	188	4\\
	189	6\\
	190	7\\
	191	6\\
	192	6\\
	193	6\\
	194	6\\
	195	3\\
	196	6\\
	197	7\\
	198	9\\
	199	12\\
	200	10\\
};
\addlegendentry{$\hat{h}_{\rm min}$};

\end{axis}
\end{tikzpicture}%
\caption{Minimum distance $d_{\rm min}$, minimum size of a noncodeword stopping set $\tilde{s}_{\rm min}$, and estimated termatiko distance $\hat{h}_{\rm min}$ versus code index for randomly generated binary measurement matrices from a protograph-based $(4,8)$-regular LDPC code ensemble.}
\label{fig:pg48_ensemble}
\end{figure*}

\section{Numerical Results} \label{sec:numerical_results}

In this section, we present numerical results for different specific measurement matrices and also for ensembles of measurement matrices, as well as simulation results of IPA performance.

\subsection{Termatiko Distance for Specific Matrices}

For all considered matrices we first find all stopping sets of size less than some threshold using the algorithm from  \cite{ros12,ros09}. Then, we exhaustively search for termatiko sets as subsets of these stopping sets \revone{according to Heuristic 1 (see Section~\ref{sec:alg_small_size}).} The results are tabulated in Table~\ref{table_of_codes} for five different measurement matrices, denoted by $A^{(1)}$, $A^{(2)}$, $A^{(3)}$, $A^{(4)}$, and $A^{(5)}$, respectively.  Due to the heuristic nature of the approach, the estimated termatiko distance is a true upper bound on the actual termatiko distance, while the estimated multiplicities are true lower bounds on the actual multiplicities. 
Measurement matrix $A^{(1)}$ is the $33 \times 121$ parity-check matrix $H(11,3)$ of the array-based LDPC code $\mathcal{C}(11,3)$ of column-weight $3$ and row-weight $11$ described in Section~\ref{sec:array_based}, $A^{(2)}$ is the parity-check matrix of the $(155,64)$ Tanner code from \cite{tan01}, $A^{(3)}$ is taken from the IEEE802.16e standard 
(it is the parity-check matrix of a rate-$3/4$, length-$1824$ LDPC code; using base model matrix A and the alternative construction, see \cite[Eq.~(1)]{ros12}),  $A^{(4)}$ is a $276 \times 552$ parity-check matrix of an irregular LDPC code, while $A^{(5)}$ is a $159 \times 265$ parity-check matrix of a $(3,5)$-regular LDPC code built from arrays of permutation matrices from Latin squares. For the matrix $A^{(1)}$, we have also compared the results with an exact enumeration of all termatiko sets of size at most $5$ \revus{obtained by an exhaustive search}. When considering all stopping sets of size at most $11$,  \revone{Heuristic 1} finds the exact multiplicities  for sizes $3$ and $4$, but it underestimates  the number of termatiko sets of size $5$ by about $7.5\%$  (the missing ones are subsets of stopping sets of size $12$ to $14$),   
%
which indicates that higher order terms (for all tabulated matrices) are likely strict lower bounds on the exact multiplicities. 
As can be seen from the table, for all matrices except $A^{(3)}$, the estimated termatiko distance is about half the stopping distance.  Also, the smallest-size termatiko sets all correspond to termatiko sets with all measurement nodes in $N$ connected to both $T$ and $S$ 
(cf.\ Theorem~\ref{thm:termatikos}). 

%
%
%
%

\subsection{Termatiko Distance for Protograph-Based Matrix Ensembles} \label{sec:protograph_based}

Now, consider the protograph-based $(3,6)$-regular LDPC code ensemble defined by the \emph{protomatrix} $H = (3,3)$. We randomly generated $200$ parity-check matrices from this ensemble using a lifting factor of $100$ (the two nonzero entries in the protomatrix are replaced by $100 \times 100$ binary matrices of row-weight $3$ in which all right-shifts of the first row (picked at random) occur in some order). For each lifted matrix, we first found all stopping sets of size at most $16$ using the algorithm from \cite{ros12,ros09}. Then, the termatiko distance was estimated for each matrix \revone{using Heuristic 1}. The results are depicted in Fig.~\ref{fig:pg36_ensemble} as a function of the code index (the blue curve shows the minimum distance $d_{\rm min}$, the red curve shows the minimum size of a noncodeword stopping set, denoted by $\tilde{s}_{\rm min}$, while the green curve shows the estimated termatiko distance $\hat{h}_{\rm min}$). 
The average $d_{\rm min}$, $s_{\rm min}$, and $\hat{h}_{\rm min}$ (over the $200$ matrices) are $6.84$, $5.92$, and $3.90$, respectively.\footnote{Note that here $5.92$ is the average stopping distance, and not the average size of the smallest noncodeword stopping sets.} We repeated a similar experiment using a lifting factor of $200$ in which case the average $d_{\rm min}$, $s_{\rm min}$, and $\hat{h}_{\rm min}$ (again over $200$ randomly generated matrices) became  $9.21$, $7.75$, and $5.80$, respectively.

Next, we repeat the same calculations for $200$ randomly generated parity-check matrices from the protograph-based $(4,8)$-regular LDPC code ensemble. For each parity-check matrix, we considered all stopping sets of size up to $14$. For some matrices, the minimum distances of the corresponding codes were larger than $14$, thus we calculated them separately. Fig.~\ref{fig:pg48_ensemble} presents the results of the calculations. The average $d_{\rm min}$, $s_{\rm min}$, and $\hat h_{\rm min}$ are $12.53$, $9.75$, and $8.41$, respectively.


\subsection{Performance of Algorithm~\ref{alg:SPLIT}}
In order to see how Algorithm~\ref{alg:SPLIT} performs, we applied it to the stopping sets of size at most $14$ for the protograph-based matrices described in the Section~\ref{sec:protograph_based}, both $(3,6)$ and $(4,8)$-regular matrices.

\begin{figure}[t]
	\centering
	\begin{tikzpicture}[font=\small]
	\begin{axis}[%
	ybar,
	height=0.3\textwidth,
	width=0.5\textwidth,
	xmin=1,
	xmax=15,
	ymin=0,
	ymax=1,
	xlabel={Stopping set size $w$},
	ymajorgrids,
	every x tick label/.append style={font=\color{black},font=\scriptsize},
	every y tick label/.append style={font=\color{black},font=\scriptsize},
	xtick={2,3,4,5,6,7,8,9,10,11,12,13,14}
	]
	\addplot[fill=black!75] table{
		2	1
		3	0
		4	1
		5	0.873386666666667
		6	0.957834578345783
		7	0.850786838340486
		8	0.935125222075764
		9	0.7743961352657
		10	0.898559607722954
		11	0.851974522292994
		12	0.870044853921687
		13	0.827807030027038
		14	0.835097935415564
	};
	\end{axis}
	\end{tikzpicture}
	\caption{Average success rate of Algorithm~\ref{alg:SPLIT} on stopping sets that allow a $(T,S)$-split for the $200$ randomly generated matrices  from the protograph-based $(3,6)$-regular LDPC code ensemble. Note that there were no splittable stopping sets of size $w=3$.}
	\label{fig:SPLIT-test-pg36}
\end{figure}
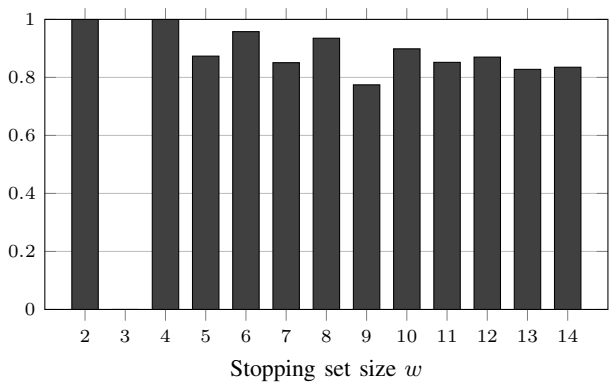

\begin{figure}[t]
	\centering
	\begin{tikzpicture}[font=\small]
	\begin{axis}[%
	ybar,
	height=0.3\textwidth,
	width=0.5\textwidth,
	xmin=1,
	xmax=15,
	ymin=0,
	ymax=1,
	xlabel={Stopping set size $w$},
	ymajorgrids,
	every x tick label/.append style={font=\color{black},font=\scriptsize},
	every y tick label/.append style={font=\color{black},font=\scriptsize},
	xtick={2,3,4,5,6,7,8,9,10,11,12,13,14}
	]
	\addplot[fill=black!75] table{
		2	1
		3	0
		4	0
		5	0
		6	0.966568338249754
		7	1
		8	0.940060698027314
		9	0.953830010493179
		10	0.954706298655343
		11	0.880172413793103
		12	0.884415133634155
		13	0.705352986811482
		14	0.621706153725858
	};
	\end{axis}
	\end{tikzpicture}
	\caption{Average success rate of Algorithm~\ref{alg:SPLIT} on stopping sets that allow a	$(T,S)$-split for the $200$ randomly generated matrices from the protograph-based $(4,8)$-regular LDPC code ensemble. Note that there were no splittable stopping sets of sizes $w=3,4,5$.}
	\label{fig:SPLIT-test-pg48}
\end{figure}
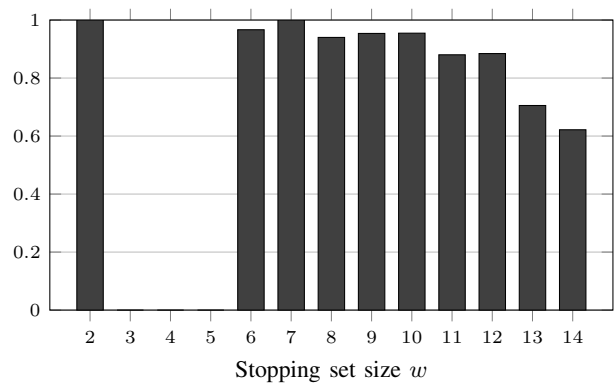

Table~\ref{tab:SPLIT_test_pg36} shows the average number of stopping sets of size $w$, for $w=1,2,\dotsc,14$ for the $200$ randomly generated $(3,6)$-regular matrices (the numbers are exact). It also presents the fraction of the matrices that have stopping sets of size $w$. In particular, all the $200$ matrices have stopping sets of size $w=13$ and $w=14$. For a fixed $w$, we also considered the total multiset of all stopping sets from all the matrices together and calculated the fraction of them that are splittable in their corresponding matrix. The last column of Table~\ref{tab:SPLIT_test_pg36} displays these numbers. Next, we built the total multiset of all splittable stopping sets from all the matrices together and repeatedly ran Algorithm~\ref{alg:SPLIT} to estimate the average success probability across the multiset. The resulting frequencies are depicted in Fig.~\ref{fig:SPLIT-test-pg36}. The aforementioned calculations were repeated for the $200$ randomly generated $(4,8)$-regular matrices. The results are presented in Table~\ref{tab:SPLIT_test_pg48} and Fig.~\ref{fig:SPLIT-test-pg48}.

\begin{table}[t]
	\scriptsize \centering \caption{Stopping sets (incl. codewords) distribution over 200 randomly generated matrices from the protograph-based $(3,6)$-regular LDPC code ensemble. Numbers are exact.}
	\label{tab:SPLIT_test_pg36}
	\vskip -2.0ex 
	\begin{tabular}{cccc}
		\toprule
		$w$ &
		\begin{tabular}{@{}c@{}}average number of \\size-$w$ stopping sets\end{tabular} &
		\begin{tabular}{@{}c@{}}fraction of codes having \\size-$w$ stopping sets\end{tabular} &
		\begin{tabular}{@{}c@{}}fraction of size-$w$ \\stopping sets \\allowing a $(T,S)$-split \end{tabular}\\
		\midrule
	     1 &    0.000 & 0.000 & - \\
	     2 &    0.080 & 0.075 & 1.000 \\
	     3 &    0.010 & 0.010 & 0.000 \\
	     4 &    0.150 & 0.125 & 0.267 \\
	     5 &    0.320 & 0.215 & 0.094 \\
	     6 &    1.350 & 0.485 & 0.222 \\
	     7 &    5.365 & 0.690 & 0.070 \\
	     8 &   10.860 & 0.925 & 0.174 \\
	     9 &   33.695 & 0.995 & 0.083 \\
	    10 &  105.935 & 1.000 & 0.099 \\
	    11 &  298.085 & 1.000 & 0.079 \\
	    12 &  953.220 & 1.000 & 0.082 \\
	    13 & 3029.230 & 1.000 & 0.070 \\
	    14 & 9887.395 & 1.000 & 0.076 \\
	    \bottomrule
	\end{tabular}
\end{table}


\begin{table}[t]
	\scriptsize \centering \caption{Stopping sets (incl. codewords) distribution over 200 randomly generated matrices from the protograph-based $(4,8)$-regular LDPC code ensemble. Numbers are exact.}
	\label{tab:SPLIT_test_pg48}
	\vskip -2.0ex 
	\begin{tabular}{cccc}
		\toprule
		$w$ &
		\begin{tabular}{@{}c@{}}average number of \\size-$w$ stopping sets\end{tabular} &
		\begin{tabular}{@{}c@{}}fraction of codes having \\size-$w$ stopping sets\end{tabular} &
		\begin{tabular}{@{}c@{}}fraction of size-$w$ \\stopping sets \\allowing a $(T,S)$-split\end{tabular}\\
		\midrule
		 1 &    0.000 & 0.000 & - \\
		 2 &    0.010 & 0.010 & 1.000 \\
		 3 &    0.000 & 0.000 & - \\
		 4 &    0.125 & 0.005 & 0.000 \\
		 5 &    0.210 & 0.020 & 0.000 \\
		 6 &    0.295 & 0.045 & 0.051\\
		 7 &    0.185 & 0.085 & 0.243 \\
		 8 &    3.415 & 0.190 & 0.013 \\
		 9 &    4.720 & 0.335 & 0.010 \\
		10 &   20.525 & 0.545 & 0.014 \\
		11 &   70.705 & 0.720 & 0.012 \\
		12 &  305.780 & 0.910 & 0.029 \\
		13 &  827.665 & 1.000 & 0.064 \\
		14 & 2219.780 & 1.000 & 0.128 \\
		\bottomrule
	\end{tabular}
\end{table}


\subsection{Adding Redundant Rows}
To illustrate the efficiency of the heuristic algorithm from  Section~\ref{sec:add-red-rows} in removing small-size termatiko sets, we
chose three out of the $200$ $(3,6)$-regular matrices (with a lifting factor of $100$) in Section~\ref{sec:protograph_based} as example matrices. More precisely, the matrices with indices $20$, $72$, and $172$, denoted by $A^{(20)}_{\textsf{PG}}$, $A^{(72)}_{\textsf{PG}}$, and $A^{(172)}_{\textsf{PG}}$, respectively, were selected. These matrices were chosen to  demonstrate different behavior patterns. \revone{As a side note, we remark that the Tanner graphs of the matrices $A^{(20)}_{\textsf{PG}}$, $A^{(72)}_{\textsf{PG}}$, and $A^{(172)}_{\textsf{PG}}$ have $126$, $22$, and $13$ cycles of length $4$, respectively (cf.\ Theorem~\ref{thm:col-a-reg}). However, both $A^{(72)}_{\textsf{PG}}$ and $A^{(172)}_{\textsf{PG}}$ have a termatiko distance of at least $3$. Therefore, the requirement in Theorem~\ref{thm:col-a-reg} to have no cycles of length $4$ is sufficient but not necessary.}

For all three matrices we applied the algorithm from Section~\ref{sec:add-red-rows} in order to remove termatiko sets by adding redundant rows. The algorithm added $30$ redundant rows to $A^{(20)}_{\textsf{PG}}$, $55$ rows to $A^{(72)}_{\textsf{PG}}$, and $68$ rows to $A^{(172)}_{\textsf{PG}}$. Due to computing limitations, we were able to tackle only a limited number of termatiko sets. $A^{(20)}_{\textsf{PG}}$ originally had the highest numbers of termatiko sets, and because of that we only processed all termatiko sets of size up to $5$ (including). For $A^{(72)}_{\textsf{PG}}$, we processed all termatiko sets of size up to $7$, and for $A^{(172)}_{\textsf{PG}}$ sizes up to $8$ were considered. Accordingly, we will occasionally denote the extended matrices by $A^{(20)}_{\textsf{EPG}^{(\mathsf 5)}}$, $A^{(72)}_{\textsf{EPG}^{(\mathsf 7)}}$, and $A^{(172)}_{\textsf{EPG}^{(\mathsf 8)}}$. The numbers of termatiko sets decreased for all matrices. Moreover, for $A^{(72)}_{\textsf{PG}}$ and $A^{(172)}_{\textsf{PG}}$ we were also able to increase their termatiko distances. Table~\ref{tab:targetST} shows the estimated termatiko set size spectra (initial part) for the original and extended matrices.

\begin{table}[t]
	\scriptsize \caption{Estimated termatiko set size spectra (initial part) for three protograph-based matrices from Fig.~\ref{fig:pg36_ensemble} before and after adding redundant rows. Numbers in angle brackets stand for termatiko distance $h_{\rm min}$, size of the smallest noncodeword stopping set $\tilde s_{\rm min}$, and minimum distance $d_{\rm min}$, respectively, for the original nonextended measurement matrices. Numbers in bold are exact. We tried to ``remove'' termatiko sets of size up to $\ell$ (including).}
	\label{tab:targetST}
	\centering
	\vskip -2.0ex
	\begin{tabular}{ccccccc}
		\toprule
		~ &
		\multicolumn{2}{c}{$A_{\textsf{PG}}^{(20)} \langle 1,4,2 \rangle$} &
		\multicolumn{2}{c}{$A_{\textsf{PG}}^{(72)} \langle 3,7,10 \rangle$} &
		\multicolumn{2}{c}{$A_{\textsf{PG}}^{(172)} \langle 6,8,6\rangle$}
		\\[1mm]
		$w$	& 
		\begin{tabular}{@{}c@{}} original \\ ($\ell=0$)	\end{tabular}	& \begin{tabular}{@{}c@{}} extended \\ ($\ell=5$)	\end{tabular} 
		&
		\begin{tabular}{@{}c@{}} original \\ ($\ell=0$)	\end{tabular}	& \begin{tabular}{@{}c@{}} extended \\ ($\ell=7$)	\end{tabular}
		&
		\begin{tabular}{@{}c@{}} original \\ ($\ell=0$)	\end{tabular}	& \begin{tabular}{@{}c@{}} extended \\ ($\ell=8$)	\end{tabular}\\
		\midrule
		1 & \textbf{2} & \textbf{2} & \textbf{0} & \textbf{0} & \textbf{0} & \textbf{0} \\
		2 & \textbf{4} & \textbf{1} & \textbf{0} & \textbf{0} & \textbf{0} & \textbf{0} \\
		3 & \textbf{11} & \textbf{0} & \textbf{1} & \textbf{0} & \textbf{0} & \textbf{0} \\
		4 & \textbf{82} & \textbf{0} & \textbf{3} & \textbf{0} & \textbf{0} & \textbf{0} \\
		5 & \textbf{837} & \textbf{16} & \textbf{19} & \textbf{2} & \textbf{0} & \textbf{0} \\
		6 & \textbf{7860} & \textbf{265} & \textbf{83} & \textbf{0} & \textbf{23} & \textbf{0} \\
		7 & 84059 & 5214 & 794 & 0 & 263 & 0 \\
		8 & 670146 & 61519 & 5204 & 98 & 1780 & 5 \\
		9 & 1885358 & 182366 & 6904 & 109 & 2134 & 10 \\
		\dots & \dots & \dots & \dots & \dots & \dots & \dots \\
		\bottomrule
	\end{tabular}	
\end{table}

%

\begin{figure*}[t!]
	\centering
	\subfloat[All three matrices, both original and extended.]{
		\label{fig:pg_3plus3_tikz-upd}
		\begin{tikzpicture}[font=\small]
		
		\begin{axis}[%
		width=0.37\textwidth,
		height=0.24\textwidth,
		at={(0.808889in,0.513333in)},
		scale only axis,
		separate axis lines,
		every outer x axis line/.append style={black},
		every x tick label/.append style={font=\color{black}},
		xmin=0,
		xmax=50,
		xlabel={Signal weight},
        		ytick={1e-9,1e-8,1e-7,1e-6,1e-5,1e-4,1e-3,1e-2,1e-1,1},
        		xtick={0,10,20,30,40,50},
        		grid style={gray,opacity=0.5,dotted},
		xmajorgrids,
		every outer y axis line/.append style={black},
		every y tick label/.append style={font=\color{black}},
		ymin=1e-9,
		ymax=1,
		ymode=log,
		ylabel={IPA FER},
		ymajorgrids,
		legend style={at={(0.97,0.03)},anchor=south east,legend cell align=left,align=left,draw=black,font=\scriptsize}
		]
		\addplot[color=blue,solid] 
		table[row sep=crcr]{%
			1 0.0099282\\
			2 0.0201012\\
			3 0.0303068\\
			4 0.0405804\\
			5 0.0508566\\
			6 0.061616\\
			7 0.0718862\\
			8 0.0827906\\
			9 0.0936194\\
			10 0.1046926\\
			11 0.1155924\\
			12 0.126773\\
			13 0.1379892\\
			14 0.1499402\\
			15 0.161467\\
			16 0.1733216\\
			17 0.1855104\\
			18 0.1980572\\
			19 0.2107016\\
			20 0.2242696\\
			21 0.2382168\\
			22 0.2528062\\
			23 0.2688286\\
			24 0.2856838\\
			25 0.3041616\\
			26 0.3249832\\
			27 0.3486072\\
			28 0.375755\\
			29 0.4064336\\
			30 0.4419196\\
			31 0.481654\\
			32 0.5265384\\
			33 0.575048\\
			34 0.6274332\\
			35 0.680695\\
			36 0.7336412\\
			37 0.7842078\\
			38 0.8306756\\
			39 0.871501\\
			40 0.9058616\\
			41 0.9333462\\
			42 0.9548056\\
			43 0.9703792\\
			44 0.9812928\\
			45 0.9888536\\
			46 0.993511\\
			47 0.9964006\\
			48 0.9981302\\
			49 0.999022\\
			50 0.999517\\
		};
		\addlegendentry{$A^{(20)}_{\textsf{PG}}$};
		
		\addplot[color=blue,dashed] 
		table[row sep=crcr]{%
			1 0.0099962\\
			2 0.0199178\\
			3 0.0297438\\
			4 0.0397218\\
			5 0.0494058\\
			6 0.0592776\\
			7 0.0690188\\
			8 0.0785044\\
			9 0.0879878\\
			10 0.0976956\\
			11 0.107273\\
			12 0.1167146\\
			13 0.1258122\\
			14 0.135476\\
			15 0.1445408\\
			16 0.154441\\
			17 0.1637342\\
			18 0.1726644\\
			19 0.1816636\\
			20 0.1910334\\
			21 0.2008596\\
			22 0.210205\\
			23 0.2207622\\
			24 0.231568\\
			25 0.243914\\
			26 0.2568908\\
			27 0.2725936\\
			28 0.291058\\
			29 0.3144856\\
			30 0.3424888\\
			31 0.376677\\
			32 0.4177474\\
			33 0.464904\\
			34 0.5180688\\
			35 0.5761188\\
			36 0.6366892\\
			37 0.697291\\
			38 0.7555142\\
			39 0.8091194\\
			40 0.8559372\\
			41 0.8948626\\
			42 0.9264364\\
			43 0.950162\\
			44 0.9675984\\
			45 0.979774\\
			46 0.9877834\\
			47 0.992982\\
			48 0.9961032\\
			49 0.9979098\\
			50 0.9989292\\
		};
		\addlegendentry{$A^{(20)}_{\textsf{EPG}^{(5)}}$};
		
		\addplot[color=red,solid] 
		table[row sep=crcr]{%
			3 8.0999999996223E-07\\
			4 3.01000000002549E-06\\
			5 7.58999999994625E-06\\
			6 1.54299999999274E-05\\
			7 2.68599999999619E-05\\
			8 4.4970000000033E-05\\
			9 6.66499999999459E-05\\
			10 9.69899999999635E-05\\
			11 0.00013428\\
			12 0.00018099\\
			13 0.00023671\\
			14 0.00030546\\
			15 0.00038936\\
			16 0.00048797\\
			17 0.00060239\\
			18 0.00073825\\
			19 0.00090755\\
			20 0.00110223\\
			21 0.0013832\\
			22 0.0016736\\
			23 0.0020912\\
			24 0.002669\\
			25 0.0035688\\
			26 0.004995\\
			27 0.0072686\\
			28 0.0110156\\
			29 0.0169658\\
			30 0.0266388\\
			31 0.0413992\\
			32 0.063252\\
			33 0.0943254\\
			34 0.1358298\\
			35 0.1903196\\
			36 0.2558564\\
			37 0.3319882\\
			38 0.4155986\\
			39 0.5031962\\
			40 0.5905548\\
			41 0.6736852\\
			42 0.7481698\\
			43 0.8127644\\
			44 0.8656986\\
			45 0.906899\\
			46 0.9379954\\
			47 0.9604332\\
			48 0.9753942\\
			49 0.9853818\\
			50 0.9916892\\
		};
		\addlegendentry{$A^{(72)}_{\textsf{PG}}$};
		
		\addplot[color=red,dashed] 
		table[row sep=crcr]{%
			8 5.99999999684187E-08\\
			9 1.00000000058387E-07\\
			10 2.29999999934449E-07\\
			11 3.29999999992836E-07\\
			12 5.59999999927285E-07\\
			13 9.69999999989035E-07\\
			14 1.60999999998523E-06\\
			15 2.48000000002691E-06\\
			16 4.0299999999327E-06\\
			17 6.33999999999357E-06\\
			18 9.55000000002482E-06\\
			19 1.47499999999523E-05\\
			20 2.41799999999293E-05\\
			21 4.8800000000071E-05\\
			22 8.45999999999902E-05\\
			23 0.0001592\\
			24 0.0003302\\
			25 0.0006774\\
			26 0.0013318\\
			27 0.0026648\\
			28 0.0051804\\
			29 0.0095668\\
			30 0.0172024\\
			31 0.0296714\\
			32 0.0487802\\
			33 0.0767528\\
			34 0.1157996\\
			35 0.1669504\\
			36 0.2304212\\
			37 0.3056012\\
			38 0.3889184\\
			39 0.4774848\\
			40 0.567058\\
			41 0.6523766\\
			42 0.7306304\\
			43 0.7978126\\
			44 0.8537698\\
			45 0.8982956\\
			46 0.931857\\
			47 0.9559438\\
			48 0.9724888\\
			49 0.983573\\
			50 0.9904782\\
		};
		\addlegendentry{$A^{(72)}_{\textsf{EPG}^{(7)}}$};
		
		\addplot[color=green,solid] 
		table[row sep=crcr]{%
			9 3.99999999789458E-08\\
			10 7.00000000186662E-08\\
			11 2.09999999944976E-07\\
			12 4.20000000000975E-07\\
			13 7.39999999943564E-07\\
			14 1.55000000001682E-06\\
			15 2.60000000007476E-06\\
			16 4.6400000000002E-06\\
			17 8.11999999994484E-06\\
			18 1.44599999999384E-05\\
			19 2.47899999999968E-05\\
			20 4.32399999999999E-05\\
			21 8.2400000000038E-05\\
			22 0.0001406\\
			23 0.0002402\\
			24 0.0004466\\
			25 0.0008692\\
			26 0.0016296\\
			27 0.0030682\\
			28 0.0057372\\
			29 0.0103774\\
			30 0.0182364\\
			31 0.0311342\\
			32 0.0503792\\
			33 0.078681\\
			34 0.1181682\\
			35 0.1696414\\
			36 0.2339106\\
			37 0.308345\\
			38 0.3921004\\
			39 0.4807468\\
			40 0.5698528\\
			41 0.6546882\\
			42 0.7325406\\
			43 0.7997056\\
			44 0.8553744\\
			45 0.8994434\\
			46 0.9325092\\
			47 0.9564436\\
			48 0.9729684\\
			49 0.983698\\
			50 0.9906886\\
		};
		\addlegendentry{$A^{(172)}_{\textsf{PG}}$};
		
		\addplot[color=green,dashed] 
		table[row sep=crcr]{%
			13 1.99999994343614E-09\\
			14 1.89999999067325E-08\\
			15 3.99999999789458E-08\\
			16 1.09999999997612E-07\\
			17 3.80000000022029E-07\\
			18 8.0999999996223E-07\\
			19 2.44999999998718E-06\\
			20 5.16999999999879E-06\\
			21 1.58000000000102E-05\\
			22 3.78000000000878E-05\\
			23 8.4799999999996E-05\\
			24 0.0001722\\
			25 0.0004072\\
			26 0.0009292\\
			27 0.0019152\\
			28 0.0039878\\
			29 0.0077052\\
			30 0.0145074\\
			31 0.0256604\\
			32 0.0433534\\
			33 0.069808\\
			34 0.1070512\\
			35 0.1564986\\
			36 0.218233\\
			37 0.2920032\\
			38 0.3757044\\
			39 0.4641992\\
			40 0.5542036\\
			41 0.6406466\\
			42 0.7200282\\
			43 0.7896106\\
			44 0.847241\\
			45 0.8934224\\
			46 0.9279766\\
			47 0.9532054\\
			48 0.9707834\\
			49 0.9823976\\
			50 0.9897482\\
		};
		\addlegendentry{$A^{(172)}_{\textsf{EPG}^{(8)}}$};
		
		\end{axis}
		\end{tikzpicture}
	}\hfill
	\subfloat[\textsf{$A^{(20)}_{\textsf{PG}}$}]{
		\label{fig:pg20_fer_tikz-upd}
		\begin{tikzpicture}[font=\small]
		
		\begin{axis}[%
		width=0.37\textwidth,
		height=0.24\textwidth,
		at={(0.808889in,0.513333in)},
		scale only axis,
		separate axis lines,
        		ytick={1e-2,1e-1,1},
        		xtick={0,10,20,30,40,50},
        		grid style={gray,opacity=0.5,dotted},
		every outer x axis line/.append style={black},
		every x tick label/.append style={font=\color{black}},
		xmin=0,
		xmax=50,
		xlabel={Signal weight},
		xmajorgrids,
		every outer y axis line/.append style={black},
		every y tick label/.append style={font=\color{black}},
		ymin=1e-2,
		ymax=1,
		ymode=log,
		ylabel={IPA FER},
		ymajorgrids,
		legend style={at={(0.97,0.03)},anchor=south east,legend cell align=left,align=left,draw=black,font=\scriptsize}
		]
		\addplot[color=blue,solid] 
		table[row sep=crcr]{%
			1 0.0099282\\
			2 0.0201012\\
			3 0.0303068\\
			4 0.0405804\\
			5 0.0508566\\
			6 0.061616\\
			7 0.0718862\\
			8 0.0827906\\
			9 0.0936194\\
			10 0.1046926\\
			11 0.1155924\\
			12 0.126773\\
			13 0.1379892\\
			14 0.1499402\\
			15 0.161467\\
			16 0.1733216\\
			17 0.1855104\\
			18 0.1980572\\
			19 0.2107016\\
			20 0.2242696\\
			21 0.2382168\\
			22 0.2528062\\
			23 0.2688286\\
			24 0.2856838\\
			25 0.3041616\\
			26 0.3249832\\
			27 0.3486072\\
			28 0.375755\\
			29 0.4064336\\
			30 0.4419196\\
			31 0.481654\\
			32 0.5265384\\
			33 0.575048\\
			34 0.6274332\\
			35 0.680695\\
			36 0.7336412\\
			37 0.7842078\\
			38 0.8306756\\
			39 0.871501\\
			40 0.9058616\\
			41 0.9333462\\
			42 0.9548056\\
			43 0.9703792\\
			44 0.9812928\\
			45 0.9888536\\
			46 0.993511\\
			47 0.9964006\\
			48 0.9981302\\
			49 0.999022\\
			50 0.999517\\
		};
		\addlegendentry{$A^{(20)}_{\textsf{PG}}$};
		
		\addplot[color=red,solid] 
		table[row sep=crcr]{%
			1 0.0099962\\
			2 0.0199178\\
			3 0.0297438\\
			4 0.0397218\\
			5 0.0494058\\
			6 0.0592776\\
			7 0.0690188\\
			8 0.0785044\\
			9 0.0879878\\
			10 0.0976956\\
			11 0.107273\\
			12 0.1167146\\
			13 0.1258122\\
			14 0.135476\\
			15 0.1445408\\
			16 0.154441\\
			17 0.1637342\\
			18 0.1726644\\
			19 0.1816636\\
			20 0.1910334\\
			21 0.2008596\\
			22 0.210205\\
			23 0.2207622\\
			24 0.231568\\
			25 0.243914\\
			26 0.2568908\\
			27 0.2725936\\
			28 0.291058\\
			29 0.3144856\\
			30 0.3424888\\
			31 0.376677\\
			32 0.4177474\\
			33 0.464904\\
			34 0.5180688\\
			35 0.5761188\\
			36 0.6366892\\
			37 0.697291\\
			38 0.7555142\\
			39 0.8091194\\
			40 0.8559372\\
			41 0.8948626\\
			42 0.9264364\\
			43 0.950162\\
			44 0.9675984\\
			45 0.979774\\
			46 0.9877834\\
			47 0.992982\\
			48 0.9961032\\
			49 0.9979098\\
			50 0.9989292\\
		};
		\addlegendentry{$A^{(20)}_{\textsf{EPG}^{(5)}}$};
		
		\addplot[color=blue,dashed] 
		table[row sep=crcr]{%
			1 0.01\\
			2 0.020100502513\\
			3 0.030300746155\\
			4 0.040599814949\\
			5 0.050996636744\\
			6 0.061489981856\\
			7 0.072078461938\\
			8 0.082760529078\\
			9 0.093534475143\\
			10 0.104398431373\\
			11 0.115350368223\\
			12 0.126388095479\\
			13 0.137509262634\\
			14 0.148711359541\\
			15 0.15999171735\\
			16 0.171347509727\\
			17 0.182775754365\\
			18 0.194273314789\\
			19 0.205836902459\\
			20 0.217463079176\\
			21 0.229148259792\\
			22 0.240888715217\\
			23 0.252680575749\\
			24 0.264519834695\\
			25 0.276402352307\\
			26 0.288323860024\\
			27 0.300279965012\\
			28 0.312266155014\\
			29 0.324277803488\\
			30 0.336310175045\\
			31 0.348358431175\\
			32 0.360417636253\\
			33 0.372482763818\\
			34 0.384548703126\\
			35 0.396610265959\\
			36 0.408662193678\\
			37 0.420699164526\\
			38 0.432715801146\\
			39 0.444706678324\\
			40 0.456666330928\\
			41 0.468589262039\\
			42 0.48046995125\\
			43 0.49230286313\\
			44 0.504082455825\\
			45 0.515803189778\\
			46 0.527459536566\\
			47 0.539045987816\\
			48 0.550557064187\\
			49 0.561987324414\\
			50 0.573331374361\\
		};
		\addlegendentry{$A^{(20)}_{\textsf{PG}}$, PIE};
		
		\addplot[color=red,dashed] 
		table[row sep=crcr]{%
			1	0.01\\
			2	0.019949748744\\
			3	0.029849246231\\
			4	0.039698492462\\
			5	0.049497493747\\
			6	0.059246268628\\
			7	0.068944853417\\
			8	0.078593307369\\
			9	0.08819171747\\
			10	0.097740202887\\
			11	0.107238919043\\
			12	0.11668806136\\
			13	0.126087868655\\
			14	0.135438626205\\
			15	0.144740668483\\
			16	0.153994381574\\
			17	0.163200205272\\
			18	0.172358634878\\
			19	0.181470222679\\
			20	0.190535579144\\
			21	0.199555373811\\
			22	0.208530335901\\
			23	0.217461254629\\
			24	0.226348979252\\
			25	0.235194418827\\
			26	0.243998541699\\
			27	0.25276237472\\
			28	0.261487002194\\
			29	0.270173564559\\
			30	0.278823256805\\
			31	0.287437326621\\
			32	0.296017072288\\
			33	0.304563840306\\
			34	0.313079022757\\
			35	0.321564054409\\
			36	0.330020409561\\
			37	0.338449598618\\
			38	0.346853164411\\
			39	0.355232678252\\
			40	0.363589735724\\
			41	0.371925952206\\
			42	0.380242958141\\
			43	0.388542394031\\
			44	0.396825905169\\
			45	0.405095136116\\
			46	0.413351724907\\
			47	0.421597296999\\
			48	0.429833458961\\
			49	0.438061791913\\
			50	0.446283844712\\
		};
		\addlegendentry{$A^{(20)}_{\textsf{EPG}^{(5)}}$, PIE};
		
		\end{axis}
		\end{tikzpicture}
	}\hfill
	
	\subfloat[$A^{(72)}_{\textsf{PG}}$]{
		\label{fig:pg72_fer_tikz-upd}
		\begin{tikzpicture}[font=\small]
		
		\begin{axis}[%
		width=0.37\textwidth,
		height=0.24\textwidth,
		at={(0.808889in,0.513333in)},
		scale only axis,
		separate axis lines,
        		ytick={1e-10,1e-9,1e-8,1e-7,1e-6,1e-5,1e-4,1e-3,1e-2,1e-1,1},
        		xtick={0,10,20,30,40,50},
        		grid style={gray,opacity=0.5,dotted},
		every outer x axis line/.append style={black},
		every x tick label/.append style={font=\color{black}},
		xmin=0,
		xmax=50,
		xlabel={Signal weight},
		xmajorgrids,
		every outer y axis line/.append style={black},
		every y tick label/.append style={font=\color{black}},
		ymin=1e-10,
		ymax=1,
		ymode=log,
		ylabel={IPA FER},
		ymajorgrids,
		legend style={at={(0.97,0.03)},anchor=south east,legend cell align=left,align=left,draw=black,font=\scriptsize}
		]
		\addplot[color=blue,solid] 
		table[row sep=crcr]{%
			3	8.0999999996223E-07\\
			4	3.01000000002549E-06\\
			5	7.58999999994625E-06\\
			6	1.54299999999274E-05\\
			7	2.68599999999619E-05\\
			8	4.4970000000033E-05\\
			9	6.66499999999459E-05\\
			10	9.69899999999635E-05\\
			11	0.00013428\\
			12	0.00018099\\
			13	0.00023671\\
			14	0.00030546\\
			15	0.00038936\\
			16	0.00048797\\
			17	0.00060239\\
			18	0.00073825\\
			19	0.00090755\\
			20	0.00110223\\
			21	0.0013832\\
			22	0.0016736\\
			23	0.0020912\\
			24	0.002669\\
			25	0.0035688\\
			26	0.004995\\
			27	0.0072686\\
			28	0.0110156\\
			29	0.0169658\\
			30	0.0266388\\
			31	0.0413992\\
			32	0.063252\\
			33	0.0943254\\
			34	0.1358298\\
			35	0.1903196\\
			36	0.2558564\\
			37	0.3319882\\
			38	0.4155986\\
			39	0.5031962\\
			40	0.5905548\\
			41	0.6736852\\
			42	0.7481698\\
			43	0.8127644\\
			44	0.8656986\\
			45	0.906899\\
			46	0.9379954\\
			47	0.9604332\\
			48	0.9753942\\
			49	0.9853818\\
			50	0.9916892\\
		};
		\addlegendentry{$A^{(72)}_{\textsf{PG}}$};
		
		\addplot[color=red,solid] 
		table[row sep=crcr]{%
			8	5.99999999684187E-08\\
			9	1.00000000058387E-07\\
			10	2.29999999934449E-07\\
			11	3.29999999992836E-07\\
			12	5.59999999927285E-07\\
			13	9.69999999989035E-07\\
			14	1.60999999998523E-06\\
			15	2.48000000002691E-06\\
			16	4.0299999999327E-06\\
			17	6.33999999999357E-06\\
			18	9.55000000002482E-06\\
			19	1.47499999999523E-05\\
			20	2.41799999999293E-05\\
			21	4.8800000000071E-05\\
			22	8.45999999999902E-05\\
			23	0.0001592\\
			24	0.0003302\\
			25	0.0006774\\
			26	0.0013318\\
			27	0.0026648\\
			28	0.0051804\\
			29	0.0095668\\
			30	0.0172024\\
			31	0.0296714\\
			32	0.0487802\\
			33	0.0767528\\
			34	0.1157996\\
			35	0.1669504\\
			36	0.2304212\\
			37	0.3056012\\
			38	0.3889184\\
			39	0.4774848\\
			40	0.567058\\
			41	0.6523766\\
			42	0.7306304\\
			43	0.7978126\\
			44	0.8537698\\
			45	0.8982956\\
			46	0.931857\\
			47	0.9559438\\
			48	0.9724888\\
			49	0.983573\\
			50	0.9904782\\
		};
		\addlegendentry{$A^{(72)}_{\textsf{EPG}^{(7)}}$};
		
		\addplot[color=blue,dashed] 
		table[row sep=crcr]{%
			1	0\\
			2	0\\
			3	7.61383E-07\\
			4	3.06099E-06\\
			5	7.695857E-06\\
			6	1.5487893E-05\\
			7	2.7288518E-05\\
			8	4.3983238E-05\\
			9	6.649617E-05\\
			10	9.5794501E-05\\
			11	0.000132892888\\
			12	0.000178857784\\
			13	0.000234811703\\
			14	0.0003019374\\
			15	0.000381481973\\
			16	0.000474760883\\
			17	0.000583161883\\
			18	0.000708148849\\
			19	0.000851265511\\
			20	0.001014139082\\
			21	0.001198483763\\
			22	0.001406104133\\
			23	0.001638898417\\
			24	0.001898861604\\
			25	0.002188088435\\
			26	0.002508776227\\
			27	0.002863227544\\
			28	0.003253852691\\
			29	0.003683172029\\
			30	0.004153818093\\
			31	0.004668537514\\
			32	0.005230192716\\
			33	0.005841763393\\
			34	0.006506347737\\
			35	0.007227163421\\
			36	0.008007548305\\
			37	0.008850960858\\
			38	0.009760980289\\
			39	0.010741306357\\
			40	0.011795758853\\
			41	0.012928276728\\
			42	0.014142916868\\
			43	0.015443852473\\
			44	0.016835371043\\
			45	0.018321871945\\
			46	0.019907863534\\
			47	0.021597959824\\
			48	0.023396876681\\
			49	0.025309427516\\
			50	0.027340518468\\
		};
		\addlegendentry{$A^{(72)}_{\textsf{PG}}$, PIE};
		
		\addplot[color=red,dashed] 
		table[row sep=crcr]{%
			5	7.89E-10\\
			6	4.733E-09\\
			7	1.6564E-08\\
			8	4.4171E-08\\
			9	9.9388E-08\\
			10	1.98793E-07\\
			11	3.64505E-07\\
			12	6.24997E-07\\
			13	1.015918E-06\\
			14	1.580933E-06\\
			15	2.37259E-06\\
			16	3.453201E-06\\
			17	4.895777E-06\\
			18	6.784985E-06\\
			19	9.21816E-06\\
			20	1.230638E-05\\
			21	1.6175589E-05\\
			22	2.0967808E-05\\
			23	2.6842422E-05\\
			24	3.3977563E-05\\
			25	4.257159E-05\\
			26	5.2844696E-05\\
			27	6.5040633E-05\\
			28	7.942858E-05\\
			29	9.6305163E-05\\
			30	0.000115996644\\
			31	0.000138861285\\
			32	0.000165291911\\
			33	0.000195718682\\
			34	0.000230612089\\
			35	0.000270486191\\
			36	0.000315902111\\
			37	0.000367471801\\
			38	0.000425862105\\
			39	0.00049179912\\
			40	0.000566072894\\
			41	0.000649542451\\
			42	0.000743141189\\
			43	0.000847882651\\
			44	0.000964866684\\
			45	0.00109528602\\
			46	0.001240433281\\
			47	0.001401708423\\
			48	0.00158062665\\
			49	0.001778826796\\
			50	0.001998080199\\
		};
		\addlegendentry{$A^{(72)}_{\textsf{EPG}^{(7)}}$, PIE};
		
		\end{axis}
		\end{tikzpicture}
	}\hfill
	\subfloat[$A^{(172)}_{\textsf{PG}}$]{
		\label{fig:pg172_fer_tikz-upd}
		\begin{tikzpicture}[font=\small]
		
		\begin{axis}[%
		width=0.37\textwidth,
		height=0.24\textwidth,
		at={(0.808889in,0.513333in)},
		scale only axis,
		separate axis lines,
        		ytick={1e-13,1e-12,1e-11,1e-10,1e-9,1e-8,1e-7,1e-6,1e-5,1e-4,1e-3,1e-2,1e-1,1},
        		xtick={0,10,20,30,40,50},
        		grid style={gray,opacity=0.5,dotted},
		every outer x axis line/.append style={black},
		every x tick label/.append style={font=\color{black}},
		xmin=0,
		xmax=50,
		xlabel={Signal weight},
		xmajorgrids,
		every outer y axis line/.append style={black},
		every y tick label/.append style={font=\color{black}},
		ymin=1e-13,
		ymax=1,
		ymode=log,
		ylabel={IPA FER},
		ymajorgrids,
		legend style={at={(0.97,0.03)},anchor=south east,legend cell align=left,align=left,draw=black,font=\scriptsize}
		]
		\addplot[color=blue,solid] 
		table[row sep=crcr]{%
			9	3.99999999789458E-08\\
			10	7.00000000186662E-08\\
			11	2.09999999944976E-07\\
			12	4.20000000000975E-07\\
			13	7.39999999943564E-07\\
			14	1.55000000001682E-06\\
			15	2.60000000007476E-06\\
			16	4.6400000000002E-06\\
			17	8.11999999994484E-06\\
			18	1.44599999999384E-05\\
			19	2.47899999999968E-05\\
			20	4.32399999999999E-05\\
			21	8.2400000000038E-05\\
			22	0.0001406\\
			23	0.0002402\\
			24	0.0004466\\
			25	0.0008692\\
			26	0.0016296\\
			27	0.0030682\\
			28	0.0057372\\
			29	0.0103774\\
			30	0.0182364\\
			31	0.0311342\\
			32	0.0503792\\
			33	0.078681\\
			34	0.1181682\\
			35	0.1696414\\
			36	0.2339106\\
			37	0.308345\\
			38	0.3921004\\
			39	0.4807468\\
			40	0.5698528\\
			41	0.6546882\\
			42	0.7325406\\
			43	0.7997056\\
			44	0.8553744\\
			45	0.8994434\\
			46	0.9325092\\
			47	0.9564436\\
			48	0.9729684\\
			49	0.983698\\
			50	0.9906886\\
		};
		\addlegendentry{$A^{(172)}_{\textsf{PG}}$};
		
		\addplot[color=red,solid] 
		table[row sep=crcr]{%
			13	1.99999994343614E-09\\
			14	1.89999999067325E-08\\
			15	3.99999999789458E-08\\
			16	1.09999999997612E-07\\
			17	3.80000000022029E-07\\
			18	8.0999999996223E-07\\
			19	2.44999999998718E-06\\
			20	5.16999999999879E-06\\
			21	1.58000000000102E-05\\
			22	3.78000000000878E-05\\
			23	8.4799999999996E-05\\
			24	0.0001722\\
			25	0.0004072\\
			26	0.0009292\\
			27	0.0019152\\
			28	0.0039878\\
			29	0.0077052\\
			30	0.0145074\\
			31	0.0256604\\
			32	0.0433534\\
			33	0.069808\\
			34	0.1070512\\
			35	0.1564986\\
			36	0.218233\\
			37	0.2920032\\
			38	0.3757044\\
			39	0.4641992\\
			40	0.5542036\\
			41	0.6406466\\
			42	0.7200282\\
			43	0.7896106\\
			44	0.847241\\
			45	0.8934224\\
			46	0.9279766\\
			47	0.9532054\\
			48	0.9707834\\
			49	0.9823976\\
			50	0.9897482\\
		};
		\addlegendentry{$A^{(172)}_{\textsf{EPG}^{(8)}}$};
		
		\addplot[color=blue,dashed] 
		table[row sep=crcr]{%
			1	0\\
			2	0\\
			3	0\\
			4	0\\
			5	0\\
			6	2.79E-10\\
			7	1.953E-09\\
			8	7.811E-09\\
			9	2.3428E-08\\
			10	5.8557E-08\\
			11	1.28795E-07\\
			12	2.57529E-07\\
			13	4.78154E-07\\
			14	8.36567E-07\\
			15	1.393937E-06\\
			16	2.229745E-06\\
			17	3.445102E-06\\
			18	5.166334E-06\\
			19	7.548842E-06\\
			20	1.0781231E-05\\
			21	1.5089704E-05\\
			22	2.0742731E-05\\
			23	2.8055983E-05\\
			24	3.7397525E-05\\
			25	4.9193278E-05\\
			26	6.3932742E-05\\
			27	8.2174974E-05\\
			28	0.000104554829\\
			29	0.000131789445\\
			30	0.000164684985\\
			31	0.000204143624\\
			32	0.000251170766\\
			33	0.000306882516\\
			34	0.000372513359\\
			35	0.000449424073\\
			36	0.000539109854\\
			37	0.000643208638\\
			38	0.000763509622\\
			39	0.000901961954\\
			40	0.001060683603\\
			41	0.00124197036\\
			42	0.001448304982\\
			43	0.001682366439\\
			44	0.001947039244\\
			45	0.002245422847\\
			46	0.002580841055\\
			47	0.002956851447\\
			48	0.003377254752\\
			49	0.003846104139\\
			50	0.004367714386\\
		};
		\addlegendentry{$A^{(172)}_{\textsf{PG}}$, PIE};
		
		\addplot[color=red,dashed] 
		table[row sep=crcr]{%
			8	9.1E-14\\
			9	8.25E-13\\
			10	4.169E-12\\
			11	1.5444E-11\\
			12	4.6811E-11\\
			13	1.22963E-10\\
			14	2.89857E-10\\
			15	6.27479E-10\\
			16	1.267766E-09\\
			17	2.419041E-09\\
			18	4.398472E-09\\
			19	7.674272E-09\\
			20	1.2919603E-08\\
			21	2.1080329E-08\\
			22	3.3459046E-08\\
			23	5.1818066E-08\\
			24	7.8504281E-08\\
			25	1.16599153E-07\\
			26	1.70097362E-07\\
			27	2.4411796E-07\\
			28	3.45152241E-07\\
			29	4.81352871E-07\\
			30	6.62869194E-07\\
			31	9.0223407E-07\\
			32	1.214807957E-06\\
			33	1.619286429E-06\\
			34	2.138277794E-06\\
			35	2.798957912E-06\\
			36	3.633809885E-06\\
			37	4.6814568E-06\\
			38	5.98759627E-06\\
			39	7.606046129E-06\\
			40	9.599911264E-06\\
			41	1.2042882215E-05\\
			42	1.5020676877E-05\\
			43	1.8632637364E-05\\
			44	2.2993494856E-05\\
			45	2.8235316052E-05\\
			46	3.4509645697E-05\\
			47	4.1989860521E-05\\
			48	5.0873750859E-05\\
			49	6.1386347169E-05\\
			50	7.3783009681E-05\\
		};
		\addlegendentry{$A^{(172)}_{\textsf{EPG}^{(8)}}$, PIE};
		
		\end{axis}
		\end{tikzpicture}
	}
	\caption{FER performance of the IPA versus the weight of the signal vector for several protograph-based measurement matrices.}
	\label{fig:pg_all_tikz-upd}
\end{figure*}
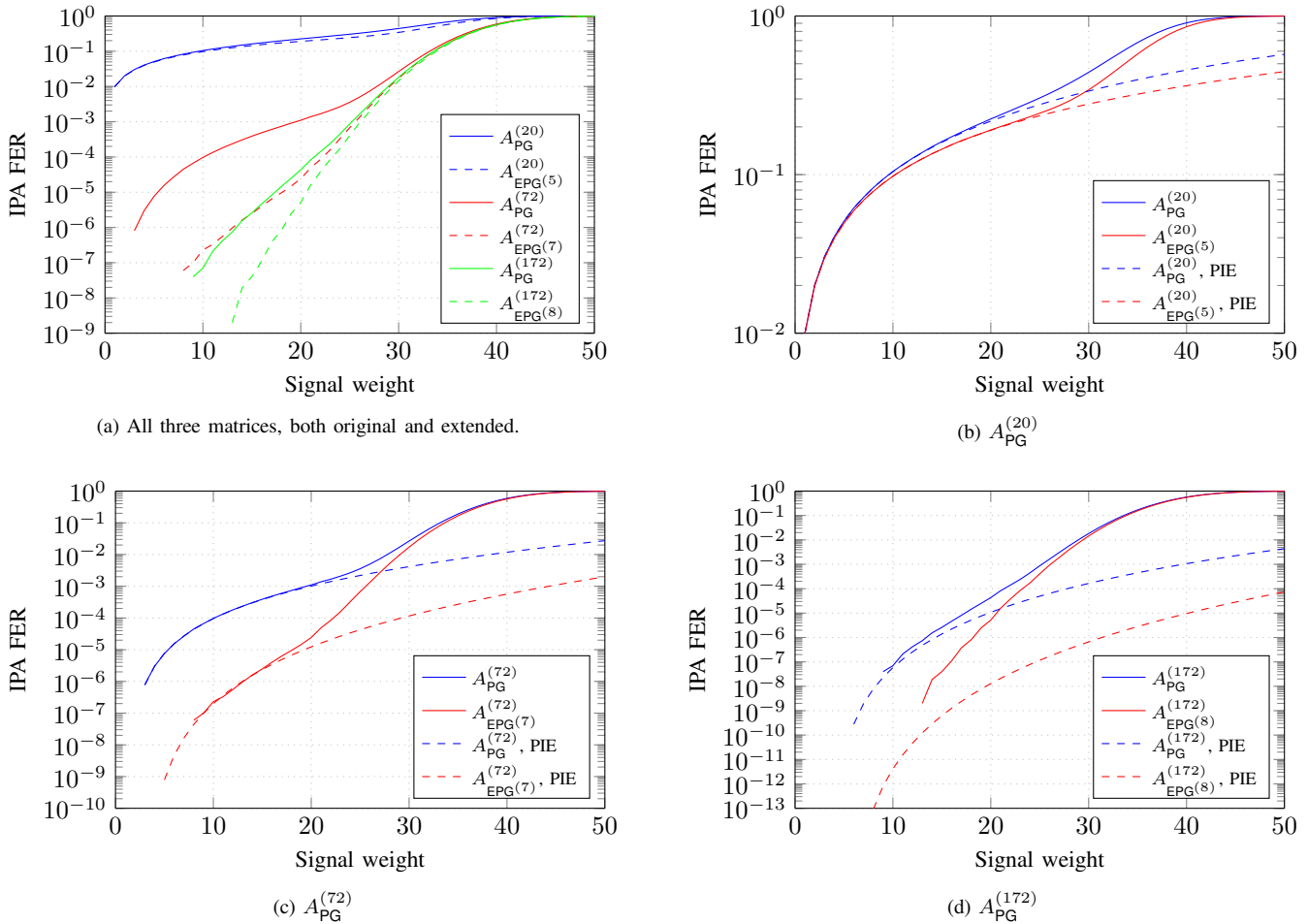

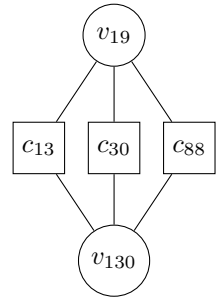
\begin{figure}
	\centering
	
	\begin{tikzpicture}
	\node at (1.0,+1.5) [circle,draw] (v19) {$v_{19}$};
	\node at (1.0,-1.5) [circle,draw] (v130) {$v_{130}$};
	
	\node at (0.0,0) [square,draw] (c13) {$c_{13}$};
	\node at (1.0,0) [square,draw] (c30) {$c_{30}$};
	\node at (2.0,0) [square,draw] (c88) {$c_{88}$};
	
	\draw (c13) -- (v19);
	\draw (c13) -- (v130);
	
	\draw (c30) -- (v19);
	\draw (c30) -- (v130);
	
	\draw (c88) -- (v19);
	\draw (c88) -- (v130);
	\end{tikzpicture}
	\caption{$\{v_{19}\}$ and $\{ v_{130} \}$ are both size-$1$ termatiko sets in $A^{(20)}_{\textsf{PG}}$.}
	\label{fig:pg20-smallest-termatikos}
\end{figure}

In order to see how changes in the termatiko set size spectra influence performance under the IPA, simulations were performed to estimate the frame-error rate (FER), i.e., the probability of failing to recover the original signal correctly for different values of its Hamming weight $w$. The results are presented in Fig.~\subref*{fig:pg_3plus3_tikz-upd}.  The three matrices $A^{(20)}_{\textsf{PG}}$, $A^{(72)}_{\textsf{PG}}$, and $A^{(172)}_{\textsf{PG}}$  represent different behavior after adding redundant rows. $A^{(20)}_{\textsf{PG}}$ is intrinsically bad and cannot be fixed as illustrated in Fig.~\ref{fig:pg20-smallest-termatikos}.  In particular, since both $v_{19}$ and $v_{130}$ are connected to $\{ c_{13}, c_{30}, c_{88} \}$ only, their values cannot be recovered. The reason being that if $v_{19} = \alpha$, $v_{130} = \beta$, and $\alpha + \beta > 0$, each of $c_{13}, c_{30}, c_{88}$ keeps only the sum $\alpha + \beta$, and there are infinitely many solutions for $\alpha$ and $\beta$. 
It is worth noting that this is not a failure of the IPA, since strictly speaking information has just been lost in the compression process (even an optimal recovery algorithm would fail here).

On the other hand, both $A^{(72)}_{\textsf{EPG}^{(\mathsf 7)}}$ and $A^{(172)}_{\textsf{EPG}^{(\mathsf 8)}}$ have increased termatiko distance (compared to $A^{(72)}_{\textsf{PG}}$ and $A^{(172)}_{\textsf{PG}}$, respectively), and show a significant improvement in the sparse region which shows the importance of designing measurement matrices with a high termatiko distance.

To better understand the curves, we also added lower bounds based on the principle of inclusion-exclusion. The following is a well-known result (see, e.g., \cite[p.~55]{bertsekas2002introduction}).
	
\begin{lemma}[Principle of Inclusion-Exclusion (PIE)]
	Assume that $\mathcal A_1, \mathcal A_2, \dotsc, \mathcal A_M$ are some arbitrary events and $\mathbb P \{ \cdot \}$ is a probability measure. Then
	\[
	 \mathbb{P} \left\{ \bigcup_{i=1}^M \mathcal A_i \right\} = \sum_{k=1}^{M} (-1)^{k-1} \left( \sum_{\scriptstyle I \subset [M] \atop \scriptstyle|I|=k} \mathbb{P} \left\{ \bigcap_{i \in I} \mathcal A_{i} \right\} \right) \,.
	\]
\end{lemma}

More precisely, we take into consideration only the $30$--$50$ smallest termatiko sets of a matrix, and then build a theoretical curve as if the matrix would contain only these termatiko sets and hence reconstruction fails if and only if the support of a signal contains any of these $30$--$50$ termatiko sets as a subset.

Assume that the termatiko sets of the matrix are $T_1, T_2, \dotsc$, and let $\mathcal A_i$ denote the event that a weight-$w$ subset of $[n]$ chosen uniformly at random is a superset of $T_i$. We remark that if $T_i \subset T_j$, then $\mathcal A_i \supset \mathcal A_j$ and $\mathcal A_i \cup \mathcal A_j = \mathcal A_i$. Therefore, if we include $T_i$ into the list of consideration, then there is no point to also include $T_j$. This pre-filtering can save computation time, as many termatiko sets are in fact subsets of others. Next, we consider only $M$ termatiko sets which we denote by $T_1, T_2, \dotsc, T_M$. Note that it is not required that the chosen termatiko sets are the $M$ smallest; any $M$ termatiko sets can be chosen and the result below will still be a correct lower bound. However, in our simulations, we took the $M$ smallest ones, for some $M$. This is also because we are particularly interested in a negative effect of the smallest termatiko sets.

With the aforementioned notation, the true FER is lower-bounded as

\ifonecolumn 
\begin{align*}
\mathsf{FER}(w) &= \mathbb{P} \left\{ \bigcup_{i} \mathcal A_i \right\} \geq \mathbb{P} \left\{ \bigcup_{i=1}^M \mathcal A_i \right\} 
\stackrel{\rm PIE}{=} \sum_{k=1}^{M} (-1)^{k-1} \left( \sum_{\scriptstyle I \subset [M] \atop \scriptstyle|I|=k} \mathbb{P} \left\{ \bigcap_{i \in I} \mathcal A_{i} \right\} \right) \\
&= \frac{1}{\binom{n}{w}} \sum_{k=1}^{M} (-1)^{k-1} \left( \sum_{\scriptstyle I \subset [M] \atop \scriptstyle|I|=k} \binom{n - \left| \bigcup_{i \in I} T_i \right|}{w - \left| \bigcup_{i \in I} T_i \right|} \right) \,.
\end{align*}
\else
\begin{align*}
\mathsf{FER}(w) &= \mathbb{P} \left\{ \bigcup_{i} \mathcal A_i \right\} \geq \mathbb{P} \left\{ \bigcup_{i=1}^M \mathcal A_i \right\} \\
&\stackrel{\rm PIE}{=} \sum_{k=1}^{M} (-1)^{k-1} \left( \sum_{\scriptstyle I \subset [M] \atop \scriptstyle|I|=k} \mathbb{P} \left\{ \bigcap_{i \in I} \mathcal A_{i} \right\} \right) \\
&= \frac{1}{\binom{n}{w}} \sum_{k=1}^{M} (-1)^{k-1} \left( \sum_{\scriptstyle I \subset [M] \atop \scriptstyle|I|=k} \binom{n - \left| \bigcup_{i \in I} T_i \right|}{w - \left| \bigcup_{i \in I} T_i \right|} \right) \,,
\end{align*}
where $\binom{a}{b}$  denotes the binomial coefficient of $a$ over $b$ and by convention $\binom{a}{b}= 0$ if $b<0$.
\fi

If the number of terms in the sum above becomes too large, we can use the truncated lower bound
\[
\mathsf{FER}(w) \geq \frac{1}{\binom{n}{w}} \sum_{k=1}^{2L} (-1)^{k-1} \left( \sum_{\scriptstyle I \subset [M] \atop \scriptstyle|I|=k} \binom{n - \left| \bigcup_{i \in I} T_i \right|}{w - \left| \bigcup_{i \in I} T_i \right|} \right) 
\]
for some $2L < M$ (the so-called Bonferroni inequality). This truncated expression becomes equal to the full inclusion-exclusion formula for weight $w$ if $\left| \bigcup_{i \in I} T_i \right| > w$ for all $I \subset [M]$, $|I| > 2L$. This simple fact allows to calculate better FER lower bounds for sparse signals faster. FER curves together with lower bounds are depicted in  Figs.~\subref*{fig:pg20_fer_tikz-upd} to \subref*{fig:pg172_fer_tikz-upd}.

Finally, we remark that the performance of the IPA and its comparison with other algorithms for efficient reconstruction of sparse signals have been investigated in \cite{RDVD12} (see Figs.~4 and 8), and we refer the interested reader to that work for such results. 

\section{Conclusion} \label{sec:conclu}
In this work, we have analyzed the failing patterns of the IPA  by introducing the concept of termatiko sets.  We have shown that the IPA fails to fully recover a nonnegative real signal $\vec x \in \Reals_{\geq 0}^n$ if and only if the support of $\vec x$ contains a nonempty termatiko set. Two heuristics to locate 
small-size termatiko sets were presented and analyzed.  Furthermore, a lower bound on the termatiko distance of column-regular binary measurement matrices with no $4$-cycles was derived. For the special case of measurement matrices equal to the parity-check matrices of array LDPC codes an upper bound on the termatiko distance equal to half of the best known upper bound on the minimum distance was given for column-weight at most $7$. For column-weight $3$ codes it was shown that the exact termatiko distance is $3$  and an explicit  formula for the multiplicity was provided.   
Adding redundant rows to the measurement matrix to improve IPA performance was considered as well, and an algorithm to efficiently search for such rows was outlined. The influence of small-size termatiko sets on IPA performance was illustrated through simulations and several numerical results for both specific and protograph-based ensembles of measurement matrices were presented, showing that having a termatiko distance strictly smaller than the stopping distance is not uncommon. In some cases, the termatiko distance can be as low as half the stopping distance. Thus, a measurement matrix (for the IPA)  should be designed to avoid small-size termatiko sets, which is considered as future work.
%
%
%
%
%
%


\section*{Appendix\\Proof of Theorem~\ref{thm:Hq3-T3}}

To prove Theorem~\ref{thm:Hq3-T3}, we need the following lemma.

\begin{lemma}\label{lem:ac-T3-struct}
	Assume $T = \{v_1, v_2, v_3\}$ is a termatiko set of size $3$ in $H(q,3)$. Define $N$ and $S$ analogously to Theorem~\ref{thm:termatikos}. Then, $S \neq \varnothing$, and for each $c \in N$, it holds that  $|\Neighbours[T]{c}| = 1$ and $|\Neighbours[S]{c}| > 0$.
\end{lemma}

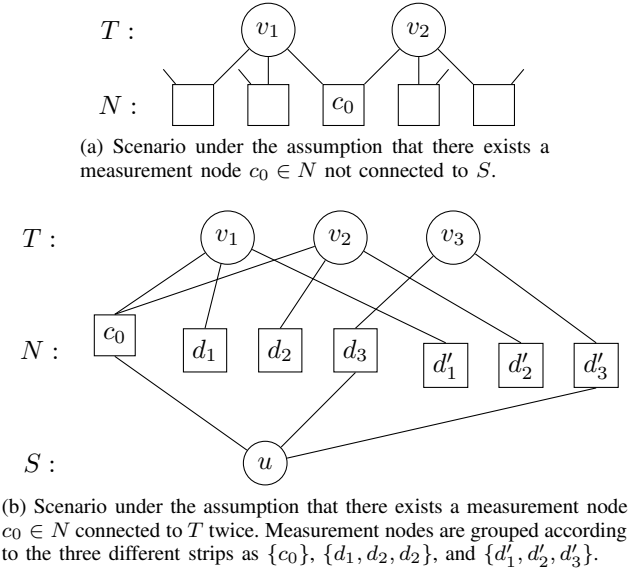
\begin{figure}
	\centering
	\subfloat[Scenario under the assumption that there exists a measurement node $c_0 \in N$ not connected to $S$.]{
		\label{fig:lem-T3-a-upd}
		\begin{tikzpicture}
		\node at (2.0,+1) [circle,draw] (v1) {$v_{1}$};
		\node at (4.0,+1) [circle,draw] (v2) {$v_{2}$};
		
		\node at (1.0,0) [square,draw] (c1) {\phantom{$c_0$}};
		\node at (2.0,0) [square,draw] (c2) {\phantom{$c_0$}};
		\node at (3.0,0) [square,draw] (c0) {$c_0$};
		\node at (4.0,0) [square,draw] (c4) {\phantom{$c_0$}};
		\node at (5.0,0) [square,draw] (c5) {\phantom{$c_0$}};
		
		\node at (0,+1) {$T:$};
		\node at (0,0) {$N:$};
		
		\draw (v1) -- (c1);
		\draw (v1) -- (c2);	
		\draw (v1) -- (c0);
		
		\draw (v2) -- (c0);
		\draw (v2) -- (c4);
		\draw (v2) -- (c5);
		
		\node at (0.5,0.6) (fn1) {};
		\draw (c1) -- (fn1);
		
		\node at (1.5,0.6) (fn2) {};
		\draw (c2) -- (fn2);
		
		\node at (4.5,0.6) (fn4) {};
		\draw (c4) -- (fn4);
		
		\node at (5.5,0.6) (fn5) {};
		\draw (c5) -- (fn5);
		\end{tikzpicture}
	}
	
	\subfloat[Scenario under the assumption that there exists a measurement node $c_0 \in N$ connected to $T$ twice. Measurement nodes are grouped according to the three different strips as $\{c_0\}$, $\{d_1,d_2,d_2\}$, and $\{d_1', d_2', d_3'\}$.]{
		\label{fig:lem-T3-b-upd}
		\begin{tikzpicture}
		\node at (2.5,+1.5) [circle,draw] (v1) {$v_1$};
		\node at (4.0,+1.5) [circle,draw] (v2) {$v_2$};
		\node at (5.5,+1.5) [circle,draw] (v3) {$v_3$};
		
		\node at (3.0,-1.5) [circle,draw] (u) {$u$};
		
		\node at (1.0,0.2) [square,draw] (c0) {$c_0$};

		\node at (2.2,0.0) [square,draw] (d1) {$d_1$};
		\node at (3.2,0.0) [square,draw] (d2) {$d_2$};
		\node at (4.2,0.0) [square,draw] (d3) {$d_3$};

		\node at (5.4,-0.2) [square,draw] (d1pr) {$d_1'$};
		\node at (6.4,-0.2) [square,draw] (d2pr) {$d_2'$};
		\node at (7.4,-0.2) [square,draw] (d3pr) {$d_3'$};
		
		\draw (v1) -- (c0.north);
		\draw (v2) -- (c0.north);
		\draw (u)  -- (c0.south);
		
		\draw (v1) -- (d1.north);
		\draw (v2) -- (d2.north);
		\draw (v3) -- (d3.north);
		\draw (u)  -- (d3.south);
		
		\draw (v1) -- (d1pr.north);
		\draw (v2) -- (d2pr.north);
		\draw (v3) -- (d3pr.north);
		\draw (u)  -- (d3pr.south);
		
		\node at (0,+1.5) {$T:$};
		\node at (0,0) {$N:$};
		\node at (0,-1.5) {$S:$};
		\end{tikzpicture}
	}
	\caption{Illustration for Lemma~\ref{lem:ac-T3-struct}.}
\end{figure}

\begin{IEEEproof}
	Assume first that some $c_0 \in N$ is not connected to $S$ (including the case $S = \varnothing$). Then, from Theorem~\ref{thm:termatikos}, $c_0$ is connected to $T$ at least twice (w.l.o.g.\ let $v_1$ and $v_2$ be these two variable nodes) and for any $c \in \Neighbours{v_1} \cup \Neighbours{v_2}$ (including $c = c_0$) it holds that $|\Neighbours[T]{c}| \geq 2$. See Fig.~\subref*{fig:lem-T3-a-upd} for illustration. As any two variable nodes share not more than one measurement node, we have $\Neighbours[]{v_1} \cap \Neighbours[]{v_2} = \{c_0\}$. Therefore, since $|\Neighbours[]{v_1}| = |\Neighbours[]{v_2}| = 3$, we have $|\Neighbours[]{v_1} \cup \Neighbours[]{v_2}| = 5$. Now, count number of edges between $T$ and $N$. On one hand, it is $|\Neighbours[]{v_1}| + |\Neighbours[]{v_2}| + |\Neighbours[]{v_3}| = 3+3+3=9$. On the other hand, it is not less than
	\[
		\sum_{c \in \Neighbours[]{v_1} \cup \Neighbours[]{v_2}} |\Neighbours[T]{c}| \geq 2 \, |\Neighbours[]{v_1} \cup \Neighbours[]{v_2}| = 10 \,.
	\]
	This contradiction shows that $S \neq \varnothing$ and that each $c \in N$ is connected to both $T$ and $S$.
	
	Now, we prove that each $c \in N$ is connected to $T$ only once. Again, assume to the contrary that some $c_0 \in N$ is connected to $T$ at least twice, w.l.o.g.\ to $v_1$ and $v_2$, and let $u \in S$ be connected  to $c_0$ (as we have just shown, such a $u$ exists). Recall that $\Neighbours[]{u} \subset N$ by definition of $S$ from Theorem~\ref{thm:termatikos}. Since $v_1$ and $v_2$ are both connected to $c_0$, they do not share any other measurement node. Also,  recall that each variable node is connected to three measurement nodes, each from a different strip. Hence, $v_1$ and $v_2$ are connected to different measurement nodes $d_1, d_2 \in N$ in another strip (different from the strip of $c_0$), and also to two different measurement nodes $d_1', d_2' \in N$ in the remaining strip. See Fig.~\subref*{fig:lem-T3-b-upd} for illustration. Now, $u$ cannot be connected to any of $d_1, d_2, d_1', d_2'$ as it already shares one measurement node with each of $v_1$ and $v_2$. Therefore, there exists a measurement node $d_3 \in \Neighbours[]{u}$ in the same strip that contains $d_1$ and $d_2$\revone{.} However, $d_3$ should be also connected to $T$. Thus, the only possibility left is that $d_3$ is connected to $v_3$. The same argument can be used for the strip that contains $d_1'$ and $d_2'$; it contains a node  $d_3'$, and $d_3'$ is connected to both $u$ and $v_3$. We have a contradiction, as $u$ and $v_3$ share two different measurement nodes (meaning that there should exist a cycle of length $4$ in the corresponding Tanner graph). Therefore, every $c \in N$ is connected to $T$ exactly once.
\end{IEEEproof}

From Lemma~\ref{lem:ac-T3-struct} it follows that $|N| = 9$ and that $v_1, v_2, v_3$ do not share any measurement nodes.

	Now, we turn to the proof of Theorem~\ref{thm:Hq3-T3}. 	From Theorem~\ref{thm:col-a-reg} we know that $h_{\rm min} \geq 3$; thus, we only need to prove the multiplicity result. Assume we have a termatiko set $T=\{v_1, v_2, v_2\}$, and denote $\Neighbours[]{v_1} = \{c_{11}, c_{21}, c_{31}\}$, where $c_{11}$, $c_{21}$, $c_{31}$ belong to the first, second, and third strip, respectively. Analogously, denote $\Neighbours[]{v_2} = \{c_{12}, c_{22}, c_{32}\}$ and $\Neighbours[]{v_3} = \{c_{13}, c_{23}, c_{33}\}$. As shown above, $|N| = |\{c_{11}, \dotsc, c_{33} \}| = 9$ (all these measurement nodes are different). As usual, we define the set $S$ as in Theorem~\ref{thm:termatikos}.
	
\begin{figure}
		\centering
		\begin{tikzpicture}[scale=.9]
		\node at (2.5,+2) [circle,draw] (v1) {$v_1$};
		\node at (4.0,+2) [circle,draw] (v2) {$v_2$};
		\node at (5.5,+2) [circle,draw] (v3) {$v_3$};
		
		\node at (3.0,-1.5) [circle,draw] (u213) {$u_{213}$};
		
		\node at (-1.0,0.2) [square,draw] (c11) {$c_{11}$};
		\node at ( 0.0,0.2) [square,draw] (c12) {$c_{12}$};
		\node at ( 1.0,0.2) [square,draw] (c13) {$c_{13}$};
		
		\node at (2.2,0.0) [square,draw] (c21) {$c_{21}$};
		\node at (3.2,0.0) [square,draw] (c22) {$c_{22}$};
		\node at (4.2,0.0) [square,draw] (c23) {$c_{23}$};
		
		\node at (5.4,-0.2) [square,draw] (c31) {$c_{31}$};
		\node at (6.4,-0.2) [square,draw] (c32) {$c_{32}$};
		\node at (7.4,-0.2) [square,draw] (c33) {$c_{33}$};
		
		\draw (v1) -- (c11.north);
		\draw (v1) -- (c21.north);
		\draw (v1) -- (c31.north);
		
		\draw (v2) -- (c12.north);
		\draw (v2) -- (c22.north);
		\draw (v2) -- (c32.north);
		
		\draw (v3) -- (c13.north);
		\draw (v3) -- (c23.north);
		\draw (v3) -- (c33.north);
		
		\draw (u213) -- (c12.south);
		\draw (u213) -- (c21.south);
		\draw (u213) -- (c33.south);
%
%
		\end{tikzpicture}
		\caption{Illustration for the proof of Theorem~\ref{thm:Hq3-T3} for $\vec \pi = (2,1,3)$ and hence $u_{213}$. Vertices $c_{11}, c_{12}, \dotsc, c_{33}$ are grouped according to the three different strips.}
		\label{fig:pi-u-explanation}
	\end{figure}
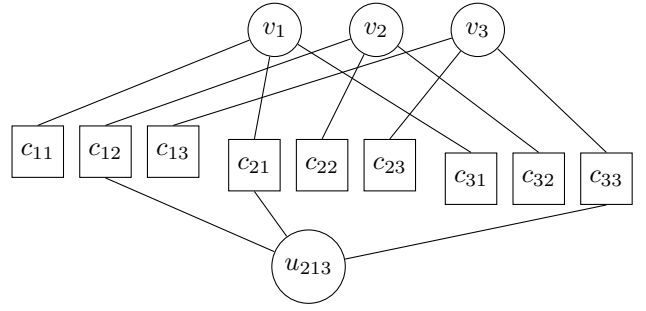

	In order not to share any two (or more) measurement nodes with any of $v_1, v_2, v_3$, each $u \in S$ should be connected to $c_{1 \pi_1}$, $c_{2 \pi_2}$, and $c_{3 \pi_3}$, where $\vec \pi = \vec \pi^{(u)} = (\pi_1, \pi_2, \pi_3)$ is some permutation of $\{1, 2, 3\}$. Thus, we will denote candidates for the set $S$ as $u_{\pi_1 \pi_2 \pi_3}$. In other words, $\Neighbours[]{u_{\pi_1 \pi_2 \pi_3}} = \{c_{1 \pi_1}, c_{2 \pi_2}, c_{3 \pi_3}\}$, from which it follows that there are $6$ candidates for $S$ and $|S| \leq 6$. Turn to Fig.~\ref{fig:pi-u-explanation} for illustration.
	
	Also, as each $c_{xy} \in N$ (for all $x,y \in \{1,2,3\}$) should be connected to $S$, $S$ should include some $u_{\vec \pi}$ with $\pi_x = y$. For example, $c_{11}$ should be connected to $S$, and thus either $u_{123}$ or $u_{132}$ (or both) should be present in $S$.
	
	Now, by applying the corresponding automorphism, we can set $v_1 = (0,0)$ and $v_2 = (2,j)$ for some $j \in \FF_q$.\footnote{Note that we have chosen the integer $2$ to make further numbers  look ``prettier'', although any nonzero value from $\FF_q$ would work here.}  With this notation, the support matrix of $T$ becomes
	\[
	\begin{bmatrix}
	0 & 2    & \cdot \\
	0 & 2+j  & \cdot \\
	0 & 2+2j & \cdot \\
	\end{bmatrix},
	\]
	where the dots stand for entries which are currently unknown.
	
	For the remainder of the proof, we exhaustively check all the cases and sub-cases, based on the assumption that some $u_{\pi_1 \pi_2 \pi_3} \in S$.
	As we noted before, since $c_{11}$ should be connected to $S$, either $u_{123}$ or $u_{132}$ (or both) should be in $S$. 
	
	\begin{enumerate}
		\item First, assume that $u_{123} \in S$, which means that $c_{11}$, $c_{22}$, and $c_{33}$ are connected to the same variable node ($u_{123}$), and thus the corresponding values in the support matrix will form an arithmetic progression. More precisely, the values $\{0, 2+j, \cdot\}$ should form an arithmetic progression,  and we immediately obtain the support matrix
		\[
		\begin{bmatrix}
		0 & 2    & \cdot \\
		0 & 2+j  & \cdot \\
		0 & 2+2j & 4+2j \\
		\end{bmatrix}.
		\]
		
		Further, $c_{12}$ should also be connected to $S$, and thus either $u_{213}$ or $u_{231}$ (or both) should be in $S$.
		
		\begin{itemize}
			\item Assuming that $u_{213} \in S$, we get that $c_{12}$, $c_{21}$, and $c_{33}$ should be connected to the same variable node $u_{213} \in S$, and hence the values $\{2, 0, 4+2j\}$ should form an arithmetic progression. From this we get that $4+2j=-2$ and then $j=-3$. The updated support matrix is
			\[
			\begin{bmatrix}
			0 & 2  & \cdot \\
			0 & -1 & \cdot \\
			0 & -4 & -2 \\
			\end{bmatrix}.
			\]
			
			\item Assuming that $u_{231} \in S$, we get that $\{2, \cdot, 0\}$ should form an arithmetic progression and then we can replace $\cdot$ by $1$. However, the values in the column of any support matrix should also form an arithmetic progression. Hence, the support matrix becomes
			\[
			\begin{bmatrix}
			0 & 2    & -2-2j \\
			0 & 2+j  & 1 \\
			0 & 2+2j & 4+2j \\
			\end{bmatrix}.
			\]
		\end{itemize}
		Other sub-cases are omitted for brevity.
		
		\item On the other hand, if we assume $u_{132} \in S$, then the values corresponding to $c_{11}$, $c_{23}$, and $c_{32}$ (i.e., $\{0, \cdot, 2+2j\}$) should form an arithmetic progression. From this we immediately obtain the updated support matrix
		\[
		\begin{bmatrix}
		0 & 2    & \cdot \\
		0 & 2+j  & 1+j \\
		0 & 2+2j & \cdot \\
		\end{bmatrix}.
		\]
		Again, we omit further sub-cases for brevity.
	\end{enumerate}
	
\begin{figure}
	\centering
	\begin{tikzpicture}
	\node at (5.5,.5) (root) {$\left[ \begin{smallmatrix}
		0 & 2    & \cdot \\
		0 & 2+j  & \cdot \\
		0 & 2+2j & \cdot \\
		\end{smallmatrix} \right]$};
	
	\node at (3,2.5) (u123) {$\left[ \begin{smallmatrix}
		0 & 2    & \cdot \\
		0 & 2+j  & \cdot \\
		0 & 2+2j & 4+2j \\
		\end{smallmatrix} \right]$};
	\draw [->] (root) -- (u123) node [midway,sloped,below,font=\scriptsize] (rootu123) {$u_{123} \in S$};
	
	\node at (8,2.5) (u132) {$\left[ \begin{smallmatrix}
		0 & 2    & \cdot \\
		0 & 2+j  & 1+j \\
		0 & 2+2j & \cdot \\
		\end{smallmatrix} \right]$};
	\draw [->] (root) -- (u132) node [midway,sloped,below,font=\scriptsize] (rootu132) {$u_{132} \in S$};
	
	\node at (2,5) (u123u213) {$\left[ \begin{smallmatrix}
		0 & 2  & \cdot \\
		0 & -1 & \cdot \\
		0 & -4 & -2    \\
		\end{smallmatrix} \right]$};
	\draw [->] (u123) -- (u123u213) node [midway,sloped,below,font=\scriptsize]  {$u_{213} \in S$};
	
	\node at (2,7.5) (u123u213u312) {$\left[ \begin{smallmatrix}
		0 & 2  & 4 \\
		0 & -1 & 1 \\
		0 & -4 & -2    \\
		\end{smallmatrix} \right]$};
	\draw [->] (u123u213) -- (u123u213u312) node [midway,sloped,below,font=\scriptsize] {$u_{312} \in S$};
	
	\node at (4,7.5) (u123u213u321) {$\left[ \begin{smallmatrix}
		0 & 2  & -2 \\
		0 & -1 & -2 \\
		0 & -4 & -2    \\
		\end{smallmatrix} \right]$};
	\draw [->] (u123u213) -- (u123u213u321) node [midway,sloped,below,font=\scriptsize] {$u_{321} \in S$};
	
	\node at (4.5,5) (u123u231) {{\boldmath $\left[ \begin{smallmatrix}
			0 & 2    & -2-2j \\
			0 & 2+j  & 1 \\
			0 & 2+2j & 4+2j \\
			\end{smallmatrix} \right]$}};
	\draw [->] (u123) -- (u123u231) node [midway,sloped,below,font=\scriptsize] {$u_{231} \in S$};
	
	\node at (7,5) (u132u213) {{\boldmath $\left[ \begin{smallmatrix}
			0 & 2    & 4+2j \\
			0 & 2+j  & 1+j \\
			0 & 2+2j & -2 \\
			\end{smallmatrix} \right]$}};
	\draw [->] (u132) -- (u132u213) node [midway,sloped,below,font=\scriptsize] {$u_{213} \in S$};
	
	\node at (9,5) (u132u231) {$\left[ \begin{smallmatrix}
		0 & 2 & \cdot \\
		0 & 2 & 1 \\
		0 & 2 & \cdot \\
		\end{smallmatrix} \right]$};
	\draw [->] (u132) -- (u132u231) node [midway,sloped,below,font=\scriptsize] {$u_{231} \in S$};
	
	\node at (7,7.5) (u132u231u312) {$\left[ \begin{smallmatrix}
		0 & 2 & -2 \\
		0 & 2 & 1 \\
		0 & 2 & 4 \\
		\end{smallmatrix} \right]$};
	\draw [->] (u132u231) -- (u132u231u312) node [midway,sloped,below,font=\scriptsize] {$u_{312} \in S$};
	
	\node at (9,7.5) (u132u231u321) {$\left[ \begin{smallmatrix}
		0 & 2 & 4 \\
		0 & 2 & 1 \\
		0 & 2 & -2 \\
		\end{smallmatrix} \right]$};
	\draw [->] (u132u231) -- (u132u231u321) node [midway,sloped,below,font=\scriptsize] {$u_{321} \in S$};
	
	\draw [->,dotted] (u123u231) -- (u123u213u312) node [at start,above,sloped,font=\scriptsize] {$j=-3$};
	\draw [->,dotted] (u123u231) -- (u132u231u312) node [at start,above,sloped,font=\scriptsize] {$j=0$};
	
	\draw [->,dotted] (u132u213) -- (u123u213u321) node [at start, above, sloped, font=\scriptsize] {$j=-3$};
	\draw [->,dotted] (u132u213) -- (u132u231u321) node [at start, above, sloped, font=\scriptsize] {$j=0$};
	\end{tikzpicture}
	
	\caption{Different cases for the proof of Theorem~\ref{thm:Hq3-T3}. Dotted arrows show special cases for particular values of the variable $j$.}
	\label{fig:cases-tree}
\end{figure}
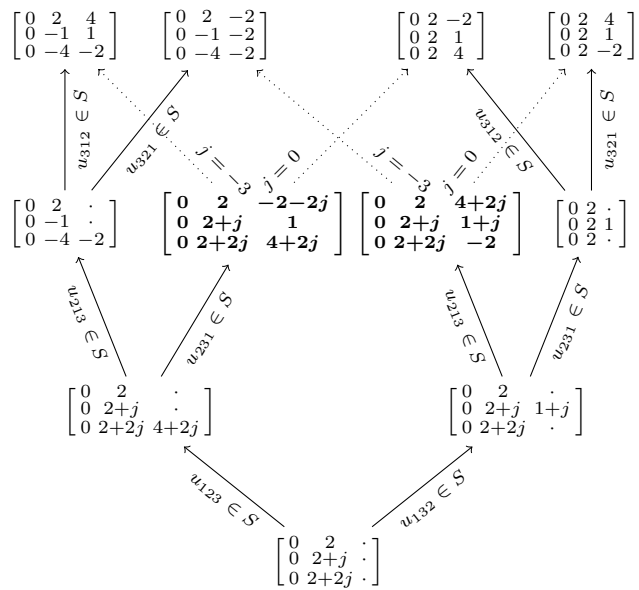

	The different cases can be represented as nodes in a search tree (see Fig.~\ref{fig:cases-tree}). Note that the branches in the tree are not mutually exclusive; but they cover all cases. This means that the same termatiko set can be obtained more than once. The two cases marked in bold in Fig.~\ref{fig:cases-tree} are general cases. Moreover, by setting $j=0$ or $j=-3$, we can obtain other particular cases (these relations are shown by dotted arrows). Note that branching stops at these general cases, as even these general forms already ensure that $\{v_1, v_2, v_3\}$ is a valid termatiko set. Other branches need to go one level deeper. Since the set of equations
	\[
	\begin{cases}
	-2-2j = 4+2j \,, \\
	1 = 1+j \,, \\
	4+2j = -2 \\
	\end{cases}
	\]
	do not have a solution for $q \geq 5$, these  two general cases do not intersect. 
However, we still need to check that the three columns are different in each of these two cases. The corresponding requirement for the first bold case is
	\[
	\begin{cases}
	0 \neq 2+j \,, \\
	0 \neq 2+2j \,, \\
	0 \neq -2-2j \,, \\
	0 \neq 4+2j \,, \\
	2 \neq -2-2j \,, \\
	2+j \neq 1 \,, \\
	\end{cases}
	\Leftrightarrow
	\begin{cases}
	j \neq -2 \,, \\
	j \neq -1\,. \\
	\end{cases}
	\]
	For the second bold case we get the requirement
	\[
	\begin{cases}
		0 \neq 2+j \,, \\
		0 \neq 2+2j \,, \\
		0 \neq 4+2j \,, \\
		0 \neq 1+j \,, \\
		2 \neq 4+2j \,, \\
		2+2j \neq -2 \,, \\
	\end{cases}
	\Leftrightarrow
	\begin{cases}
	j \neq -2 \,, \\
	j \neq -1\,. \\
	\end{cases}
	\]
	Therefore, in total there are $q-2$ choices for $j$ in each of the cases. This means that there are exactly $2(q-2)$ termatiko sets with fixed $v_1 = (0,0)$ and $v_2 = (2,\cdot)$. Any other termatiko set of size $3$ in $H(q,3)$ can be obtained by applying an automorphism (there are $q^2(q-1)$ such automorphisms). However, in this manner, we count each termatiko set $3!=6$ times. Thus, the total number of distinct size-$3$ termatiko sets in $H(q,3)$ is $q^2(q-1)(q-2)/3$.

\balance

\end{document}